\documentclass[aps,rmp,reprint,amsmath,amssymb,graphicx,longbibliography,floatfix]{revtex4-1}

\usepackage{hyperref}
\usepackage{times}
\usepackage{microtype}
\usepackage{graphicx}
\usepackage{amsfonts}
\usepackage{amsmath}

\usepackage{amsthm}
\usepackage{amssymb}
\usepackage{dsfont}
\usepackage{color}
\usepackage{soul}
\usepackage{bm}
\usepackage{mathptm}

\usepackage{tikz}
\usetikzlibrary{matrix,decorations.pathreplacing}

\definecolor{DarkRed}{rgb}{0.9,0,0}
\definecolor{DarkBlue}{rgb}{0,0,0.9}

\newcommand{\ve}[1]{{\boldsymbol{#1}}}

\newcommand{\boldsymbolB}{\boldsymbol{B}}

\newcommand{\boldsymbold}{\ve{d}} 
\newcommand{\boldsymbolf}{\ve{f}} 
\newcommand{\boldsymbolh}{\ve{h}} 
\newcommand{\boldsymbolj}{\ve{j}} 
\newcommand{\boldsymbolk}{{\ve{k}}} 
\newcommand{\vk}{{\ve{k}}} 
\newcommand{\boldsymbolm}{\ve{m}} 
\newcommand{\boldsymboln}{\ve{n}} 
\newcommand{\boldsymbolp}{\ve{p}} 
\newcommand{\vp}{{\ve{p}}} 
\newcommand{\boldsymbolq}{{\ve{q}}} 
\newcommand{\vq}{\ve{q}} 
\newcommand{\boldsymbolr}{\ve{r}} 
\newcommand{\vecr}{\ve{r}} 
\newcommand{\vecR}{\ve{R}} 
\newcommand{\boldsymbolR}{\ve{R}} 

\newcommand{\boldsymbolQ}{\ve{Q}} 

\newcommand{\boldsymbolJ}{\ve{J}} 
\newcommand{\boldsymbolM}{\ve{M}} 
\newcommand{\boldsymbolA}{\ve{A}} 
\newcommand{\boldsymbolS}{\ve{S}} 

\newcommand{\boldsymbolz}{\ve{z}} 

\newcommand{\vecj}{\ve{j}} 
\newcommand{\vecA}{\ve{A}} 

\newcommand{\vecm}{\ve{m}} 
\newcommand{\vecM}{\ve{M}} 

\newcommand{\boldsymbolsigma}{\boldsymbol{\sigma}} 

\newcommand{\e}[1]{\mathrm{e}^{#1}}

\newcommand{\eg}{\textit{e.g. }}
\newcommand{\etal}{\emph{et al.}}
\def\i{\mathrm{i}}

\newcommand{\Ow}{Odd-$\omega$~}
\newcommand{\ow}{odd-$\omega$~}
\newcommand{\Ew}{Even-$\omega$~}
\newcommand{\ew}{even-$\omega$~}

\newcommand{\bk}{{\bf k}}

\begin{document}

\title{Odd-frequency superconductivity}

\author{Jacob Linder}
\affiliation{Center for Quantum Spintronics, Department of Physics, Norwegian University of Science and Technology, NO-7491 Trondheim, Norway}
\author{Alexander V. Balatsky}
\affiliation{NORDITA, Roslagstullsbacken 23, SE-106 91 Stockholm, Sweden}
\affiliation{and Institute for Materials Science, Los Alamos National Laboratory, Los Alamos, NM 87545, USA}
\affiliation{and Department of Physics, University of Connecticut, Storrs, CT 06269, USA}

\date{\today{}}

\begin{abstract}
This article reviews  odd-frequency (odd-$\omega$) pairing with focus on superconducting systems. Since Berezinskii introduced the concept of odd frequency order in 1974 it has been viewed as an exotic and rarely occurring in nature. Here, we present a view that the Berezinskii state is in fact a ubiquitous superconducting order that is both non-local and odd in time. It appears under quite general circumstances in many physical settings including bulk materials, heterostructures and dynamically driven superconducting states, and it is therefore important to understand the nature of \ow pairing. We present the properties of \ow pairing in bulk materials, including possible microscopic mechanisms, discuss definitions of the \ow superconducting order parameter, and the unusual Meissner response of odd-frequency superconductors. Next, we present how \ow pairing is generated in hybrid structures of nearly any sort and focus on its relation to Andreev bound states, spin polarized Cooper pairs, and Majorana states. We overview how \ow pairing can be applied to non-superconducting systems such as ultracold Fermi gases, Bose-Einstein condensates, and chiral spin-nematics.  Due to the growing importance of dynamic orders in quantum systems we also discuss the emergent view that the \ow state is an example of phase coherent dynamic order. We  summarize the  recent progress made in understanding the emergence of \ow states in driven superconducting systems. More general view of \ow superconductivity suggests an interesting  approach to this state as a realization of the hidden order with inherently dynamic correlations that have no counterpart in conventional orders discussed earlier. We overview  progress made in this rapidly evolving field and  illustrate the ubiquity of the \ow states  and potential for future discoveries of these states in variety of settings. We sum up the general rules or, as we call them, design principles, to induce \ow components in various settings, using the SPOT rule. Since the pioneering prediction of \ow superconductivity by Berezinskii, this state has become a part of every-day conversations on superconductivity. To acknowledge this, we will call the \ow state a Berezinskii pairing as well in this article.
\end{abstract}

\maketitle

\tableofcontents{}


\section{Introduction}\label{sec:intro}

\subsection{Berezinskii symmetry relation}

The phenomenon of superconductivity, discovered more than 100 years ago, has stood the test of time. It remains today one of the most important and flourishing research areas of quantum condensed matter physics due to its allure both from a fundamental physics viewpoint and from a technological perspective. One fact which presumably has been a key reason for the sustained interest in this field is that superconductors demonstrate the unique quantum phenomena of a condensate in a macroworld. Superconductors discovered to date come in a variety of exotic forms. Conventional low-$T_c$ superconductors such as Al and Nb are well described by the seminal theory of Bardeen, Cooper, and Schrieffer (BCS) \cite{bardeen_pr_57} which is widely regarded as one of the major accomplishments in theoretical condensed matter physics.

As so often is the case in physics, symmetry is a cornerstone in the theory of superconductivity and in fact dictates the properties of the basic constituents of superconductors, the Cooper pairs. We will return in Sec. \ref{sec:symmetry} to the issue of symmetry in superconductors and why it is important. For now, we will be content with noting that the function which mathematically describes how the two electrons making up the Cooper pair correlate to each other depends on the position, spin, and time coordinate of these electrons. The time coordinate is usually disregarded, as in BCS theory. However, the symmetry property of a paired state  allows for the interesting possibility that the two electrons are not correlated at equal times and that they are instead correlated as the time separation grows. This is indeed accomplished if the correlation function is odd in time. For historic reasons  this novel type of superconducting correlations that are odd in relative time or frequency, is known as odd-frequency (odd-$\omega$) pairing.

To illustrate the richness of the universe of superconducting states  we start with the Berezinskii classification \cite{berezinskii_pisma_74,balatsky_prb_92}. A key object in discussion of superconductivity is the two-fermion correlation function $\Delta_{\alpha\beta, ab}({\bf r},t) = \langle \mathcal{T}_{t} c_{\alpha,a}({\bf r}, t)c_{\beta,b}(0,0)\rangle$  that describes the pairing correlations in superconductors. Here, $\mathcal{T}$ is the time-ordering operator, $\boldsymbolr$ and $t$ are the relative spatial and time coordinates of the electrons comprising the Cooper pair, $\{a,b\}$ denote any orbital/band degree of freedom, while $\{\alpha, \beta\}$ are spin indices of the two fermions in the correlator, respectively. This anomalous  two-fermion pairing amplitude will occasionally be referred to as a "Cooper pair amplitude" for simplicity. 

Berezinskii was the first \cite{berezinskii_pisma_74}, to our knowledge, to point out that due to the Fermi statistics of the operators that enter into a fermionic pairing state amplitude, there are symmetry constraints on the permutation properties of the two operators in the pairing state. More technical details will be given in the next section. We here introduce the parity of the Cooper pair with respect to relative coordinate inversion $P^*$:
\begin{align}
P^*\Delta_{\alpha\beta, ab}(\vecr,t)P^{*-1} = \Delta_{\alpha\beta, ab}(-\vecr,t)
\end{align}
with respect to time coordinate permutation $T^*$, resulting in a sign change of the relative time $t$:
\begin{align}
T^*\Delta_{\alpha\beta, ab}(\vecr,t)T^{*-1} = \Delta_{\alpha\beta, ab}(\vecr,-t)
\end{align}
with respect to spin permutation $S$:
\begin{align}
S\Delta_{\alpha\beta, ab}(\vecr,t)S^{-1} = \Delta_{\beta\alpha, ab}(\vecr,t)
\end{align}
and finally with respect to orbital index permutation $O$:
\begin{align}
O\Delta_{\alpha\beta, ab}(\vecr,t)O^{-1} = \Delta_{\alpha\beta, ba}(\vecr,t).
\end{align}
Using the permutation operations acting on spatial, time, spin and if present, orbital indices of the pair correlation (Cooper pairs), following Berezinskii \cite{berezinskii_pisma_74}, one can  show that the combined action of spin permutation, orbital index permutation,  orbital parity, and time permutation on the pairing amplitude $\Delta$ leads to a change in sign: $S P^* O T^* \Delta_{\alpha\beta,ab}({\bf r}, t) = - \Delta_{\alpha\beta,ab}({\bf r}, t) $. We write this condition symbolically as
 \begin{eqnarray}
 SP^*OT^* = -1
 \label{eq:SPOT1}
 \end{eqnarray}

We note that $P^*$ and $T^*$ are not the full space and time inversions. These operations merely permute the relative coordinates and times of the pairing correlator. The fact that operation of permuting $t \rightarrow -t$ is not equivalent to time reversal can be seen from the fact that if we apply true time reversal $T$ to $\Delta$ in above equations, we would convert  $\Delta$ to  $\Delta^{\dag}$. This is not the case for the Berezinskii constraint. Instead, $T^*$ is merely permuting the times of two particles in the pair. By same logic $P^*$ is not the full space inversion but the permutation of two coordinates of particles.

 With the binary possibilities for each of the symmetries $P^{*2} = T^{*2} = S^2 =1, $ (here we deal with integer spin systems) we find for a single band model there are $2^2 = 4$  possible superconducting states possible. For completeness we also give a table for the interorbit odd states $O = -1$.   With the inclusion of the multiorbital pairing one finds that there are $2^3 = 8$ overall pairing states possible. All possible superconducting states can are enumerated in this 8-fold classification. \Ow states have $T^*=-1$ and form a class that is distinct from the \ew class where $T^*=+1$.  For example, \ow superconductors include singlet $p$-wave and triplet $s$-wave pairing states.

For simplicity, in the rest of the paper we will drop the asterisk by $T^*$ and $P^*$. When we come to the cases where it is especially important to distinguish them from true parity and time reversal, we will explicity call them out to again highlight the difference.

To illustrate the symmetry relations between \ow and \ew pairing for now we will consider a single band (single orbit $(O=+1$ fixed) case. The resulting possible pairing states are shown in Table \ref{tab:spot}. An immediate consequence of this table is that, within same spin pairing state, one can use an external field, interface scattering, or external time dependent drive to convert the pairing symmetry from \ow into \ew and from \ow state to \ew states. The basic rule of conversion is to change the parity of two binary indices in the table at the same time so as to preserve the overall product $SPOT = -1$ that is fixed by Fermi statistics (Berezinskii rule). \footnote{It is often said that the \ow or  Berezinskii pairing is the consequence of the Pauli principle.  We here simply point out there are no simple commutation or anticommutation rules for operators taken at different times. Hence, the \ow state is possible due to a constraint on the time (or contour) ordered propagator and not due to Pauli principle.}.  This simple rule points to variety of ways to create Berezinskii states and the ubiquity of these states that result.  As will be discussed, one efficient way to generate \ow states is to {\it induce} \ow amplitudes as a result of scattering of conventional Cooper pairs.
 There are also scenarios which allow an \ow state as the global minimum of the free energy. Considering, for instance, a spin-triplet state $S = +1$, one can convert an \ew odd-parity state into an \ow even-parity state. A complete and interactive table demonstrating possible conversions including the orbital index is available as Supplementary Information to this review.

 \begin{center}
 	\begin{table}[]
 		
 		\caption{Symmetry properties of the anomalous two-fermion correlator also known as superconducting Gorkov function,  $\Delta_{\alpha \beta}$ under the operators $SPOT$ where we have fixed $O=+1$. The \ow states are those where $T\Delta
 = - \Delta$. Adapted from \cite{triola_prb_16}. \\}
 		
 		\label{tab:spot}
 		
 		\begin{tabular}{ccccc}
 			
 			\hline
 			
 			$S$ & $P^*$ & $O$ & $T^*$ & Total \\
 			
 			\hline
 			
 			+1 & +1 & +1 & -1 & -1\\
 			
 			+1 & -1 & +1 &  +1 & -1\\

 			-1 & +1 & +1 & +1 & -1\\
 			
 			-1 & -1 & +1 & -1 & -1\\
 			
 		\end{tabular}
 		
 	\end{table}
 \end{center}

\begin{center}
 	\begin{table}[]
 		
 		\caption{Symmetry properties of the  superconductor with $O = -1$. \\}
 		
 		\label{tab:spot2}
 		
 		\begin{tabular}{ccccc}
 			
 			\hline
 			
 			$S$ & $P^*$ & $O$ & $T^*$ & Total \\
 			
 			\hline
 			
 			+1 & -1 & -1 & -1 & -1\\
 			
 			+1 & +1 & -1 &  +1 & -1\\

 			-1 & -1 & -1 & +1 & -1\\
 			
 			-1 & +1 & -1 & -1 & -1\\
 			
 		\end{tabular}
 		
 	\end{table}
 \end{center}

Such a non-local pairing in time seems rather unusual at first glance. It essentially implies that the electrons must avoid each other in time so that there exists no correlation between them when their time-coordinates are equal. It is interesting to note that such a retardation effect in time is in fact also present in the microscopic mechanism underlying superconductivity in BCS-theory, namely electron-phonon scattering. It is responsible for two electrons ultimately attracting each other by interacting with the lattice and avoiding each other in time. However, it turns out that one can (somewhat miraculously)  get most of the properties of BCS superconductors by disregarding this retardation effect in BCS theory. In many cases, one obtains very good agreement with experimental data ignoring the time dependence of the pair correlations in BCS. In contrast, the retardation effect is inherent to the nature of \ow pairing that one simply can not ignore it for such a state. These strong retardation correlations  need to be captured to reveal the \ow state. It is arguably this aspect that makes it challenging to see \ow state using conventional computational and experimental tools.

With  the premise that \ow pairing is theoretically possible, a number of question rise to the surface. What is the underlying microscopic mechanism that can provide a pairing between electrons that is odd and non-local in time? In which materials could this be realized? Are the properties of \ow superconductivity the same as conventional superconductors? We will answer these questions and discuss other possible \ow states beyond superconductivity in this review.

\subsection{ Historic perspective}

Before proceeding to a detailed exposition of each of the topics related to \ow pairing, we now provide, to the best of our knowledge,  a timeline from the very conception of \ow pairing as a theoretical idea in 1974 to present-day state-of-the-art experiments.  Numerous experiments will be discussed later in the review.

It has been a  privilege to follow the evolution of this field from a stage where \ow Berezinskii pairing was considered rare and exotic to the present understanding where it has been realized that \ow pairing is generated under many circumstances:  nearly any type of hybrid structure involving a superconductor, in multiband superconductors, in driven superconductors with time dependent pairing states - in fact, as will be explained in this review, it seems harder to avoid it than to generate it. The abundant occurrence of \ow states is an important reason for why a solid understanding has become increasingly relevant. Our understanding of this concept has reached the point where we can make predictions and suggest new designs to create Berezinskii states.

Berezinskii \cite{berezinskii_pisma_74} was the first who  realized that a two-electron pairing correlation, with temporal coordinates $t_1$ and $t_2$, could be odd in $t_1-t_2$ or, as he introduced it, odd in frequency (the Fourier-transform of the relative coordinate $t_1-t_2$). This suggestion was motivated by the discovery of the superfluid phase in $^3$He, where he hypothesized that for sufficiently large spin-density fluctuations a pairing state with spin $S=1$ and even orbital angular momentum $L$ could arise. An example of an even orbital angular momentum pairing is the isotropic $s$-wave phase where $L=0$. It eventually turned out that this was not the superfluid pairing state that was realized in $^3$He (instead, it was an odd orbital angular momentum state $L=1$), but an idea had been sown. We note that the often mentioned Pauli principle for two operators taken at different times is misleading. There is no simple commutation or anticommutation rules for operators creating states at different times. However, the notion that the overall correlator can be odd in time to be consistent with the Fermi statistics, and hence to a degree is a consequence of a Pauli principle, is correct.

Further explorations of \ow pairing began in the beginning of the 1990s when Kirkpatrick and Belitz \cite{kirkpatrick_prl_91, belitz_prb_92} and Balatsky and Abrahams \cite{balatsky_prb_92} rekindled the interest in this type of superconductivity. A purely electronic mechanism that could generate spin-triplet \ow pairing of the same kind as Berezinskii suggested for $^3$He, in two-dimensional and disordered systems with strong quasiparticle interactions  was suggested in \cite{kirkpatrick_prl_91}. A new class of spin-singlet \ow superconductors was introduced in \cite{balatsky_prb_92} and their corresponding physical properties were enumerated. This included features which were diametrically opposite to the behavior of BCS superconductors, such as a finite zero-energy density of states that is enhanced beyond the value of the normal state instead of a gapped and fully suppressed density of states. The authors proposed that electron-phonon interaction might be sufficient to, in principle, provide the pairing glue required for \ow-pairing, but later showed that renormalization effects would prevent this unless a spin-dependence, such as antiferromagnetic fluctuations, was taken into account \cite{abrahams_prb_93, abrahams_prberratum_95}.

Other works soon appeared, where the existence of \ow pairing was discussed in the context of a two-channel Kondo system \cite{emery_prb_92}, the one-dimensional $t-J-h$ model \cite{balatsky_prb_93}, and the two-dimensional Hubbard model \cite{bulut_prb_93}. However, a severe problem with \ow superconductors was brought into evidence by Abrahams \etal~who pointed out that there was a sign problem \cite{abrahams_prb_95} with the Meissner effect: the superfluid density appeared to be negative and thus indicating an instability of the entire homogeneous \ow pairing state.

An exception to the Meissner problem was the works by Coleman, Miranda, and Tsvelik \cite {coleman_prl_93, coleman_prb_94, coleman_prl_95} who studied \ow-pairing in a Kondo lattice and heavy fermion compounds. Their idea was built on the interesting twist that \ow superconductivity was driven by an anomalous composite, staggered three-body scattering amplitude which turned out to provide a stable superconducting phase with a diamagnetic Meissner response. A similar resolution was also proposed in \cite{abrahams_prb_95}, who suggested that a stable Meissner state could be achieved by introducing a composite condensate (see Sec \ref{sec:composite}) where there existed a joint condensation of Cooper pairs and density fluctuations. Their work also addressed the subtle issue of how to define an appropriate order parameter for a condensate whose correlation function vanishes at equal times, as will be discussed in more detail later. On general grounds, for any  quantum mechanical system where a broken symmetry exists, it should be possible to describe it by a many-body Schr{\"o}dinger equation that is first order in time. Thus, for the stationary  broken symmetry state there should exist some equal time order encoded in the corresponding wavefunction.

During the end of the 90s, there was less activity in the field of \ow superconductivity with only a few works emerging \cite{coleman_jpcm_97, zachar_prb_01, hashimoto_jpsj_00, hashimoto_prb_01}, yet one of these \cite{belitz_prb_99} solved a crucial problem that had haunted the stability of the \ow-superconducting state. Belitz and Kirkpatrick showed that the sign problem with the Meissner effect in a bulk \ow-state could be resolved by carefully considering the the reality properties of the gap function (its real and imaginary parts), beyond what was possible to manipulate via global gauge transformations. In doing so, they identified the origin of an extra minus sign which would restore the thermodynamic stability of the \ow superconducting state and provide the usual Meissner response. This stability was confirmed in a later work by Solenov \etal~\cite{solenov_prb_09}.

\subsection{Design principles for Berezinskii state}

The field changed drastically in 2001 after a pioneering work by Bergeret, Volkov, and Efetov \cite{bergeret_prl_01} where they showed that \ow pairing would arise by placing a conventional BCS superconductor in contact with a ferromagnet. The approach by Bergeret \etal~was different from previous literature in that Bergeret \etal ~had focused on the possibility of having \ow pairing as a proximity effect instead of arising as an intrinsic bulk effect. It also had the desirable consequence that it demonstrated how it is possible to design \ow spin-triplet pairing systems by combining conventional  superconductors and ferromagnets in an appropriate fashion \cite{volkov_prl_03}. This work had an important impact on the field. It opened up a new vistas on routes to realize the \ow Berezinskii state as a result of scattering of conventional Cooper pairs into \ow correlations. Other groups soon followed and the number of publications on \ow pairing arising in hybrid structures underwent a sharp rise. We mention in particular that early key theoretical advances regarding the consequences of spin-triplet pairing with an \ow symmetry in superconductor/ferromagnet (S/F) structures were provided by Belzig, Buzdin, Eschrig, Nazarov, Volkov and co-workers with respect to for instance the density of states \cite{buzdin_prb_00,zareyan_prl_01}, superconducting spin-valve effects \cite{huertashernando_prl_02, bergeret_prb_03}, and supercurrents \cite{eschrig_prl_03}. The reader is referred to \cite{buzdin_rmp_05} for additional references.

Another key insight was provided in 2005 when Tanaka, Golubov and co-workers showed that \ow pairing could in fact be provided in proximity structures without any magnetism present. This was accomplished by utilizing $p$-wave superconductors instead of conventional BCS ones \cite{tanaka_prb_05a, tanaka_prb_05b, tanaka_prberratum_06}. Such superconductors are more scarce than the garden variety superconductors Al and Nb, and their pairing symmetry is often subject to a vigorous debate. However, the principle was clear: one did not necessarily have to break spin-rotational symmetry by an exchange field in a proximity structure to generate \ow pairing as suggested in \cite{bergeret_prl_01}. It would be sufficient to break translational symmetry simply by means of an interface in a heterostructure.

This insight had profound consequences as it also meant that phenomena such as Andreev bound-states occuring for certain crystallographic orientations of high-$T_c$ superconductors, widely regarded as clear evidence of the $d$-wave symmetry of these compounds, could be interpreted as a direct manifestation of \ow pairing. It also meant that \ow Berezinskii pairing would in fact appear in arguably the simplest conceivable superconducting hybrid structure: a ballistic normal metal coupled to a superconductor \cite{tanaka_prb_07a, tanaka_prl_07b, eschrig_jltp_07} due to broken translational symmetry.

A decade after the prediction of \ow pairing in S/F structures, the field was enjoying much attention and several proposals were put forth in terms of how one would be able to apply external control over \ow pairing, dictating when it would appear or not, by utilizing for instance spin-active interfaces \cite{linder_prl_09} or multilayered magnetic structures \cite{houzet_prb_07}. One of the key aspects fuelling the increasing interest in \ow pairing was the fact that its combined robustness toward impurity scattering and spin-polarized nature opened an intriguing possibility of utilizing it as a resilient way to achieve spintronics with superconductors \cite{eschrig_phystoday_11, linder_nphys_15}. 

Activity regarding the realization of \ow pairing in the bulk of a material was also revitalized, with authors investigating quasi-1D systems \cite{shigeta_prb_11, ebisu_prb_15}, strong-coupling superconductivity \cite{kusunose_jpsj_11a}, and systems with broken time-reversal symmetry \cite{matsumoto_jpsj_12}.

It has been realized that \ow pairing can also generally appear in superconductors where the fermions are characterized by an additional index, such as which band/orbital they belong to. This quantum number must consequentially be accounted for in the Pauli principle on equal footing as \eg the spin index. A series of works investigated this effect \cite{blackschaffer_prb_13b, asano_prb_15, aperis_prb_15,balatsky_unpub_17}, highlighting in particular the role played by hybridization between different bands,  orbitals or even leads of a heterostructures.

Another important research  direction recently formed that focuses on superconducting heterostructures with topological materials where \ow states  are also predicted \cite{blackschaffer_prb_12}. These structures were also shown to host \ow superconductivity due to an interplay of the proximity effect, spatial inhomogeneity and spin dependent interfaces \cite{triola_prb_14, triola_prl_16}.

The above discussion clearly points to the {\em design principles} for the \ow Berezinskii state. In all of the above examples conventional Cooper pairs are "converted" into Berezinskii pairs. We thus would expect that {\em any heterostructure} in the presence of conventional Cooper pairs will, with a certain probability, convert them into \ow pairs. For example,  the FM/SC heterostructures convert conventional s-wave singlet pairs ($ S = -1, P = +1, T  = +1, O = +1$) into spin triplet s-wave Berezinkii pairs  ($S = +1, P = +1, T = -1, O = +1 $).     If we are looking for conversion of even frequency pair to \ow pairs we would need to : a) start with conventional pairs, b) design the scattering process that changes one of the quantum numbers of the pair, and finally c) allow for retarded pairing in the analysis in order for Berezinskii state be probed. The only constraint on this "design approach" is the requirement $(SPOT)_{initial} = (SPOT)_{final} = -1$ as demanded by Berezinski state.  To keep the $SPOT$ product same one would need to change at least two parities simultaneously.   The only requirement is for macrostructure to induce matrix elements in the scattering to mix up states with different quantum numbers, \eg of different parity or spin or orbital index. One thus requires a change not only in $T$ parity, but also in other quantum numbers like $P$ (\eg for a SC heterostructure with a disordered metal) or $S$ (\eg for magnetically active interfaces).  Any known  examples of heterostructures and bulk \ow Berezinskii state induction given here obey these {\it design rules}.  The wealth of possibilities is indeed larger than what was considered to date. As we review specific examples, we will comment on that quantum numbers of the SPOT are changed on a case-by case basis. To illustrate this point, we can apply this principle to Josephson junctions. In that case, we can convert Cooper pairs ($S = -1, P = +1, T = +1, O = +1$) into Berezinskii spin singlet pairs ($S = -1, P = +1, T = -1, O = -1$) by considering the left and right lead as effective orbital indices. Hence it is possible to introduce the \ow pairs in  {\em conventional} Josephson junctions, as explained in more detail in Sec. \ref{sec:josephson} below.

\subsection{Berezinskii pairing as a dynamic quantum order}

Aside from heterostructures as a way to induce \ow states, a new direction for the design of \ow states is clear: the time domain. The proposal is to induce \ow Berezinskii states by driving the quantum systems dynamically  with external fields. Driven quantum matter  provides an interesting new possibility to create on-demand new quantum states. It is known that quantum states can develop nontrivial orders in time, as was shown \eg to be the case for time crystals (tX) \cite{wilczek_prl_12, zhang_nature_17, choi_nature_17}. We also know that the Berezinskii state, due to its intrinsic time-dependence, is a state where dynamics can be essential. Hence it is natural to expect a formation of the Berezinskii state in driven quantum systems.

Time dynamics is crucial for both \ow Berezinski pairing and tX. Yet how it enters into a description of the respective orders differ.  In case of the \ow state, one considers a two particle condensate $\langle \mathcal{T}c(t_1)c(t_2)\rangle$,  where correlations are {\em odd in relative time} $t = t_1 - t_2$. In the tX state, order-in-time occurs  in the mass or spin density. These quantities can be expressed as a  two fermion  correlation representing local spin or density.  As a result, the  tX state  exhibits dynamic order in the {\em "center of mass" time} $T_\text{cm} = t_1 + t_2$. The tX and Berezinskii states thus correspond to dynamic quantum order forming in the {\em center vs. relative} time. It is nevertheless important to emphasize that a tX state breaks time translational symmetry, whereas an \ow Berezinskii state does not necessarily do so.
 A more detailed discussion concerning the possible connections between tX and \ow Berezinskii pairing is given  in the section on Josephson effect, where one can  demonstrate the generation of a \ow cross junction pair amplitude that exhibits periodic Rabi-like oscillations \cite{balatsky_unpub_17}. The Berezinskii pairing state can also be induced in {\em any conventional superconductor} by applying time dependent drives \cite{triola_prb_16, triola_prb_17}.

\subsection{Berezinskii pairing and relation to other quantum order}

There is a priori no reason to expect that the \ow states are confined only to superconducting states. Hence the exploration of other \ow pairing states is only natural. We mention here briefly some possible connections of the \ow Berezinskii state to other unusual states of matter. One natural connection is to hidden order states. The prototypical example include the hidden order state in heavy fermion compounds like URu$_2$Si$_2$ \cite{mydosh_rmp_11}. Another example of the possible hidden order is the so-called pseudogap states of high-$T_c$ oxide superconductors \cite{norman_aip_05}. In both of these cases, we see well defined spectroscopic and thermodynamic features while lacking an understanding of what the possible order parameter is in the (pre)ordered phase. We know "conventional" orders described by equal spin-spin or charge-charge correlations functions that have equal time correlations can be easily measured. On the other hand, a state where conventional probes of equal-time spin and charge correlations fail to detect any order could posses an unconventional order. One possible explanation of hidden orders is to assume that these orders exhibit composite order or \ow order just like \ow superconductors. Thus one might take a broader view that any \ow state  represents a class of {\em hidden order}  states in that there are no equal time correlations.  Such a viewpoint has indeed been explored and led to the prediction that \ow pairing may occur in Bose-Einstein condensates \cite{balatsky_arxiv_14}, density waves \cite{kedem_arxiv_15}, Kondo systems \cite{coleman_physica_93, flint_prl_10, flint_prb_11, erten_prl_17} and spin nematics \cite{balatsky_prl_95} and will be summarized in Sec. \ref{sec:nonsc}.
Another intriguing observation, again demonstrating the fundamental relevance of \ow pairing in a variety of contexts, was that Majorana bound-states in superconducting structures inevitably would have to be accompanied by the presence of \ow correlations, indicating a strong relationship between them \cite{asano_prb_13, huang_prb_15}.

\subsection{Observables related to \ow pairing}

There are multiple features in \ow superconductivity that can help us identify \ow Berezinskii phase experimentally. Some earlier observations done at a time where their relation to \ow pairing was not known theoretically can today be taken as evidence of \ow superconductivity. An example of this, already alluded to above, is the observation of zero-bias conductance peaks in [110]-oriented YBCO (see \eg \cite{wei_prl_98}). At the time, it was taken as direct evidence of Andreev surface-states of $d$-wave superconductors, but today we know that it is also to be taken as evidence of \ow pairing due to the realization that Andreev surface-states are a manifestation of \ow superconductivity. In this sense, one could argue that \ow pairing was experimentally observed as early as 1966 by Rowell and McMillan \cite{rowell_prl_66, rowell_prl_73} who observed sharp resonances in the density of states in ballistic S/N bilayers. These resonances were 40 years later shown \cite{tanaka_prb_07a} to be a direct manifestation of \ow pairing.

More indirect evidence has also been put forth in terms of long-ranged supercurrents \cite{keizer_nature_06, khaire_prl_10, robinson_science_10} through strongly polarized and diffusive materials, which can only exist if carried by \ow Cooper pairs since these are immune precisely toward both impurity scattering and pair-breaking due to the Zeeman-field of a ferromagnet.
However, two recent advances have been made on the experimental arena regarding the direct observation of \ow pairing. The spectroscopic signatures of \ow Cooper pairs induced in a superconductor as seen in the density of states via STM-measurements were reported in \cite{dibernardo_natcom_15}, while the much debated paramagnetic Meissner response characteristic of \ow superconductivity was reported in \cite{dibernardo_prx_15}.

The development of the understanding, and not the least relevance, of \ow pairing since the proposition of Berezinskii has been adventurous. Not only do \ow states continue to intrigue us due to their unusual temporal properties, being non-local and odd in time, but their fundamental influence on both the electromagnetic response and spin properties of superconductors has made them more relevant for possible technological applicatons. There are previous reviews in the field which have dealt with various aspects of \ow pairing, such as its existence in S/F structures \cite{bergeret_rmp_05}, more general superconducting proximity systems \cite{golubov_jpcm_09}, and its relation to topology  \cite{tanaka_jpsj_12}. In this review, we aim to provide a comprehensive treatment of all known aspects of \ow pairing, be it bulk or proximity systems, and also cover the most recent activity in the field, not the least on the experimental arena.  At the same time, we are aware that the field of \ow Berezinskii pairing is a rapidly developing one and there are new examples and aspects of this unusual state that are continuously being discovered. We acknowledge this while attempting to provide a comprehensive review based on the accumulated knowledge and material  available  to date.



\section{Symmetries of superconducting states}\label{sec:symmetry}

\subsection{Why does the superconducting symmetry matter?}\label{sec:whysym}

Symmetry is a profound tool in physics which allows us to summarize the  information about how a system behaves, down to the microscopic level. Superconductivity is no exception and the symmetry characterizing the superconducting state of a material or composite system is of crucial importance. The main reason for this is that the so-called order parameter $\Delta$ characterizing the state must be a reflection of its environment, both in terms of the crystal lattice in which the electrons reside and the pairing interaction which allows them to form Cooper pairs. The order parameter symmetry thus provides direct information about the physical origin of superconductivity.

An example of this is Cooper pairs where the electrons have a relative angular momentum $L$ to each other, such as $p$-wave ($L=1$) pairing which allows the electrons to avoid each other more effectively in space. In this way, the Coulomb repulsion between the electrons can be partially mitigated and $p$-wave pairing is thus a relevant candidate for strongly interacting systems. When the electrons are correlated via \ow pairing, it means that they avoid each other in time instead of in space. This is also a viable way to reduce the Coulomb repulsion and strongly interacting systems have thus indeed over the years been investigated as potential hosts for \ow superconductivity \cite{balatsky_prb_93, coleman_prl_93}.

Since $\Delta$ also determines the gap of the quasiparticle excitations in a superconducting system, its symmetry properties can also be probed by how the quasiparticles behave. An example of this can be the manner in which the excitations transport charge or how they magnetically respond to external fields. \Ow superconductivity is ununsual in this regard as it not only can be gapless, but as it can even increase the Fermi level density of states of the superconducting state above its normal-state value. Determining the symmetry of the order parameter $\Delta$ is thus one of, if not the most important task that should be undertaken to understand the physics of a superconducting state.

\subsection{Berezinskii classification scheme}\label{sec:classification}



A superconducting two-fermion condensate is in general characterized by the
time-ordered expectation value
\begin{align}\label{eq:fundamentalf}
f_{\alpha\beta, ab}(\boldsymbolr_1,\boldsymbolr_2;t_1,t_2) = \mathcal{T}\langle \psi_{\alpha,a}(\boldsymbolr_1;t_1)\psi_{\beta,b}(\boldsymbolr_2;t_2)\rangle
\end{align}
known as the anomalous Green function which may be taken as a superconducting order parameter. Here, $\{\alpha,\beta\}$ denote the spin indices of the fermion annihilation field operators
$\psi_\alpha$ and $\psi_\beta$ whereas $\{(\boldsymbolr_i;t_i)\}$ denotes the position and time coordinate of field $i=1,2$. We have incorporated the indices $\{a,b\}$ which refer to
any other degrees of freedom characterizing the fermions, such as their band index in multiband systems, and we take $\{a,b\}$ to be precisely this band index in what follows for the sake of
concreteness. At equal times, $\mathcal{T}$ is to be understood as a normal ordering operator.

Superconducting order  that spontaneously breaks only the U(1) gauge symmetry below the critical temperature is known as
conventional superconductivity. Any other type of superconducting order is usually referred to as unconventional superconductivity.
A common example is superconducting order parameters that transform according to a non-trivial representation
of the point-group symmetry of the crystal for a given material. An $s$-wave order parameter is fully isotropic in $\boldsymbolk$-space
and thus is invariant under any symmetry operations of the crystal, causing the order parameter to transform according to the
trivial representation (identity transformation) of the point-group. A $d$-wave order parameter, on the other hand, transforms according to a non-trivial
representation. If the crystal structure lacks an inversion center, it is no longer possible to characterize the superconducting states in terms of their parity symmetry and the allowed order parameter symmetries in general become mixtures of even and odd parity components.

Now, the Pauli exclusion principle places restrictions on the symmetry properties of the anomalous Green function $f_{\alpha\beta,ab}(\boldsymbolr_1,\boldsymbolr_2;t_1,t_2)$ at equal times $t_1=t_2$.
It states that two half-integer spin fermions that are identical cannot simultaneously reside in the same quantum state and that the function characterizing
the state of the fermions must be \textit{odd} under an exchange of the particles
at equal times. This means that the anomalous Green function must always satisfy the following relation:
\begin{align}\label{eq:pauli}
f_{\alpha\beta,ab}(\boldsymbolr_1,\boldsymbolr_2;t_1,t_1) = -f_{\beta\alpha,ba}(\boldsymbolr_2,\boldsymbolr_1;t_1,t_1).
\end{align}
The symmetry of a superconducting state may thus be classified according to whether $f$ remains invariant or acquires a sign change upon exchanging the electron spins $\{\alpha,\beta\}$,
spatial coordinates $\{\boldsymbolr_1,\boldsymbolr_2\}$, or the band
indices $\{a,b\}$, at equal times $t_1=t_2$. For instance, a conventional BCS superconductor is invariant under an
exchange of the electron spatial coordinates:
\begin{align}
f_{\alpha\beta,ab}(\boldsymbolr_1,\boldsymbolr_2;t_1,t_1) =
f_{\alpha\beta,ab}(\boldsymbolr_2,\boldsymbolr_1;t_1,t_1)
\end{align}
but acquires a sign change under an exchange of
the spin coordinates:
\begin{align}
f_{\alpha\beta,ab}(\boldsymbolr_1,\boldsymbolr_2;t_1,t_1) =
-f_{\beta\alpha,ab}(\boldsymbolr_1,\boldsymbolr_2;t_1,t_1).
\end{align}

The complete set of possible symmetry combinations that are consistent with Eq. (\ref{eq:pauli})
are listed in Table \ref{tab:symmetries}. The \ow class of superconducting states are defined as those that have
an anomalous Green function acquiring a sign change upon interchanging the time-coordinates of
the Cooper pair, i.e. $f_{\alpha\beta,ab}(\boldsymbolr_1,\boldsymbolr_2;t_1,t_2) =
-f_{\alpha\beta,ab}(\boldsymbolr_1,\boldsymbolr_2;t_2,t_1)$. This means that the pairing correlation in fact vanishes at equal times $t_1=t_2$ since $f=-f$ is solved by $f=0$.

\begin{widetext}

\begin{center}

\begin{table}[]

\centering

\caption{Superconducting symmetries and their realization in materials and hybrid structures. S denotes a conventional BCS $s$-wave singlet superconductor, N denotes a normal metal, while F denotes a ferromagnetic metal. In the hybrid structure case, the table lists the symmetry of the superconducting correlations induced in the part of the structure that is not superconducting on its own, \eg in the N part of an S/N bilayer, as unconventional superconducting pairing can be generated by proximity to a fully conventional superconductor. Below, the examples for the odd spin symmetry are singlet whereas the even spin symmetry are triplets. Similarly, the examples for the even parity symmetry are $s$-wave while the odd parity symmetry examples are $p$-wave.\\}

\label{tab:symmetries}

\begin{tabular}{cccccc}

\hline

Spin $(S)$ & Parity $(P)$ &  Band $(O)$ &  Frequency $(T)$  &  Example: bulk & Example: hybrid \\

\hline

Odd & Even & Even & \Ew & Al, Nb \cite{bardeen_pr_57} & S/N  \cite{tanaka_prl_07b} \\

Odd & Even & Odd & \Ow & -  & Multiband S, JJ \cite{komendova_prb_15,balatsky_unpub_17}\\

Odd & Odd & Even & \Ow & -  & S/N \cite{tanaka_prb_07a} \\

Odd & Odd & Odd &  \Ew & -  & - \\

\hline
Even & Even & Odd & \Ew & -  & - \\

Even & Even & Even & \Ow & MgB$_2$ \cite{aperis_prb_15}  & S/F \cite{bergeret_prl_01} \\

Even & Odd & Odd & \Ow & Sr$_2$RuO$_4$ \cite{komendova_prl_17}  & - \\

Even & Odd & Even & \Ew  & Sr$_2$RuO$_4$ \cite{maeno_nature_94}  & S/F \cite{yokoyama_prb_07}\\

\end{tabular}

\end{table}

\end{center}

\end{widetext}

Rather than expressing the
anomalous Green function in terms of the individual space and time coordinates, it is common
in the literature to introduce a mixed representation with new center of mass and relative coordinates:
\begin{align}
f_{\alpha\beta,ab}(\boldsymbolr_1,\boldsymbolr_2;t_1,t_2) = f_{\alpha\beta,ab}(\boldsymbolr,\boldsymbolR; t, T)
\end{align}
where we introduced
\begin{align}
\boldsymbolr &= \boldsymbolr_1-\boldsymbolr_2,\; \boldsymbolR = (\boldsymbolr_1+\boldsymbolr_2)/2,\notag\\
t &= t_1-t_2, T = (t_1+t_2)/2.
\end{align}
For brevity of notation, assume in what follows that there is no dependence on the center of mass coordinate $\boldsymbolR$ or $T$ in the problem. The following argumentation is valid even if this simplification is not made, and the equations then hold true for each set of points $(\boldsymbolR,T)$. By Fourier-transforming the relative coordinates, one acquires a momentum dependent anomalous Green function via:
\begin{align}
f_{\alpha\beta,ab}(\boldsymbolp;t) = \int d\boldsymbolr \e{-\i\boldsymbolp\boldsymbolr} f_{\alpha\beta,ab}(\boldsymbolr;t).
\end{align}
In this mixed representation, the Pauli principle is expressed as
\begin{align}\label{eq:paulit}
f_{\alpha\beta,ab}(\boldsymbolp;0) = -f_{\beta\alpha,ba}(-\boldsymbolp,0)
\end{align}
since equal times $t_1=t_2$ give $t=0$. At first glance, this seems to indicate that the Green function must be odd under inversion of
momentum or exchange of spin coordinates. However, another
possibility exists, as may be seen by Fourier transforming the relative time coordinate and thus obtain an energy-dependent Green function
\begin{align}
f_{\alpha\beta,ab}(\boldsymbolp;E) = \int dt \e{\i Et} f_{\alpha\beta,ab}(\boldsymbolp;t).
\end{align}
Eq. (\ref{eq:paulit}) then reads:
\begin{align}\label{eq:pauliE}
\int dE f_{\alpha\beta,ab}(\boldsymbolp;E) = - \int dE f_{\beta\alpha,ba}(-\boldsymbolp,E).
\end{align}
Note that in all integrals, the limits are $[-\infty,\infty]$. This provides two ways to satisfy Eq. (\ref{eq:pauliE}). Either
\begin{align}
f_{\alpha\beta,ab}(\boldsymbolp;E) = -f_{\beta\alpha,ba}(-\boldsymbolp;E)
\end{align}
or
\begin{align}
f_{\alpha\beta,ab}(\boldsymbolp;E) = -f_{\beta\alpha,ba}(-\boldsymbolp;-E).
\end{align}
The latter possibility is referred to as \textit{odd-frequency pairing} or Berezinskii pairing. It is seen from the above equations that if the anomalous Green function is odd under
exchange of time coordinates [$t \to (-t)]$, it is also odd under a sign change of $E$.

The majority of the literature works with either Matsubara Green functions or retarded/advanced Green functions when dealing with \ow pairing, so we here explain the relation between these two approaches briefly. To simplify the notation, we here omit the band indices. In the Matsubara formalism, one defines
\begin{equation}
f^\text{M}_{\alpha\beta}(\mathbf{r}_1,\mathbf{r}_2;\tau_1,\tau_2) = \mathcal{T}\{ \langle \psi_\alpha(\mathbf{r}_1;\tau_1) \psi_\beta(\mathbf{r}_2;\tau_2)\rangle\},
\end{equation}
and after a Fourier-transformation to the mixed representation one has
\begin{align}
f^\text{M}_{\alpha\beta}(\vp;\i\omega_n) &= \int^\beta_0 \text{d}\tau \e{\i\omega_n\tau} f^\text{M}_{\alpha\beta}(\vp;\tau),\notag\\
f^\text{M}_{\alpha\beta}(\vp;\tau) &= \frac{1}{\beta} \sum_n \e{-\i\omega_n\tau} f^\text{M}_{\alpha\beta}(\vp;\i\omega_n),
\end{align}
with $\tau$ as a complex time, $\beta$ as inverse temperature, and frequencies $\omega_n = (2n+1)\pi/\beta$. In this technique, one may apply the same procedure as for the real-time Green functions and arrive at
\begin{align}
\sum_n [f^\text{M}_{\alpha\beta}(\vp;\i\omega_n) + f^\text{M}_{\beta\alpha}(-\vp;\i\omega_n)] = 0,
\end{align}
which also leads to the requirement that
\begin{equation}\label{eq:matsubara}
f^\text{M}_{\alpha\beta}(\vp;\i\omega_n) = -f^\text{M}_{\beta\alpha}(-\vp;-\i\omega_n).
\end{equation}
The real-time retarded and advanced Green functions may be obtained from the Matsubara Green function by analytical continuation as follows $(\delta\to0)$:
\begin{align}
\lim_{\i\omega_n\to E \pm\i\delta} f^\text{M}_{\alpha\beta}(\vp;\i\omega_n) = f^\text{R(A)}_{\alpha\beta}(\vp;E).
\end{align}

The Pauli-principle can also be expressed by the retarded and advanced anomalous Green functions by using Eq. (\ref{eq:matsubara}). To see this, we perform an analytical continuation on the right hand side of Eq. (\ref{eq:matsubara}), yielding
\begin{align}
\lim_{\i\omega_n \to E+\i\delta} f^\text{M}_{\alpha\beta}(\vp;\i\omega_n) &= f^\text{M}_{\alpha\beta}(\vp;E+\i\delta) \notag\\
&= f^\text{R}_{\alpha\beta}(\vp;E),
\end{align}
while the same operation on the left-hand side produces
\begin{align}
\lim_{\i\omega_n \to E +\i\delta} [-f^\text{M}_{\beta\alpha}(-\vp;-\i\omega_n)] &= -f^\text{M}_{\beta\alpha}(\vp;-E-\i\delta) \notag\\
&= -f^\text{A}_{\beta\alpha}(-\vp;-E).
\end{align}
Equating the two sides, we finally arrive at
\begin{equation}\label{eq:pauliRA}
f^\text{R}_{\alpha\beta}(\vp;E) = -f^\text{A}_{\beta\alpha}(-\vp;-E).
\end{equation}
Actually, this information is embedded already in the definitions of the retarded and advanced Green functions, and Eq. (\ref{eq:pauliRA}) may be verified by direct Fourier-transformation without going via Eq. (\ref{eq:matsubara}). It is also worth underscoring that the Matsubara technique is only valid for equilibrium situations, while the Keldysh formalism and the corresponding Green functions are viable also for non-equilibrium situations.
The distinction between odd- and even-frequency correlations for the retarded and advanced Green functions is now as follows:
\begin{align}\label{eq:defevenodd}
\text{Odd-frequency:}&\; f^\text{R}_{\alpha\beta}(\vp;E) = -f^\text{A}_{\alpha\beta}(\vp;-E),\notag\\
\text{Even-frequency:}&\; f^\text{R}_{\alpha\beta}(\vp;E) = f^\text{A}_{\alpha\beta}(\vp;-E).
\end{align}

\section{Symmetry classification of the \ow states}
\subsection{Symmetry properties of the linearized gap equation}\label{subsec:linearizedgap}
The symmetry classification of superconducting \ew states can be extended to include Berezinskii states \cite{Geilhufe2018}. Such a symmetry classification is usually done for the linearized gap equation, which holds close to the superconducting transition temperature. To incorporate retardation effects and with that an integration in $\omega$-space the Bethe-Salpeter equation or linearized Eliashberg equation is considered, which can be written in most general form \cite{riseborough2004heavy} as
\begin{multline}
 v \Delta_{\alpha\beta}(\vk,i\omega_n) = - \sum_{\gamma,\delta} \sum_{\vk'} \sum_m \Gamma_{\alpha\beta\gamma\delta}(\vk,\vk',i\omega_m,i\omega_n)\\ \times G_\gamma(\vk',i\omega_m)G_\delta(-\vk',-i\omega_m) \Delta_{\gamma\delta}(\vk',i\omega_m).
 \label{gap_equation}
\end{multline}
Equation \eqref{gap_equation} represents a linear eigenvalue equation of the form $v \Delta = \hat{V}\,\Delta$, where $\hat{V}$ denotes integration including the kernel
\begin{multline}
V_{\alpha\beta\gamma\delta}(\vk,\vk',i\omega_m,i\omega_n) = \Gamma_{\alpha\beta\gamma\delta}(\vk,\vk',i\omega_m,i\omega_n) \\ \times G_\gamma(\vk',i\omega_m)G_\delta(-\vk',-i\omega_m).
\end{multline}
$G_\gamma$ is a normal Green function for an electron with spin $\gamma$ and $\Gamma$ is the interaction vertex that depends on momenta, frequencies, spin and orbital indices. It is assumed that the symmetry of the crystal is reflected in the kernel $V$ and described by the symmetry group $\mathcal{G}$. Each eigenvector of \eqref{gap_equation} transforms as a basis function of an irreducible representation $\Gamma^p$ of $\mathcal{G}$ and the degeneracy of the corresponding eigenvalue is determined by the dimension of $\Gamma^p$, which will be denoted by $d_p$. Hence, the linearized gap equation can be reformulated as
\begin{equation}
 v^{p,\nu} \hat{\Delta}_{m}^{p,\nu} = \hat{V} \tilde{\Delta}_{m}^{p,\nu},
 \label{gap_equation2}
\end{equation}
where $m=1,\dots,d_p$ and $\nu=1,2,\dots$ counts over the multiple non-equivalent subspaces transforming as the same irreducible representation. The superconducting instability occurs when the largest eigenvalue $v^{p,\nu}$ is equal to unity. Even though the pairing potential is invariant under every symmetry transformation of the group $\mathcal{G}$, the dominating gap function itself is only invariant under a subgroup, represented by one of the irreducible representations of $\mathcal{G}$.
It is assumed that the gap function transforms similarly to a pairing wave function.
Considering spin-orbit coupling, each rotation in space (proper or improper) is connected to a specific rotation in spin space. Applying the transformation operator associated to a specific symmetry transformation $g\in\mathcal{G}$ gives
\begin{equation}
g \hat{\Delta}(\vk) = \hat{u}^T(g) \hat{\Delta}\left(\hat{R}^{-1}(g)\vk\right) \hat{u}(g).
  \label{rot_gap}
\end{equation}
Here, $\hat{R}(g)\in O(3)$ denotes the three-dimensional rotation matrix and $\hat{u}(g)\in SU(2)$ the corresponding rotation matrix in spin space for the transformation $g\in\mathcal{G}$.

To capture the symmetry of odd-frequency states, we make use of the operator $\hat{T}^*$ which corresponds to a permutation of the two times present in a particle-particle correlation function (here we reinstate the asterisk in $\hat{T}^*$ to distinguish it from true time reversal $\hat{T}$).  We discuss the transformation behavior under $\hat{T}^*$ for the anomalous Green function $F$, given by
\begin{figure}
\includegraphics[width=8cm]{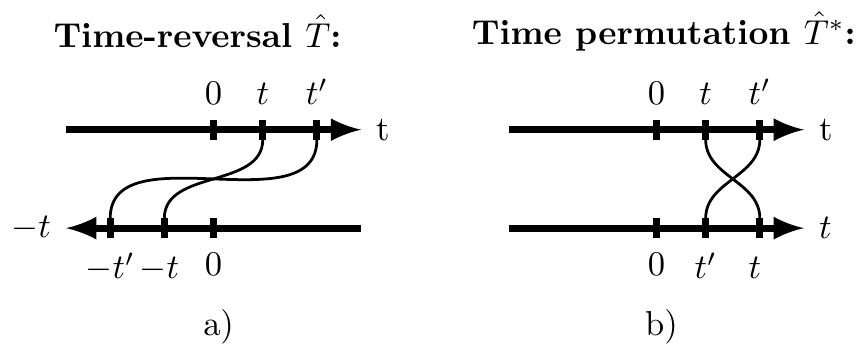}
\caption{(a) Time-reversal $\hat{T}$ and (b) time-permutation $\hat{T}^*$ symmetries for two
times; an odd-frequency superconductor breaks the time-permutation symmetry $\hat{T}^*$.}
\end{figure}
\begin{equation}
 F_{\sigma\sigma'}\left(\boldsymbol{k},t_1,t_2\right) = \left<\mathcal{T} c_\sigma\left(\boldsymbol{k},t_1\right) c_{\sigma'}\left(-\boldsymbol{k},t_2\right) \right>.
\end{equation}
Here, the operator $\mathcal{T}$ denotes the time-ordering operator, i.e.,
\begin{multline}
 F_{\sigma\sigma'}\left(\boldsymbol{k},t_1,t_2\right) =
\left<\theta(t_1-t_2) c_\sigma\left(\boldsymbol{k},t_1\right) c_{\sigma'}\left(-\boldsymbol{k},t_2\right) \right. \\- \left. \theta(t_2-t_1) c_{\sigma'}\left(-\boldsymbol{k},t_2\right)c_\sigma\left(\boldsymbol{k},t_1\right) \right>
\label{appb:eq1}
\end{multline}
Reversing $t_1$ and $t_2$ leads to
\begin{multline}
 F_{\sigma\sigma'}\left(\boldsymbol{k},t_2,t_1\right) =
\left<\theta(t_2-t_1) c_\sigma\left(\boldsymbol{k},t_2\right) c_{\sigma'}\left(-\boldsymbol{k},t_1\right) \right. \\- \left. \theta(t_1-t_2) c_{\sigma'}\left(-\boldsymbol{k},t_1\right)c_\sigma\left(\boldsymbol{k},t_2\right) \right>.
\label{appb:eq2}
\end{multline}
Hence, by comparing \eqref{appb:eq1} and \eqref{appb:eq2}, one obtains
\begin{equation}
F_{\sigma\sigma'}\left(\boldsymbol{k},t_2,t_1\right) = -F_{\sigma'\sigma}\left(-\boldsymbol{k},t_1,t_2\right).
\end{equation}
Since the gap $\hat{\Delta}$ is related to $\hat{F}$, a similar transformation behavior is present,
\begin{equation}
\Delta_{\sigma\sigma'}\left(\boldsymbol{k},t_2,t_1\right) = -\Delta_{\sigma'\sigma}\left(-\boldsymbol{k},t_1,t_2\right).
\label{tr_gap}
\end{equation}
It follows that $\hat{T}^*$ can be discussed without explicitly taking into account the times $t_1$ and $t_2$.

With respect to the interchange of the spin indices within the gap function, mediated by the operator $\hat{S}$, the gap function can be considered to be odd (singlet) or even (triplet). The resulting form of the gap in these cases is given by the antisymmetric matrix
\begin{equation}
 \hat{\Delta}(\vk) = i \Psi(\vk) \hat{\sigma}^y,
\end{equation}
for the spin singlet and by the symmetric matrix
\begin{equation}
 \hat{\Delta}(\vk) = i \left(\boldsymbol{d}(\vk)\cdot\boldsymbol{\sigma}\right) \hat{\sigma}^y,
\end{equation}
for the spin triplet. Following equations \eqref{rot_gap} and \eqref{tr_gap}, the transformation under group elements $g$ and under time-permutation $\hat{T}^*$ can be expressed in terms of transformations of $\Psi$ and $\boldsymbol{d}$ via
\begin{align}
\hat{g} \Psi(\boldsymbol{k}) &= \Psi\left(\hat{R}^{-1}(g)\boldsymbol{k}\right),\\
\hat{T}^* \Psi(\boldsymbol{k}) &= \Psi(-\boldsymbol{k}),
\end{align}
and
\begin{align}
\hat{g} \boldsymbol{d}(\boldsymbol{k}) &= \det\left(\hat{R}(g)\right)\hat{R}(g)\boldsymbol{d}\left(\hat{R}^{-1}(g)\boldsymbol{k}\right),\\
\hat{T}^* \boldsymbol{d}(\boldsymbol{k})&= -\boldsymbol{d}(-\boldsymbol{k}).
\end{align}
The gap function has to be odd under the application of a combination of parity operator ($\hat{P}$), spin interchange ($\hat{S}$) and time-permutation ($\hat{T}^*$),
\begin{equation}
\hat{P}\hat{S}\hat{T}^* = -1.
\label{ist}
\end{equation}
Therefore, by considering an even behavior under time-permutation $\hat{T}^*\hat{\Delta} = \hat{\Delta}$, a spin singlet gap (odd under spin interchange) restricts the gap function to be even under parity, whereas a spin triplet gap (even under spin interchange) has to come with an odd parity. However, allowing for an odd-time (or odd-frequency) dependence of the gap function, $\hat{T}^*\hat{\Delta} = -\hat{\Delta}$, brings the options of constructing an odd-parity spin singlet and an even-parity spin triplet gap.

Superconductivity is mediated by a pairing of electrons in $\vk$-space. In three dimensions it is possible to define 7 crystal systems and 32 crystal classes. The latter are connected to the 32 point groups. According to \eqref{tr_gap}, time-permutation $\hat{T}^*$ is a symmetry element of order 2, i.e., $(\hat{T}^*)^2=1$. Hence, incorporating $\hat{T}^*$, the symmetry group of the interaction kernel $\mathcal{G}$ can be extendes as follows,
\begin{equation}
\mathcal{G}^{\text{II}} = \mathcal{G} \oplus \hat{T}^* \mathcal{G},
\label{schub}
\end{equation}
where $\oplus$ denotes the set sum or unification of the two sets $\mathcal{G}$ and $\hat{T}^* \mathcal{G}$ ($\hat{T}^* \mathcal{G}$ is the element wise product of $\hat{T}^*$ and $g\in\mathcal{G}$).
If the pairing potential in Eq. \eqref{gap_equation} is invariant under $\hat{T}^*$, it is also invariant under every transformation contained in $\mathcal{G}^{\text{II}}$.

For the group order we obtain $\operatorname{ord}{\mathcal{G}^{\text{II}}} = 2\operatorname{ord}\mathcal{G}$. Furthermore, $\hat{T}^*$ commutes with every element $g\in\mathcal{G}$ and $\left\{E,T\right\}$ is an Abelian invariant subgroup of $\mathcal{G}^{\text{II}}$. $\mathcal{G}^{\text{II}}$ can be written as a semi-direct product of $\mathcal{G}$ and $\left\{E,\hat{T}^*\right\}$. It follows by induction \cite{hergert2018} that twice as many irreducible representations occur for $\mathcal{G}^{\text{II}}$ as they occur for $\mathcal{G}$. If $\Gamma_i$ is an irreducible representation of $\mathcal{G}$, then $\Gamma_i^+$ and $\Gamma_i^-$ are irreducible representations of $\mathcal{G}^{\text{II}}$, where the characters are given by
\begin{align}
 \chi_i^+(\hat{T}^*g) &= \chi_i(g),\label{trchi1} \\
 \chi_i^-(\hat{T}^*g) &= -\chi_i(g),\label{trchi2} \\
\end{align}
for all $g \in \mathcal{G}^{\text{II}}$.

\subsection{An example for the square lattice}
\begin{table*}[t!]
\begin{tikzpicture}
\draw[] (0,0) node{
\begin{tabular}{c cccccccccc|cccccccccc}
\hline\hline\\[-2ex]
  & $E$ & $2 C_2'$ & $2\sigma_v$ & $2C_2''$ & $2 \sigma_d$ & $2S_4$ & $2C_4$ & $I$ & $C_2$ & $\sigma_h$ & $T$ & $2 TC_2'$ & $2T\sigma_v$ & $2TC_2''$ & $2 T\sigma_d$ & $2TS_4$ & $2TC_4$ & $TI$ & $TC_2$ & $T\sigma_h$\\
  \hline
 $\text{A}^{+}_{\text{1g}}$ & 1 & 1 & 1 & 1 & 1 & 1 & 1 & 1 & 1 & 1& 1 & 1 & 1 & 1 & 1 & 1 & 1 & 1 & 1 & 1 \\
 $\text{A}^{+}_{\text{2g}}$ & 1 & -1 & -1 & -1 & -1 & 1 & 1 & 1 & 1 & 1 & 1 & -1 & -1 & -1 & -1 & 1 & 1 & 1 & 1 & 1\\
 $\text{B}^{+}_{\text{1g}}$ & 1 & 1 & 1 & -1 & -1 & -1 & -1 & 1 & 1 & 1 & 1 & 1 & 1 & -1 & -1 & -1 & -1 & 1 & 1 & 1\\
 $\text{B}^{+}_{\text{2g}}$ & 1 & -1 & -1 & 1 & 1 & -1 & -1 & 1 & 1 & 1 & 1 & -1 & -1 & 1 & 1 & -1 & -1 & 1 & 1 & 1\\
 $\text{E}^{+}_{\text{g}}$ & 2 & 0 & 0 & 0 & 0 & 0 & 0 & 2 & -2 & -2 & 2 & 0 & 0 & 0 & 0 & 0 & 0 & 2 & -2 & -2 \\
 $\text{A}^{+}_{\text{1u}}$ & 1 & 1 & -1 & 1 & -1 & -1 & 1 & -1 & 1 & -1 & 1 & 1 & -1 & 1 & -1 & -1 & 1 & -1 & 1 & -1\\
 $\text{A}^{+}_{\text{2u}}$ & 1 & -1 & 1 & -1 & 1 & -1 & 1 & -1 & 1 & -1& 1 & -1 & 1 & -1 & 1 & -1 & 1 & -1 & 1 & -1 \\
 $\text{B}^{+}_{\text{1u}}$ & 1 & 1 & -1 & -1 & 1 & 1 & -1 & -1 & 1 & -1 & 1 & 1 & -1 & -1 & 1 & 1 & -1 & -1 & 1 & -1 \\
 $\text{B}^{+}_{\text{2u}}$ & 1 & -1 & 1 & 1 & -1 & 1 & -1 & -1 & 1 & -1  & 1 & -1 & 1 & 1 & -1 & 1 & -1 & -1 & 1 & -1 \\
 $\text{E}^{+}_{\text{u}}$ & 2 & 0 & 0 & 0 & 0 & 0 & 0 & -2 & -2 & 2 & 2 & 0 & 0 & 0 & 0 & 0 & 0 & -2 & -2 & 2\\
  \hline
  $\text{A}^{-}_{\text{1g}}$ & 1 & 1 & 1 & 1 & 1 & 1 & 1 & 1 & 1 & 1  & -1 & -1 & -1 & -1 & -1 & -1 & -1 & -1 & -1 & -1\\
  $\text{A}^{-}_{\text{2g}}$ & 1 & -1 & -1 & -1 & -1 & 1 & 1 & 1 & 1 & 1
& -1 & 1 & 1 & 1 & 1 & -1 & -1 & -1 & -1 & -1\\
  $\text{B}^{-}_{\text{1g}}$ & 1 & 1 & 1 & -1 & -1 & -1 & -1 & 1 & 1 & 1
& -1 & -1 & -1 & 1 & 1 & 1 & 1 & -1 & -1 & -1 \\
  $\text{B}^{-}_{\text{2g}}$ & 1 & -1 & -1 & 1 & 1 & -1 & -1 & 1 & 1 & 1
 & -1 & 1 & 1 & -1 & -1 & 1 & 1 & -1 & -1 & -1 \\
  $\text{E}^{-}_{\text{g}}$ & 2 & 0 & 0 & 0 & 0 & 0 & 0 & 2 & -2 & -2
 & -2 & 0 & 0 & 0 & 0 & 0 & 0 & -2 & 2 & 2\\
  $\text{A}^{-}_{\text{1u}}$ & 1 & 1 & -1 & 1 & -1 & -1 & 1 & -1 & 1 & -1
& -1 & -1 & 1 & -1 & 1 & 1 & -1 & 1 & -1 & 1  \\
  $\text{A}^{-}_{\text{2u}}$ & 1 & -1 & 1 & -1 & 1 & -1 & 1 & -1 & 1 & -1
& -1 & 1 & -1 & 1 & -1 & 1 & -1 & 1 & -1 & 1 \\
  $\text{B}^{-}_{\text{1u}}$ & 1 & 1 & -1 & -1 & 1 & 1 & -1 & -1 & 1 & -1
& -1 & -1 & 1 & 1 & -1 & -1 & 1 & 1 & -1 & 1 \\
  $\text{B}^{-}_{\text{2u}}$ & 1 & -1 & 1 & 1 & -1 & 1 & -1 & -1 & 1 & -1
 & -1 & 1 & -1 & -1 & 1 & -1 & 1 & 1 & -1 & 1 \\
  $\text{E}^{-}_{\text{u}}$ & 2 & 0 & 0 & 0 & 0 & 0 & 0 & -2 & -2 & 2
& -2 & 0 & 0 & 0 & 0 & 0 & 0 & 2 & 2 & -2\\
 \hline\hline
\end{tabular}
};
\draw [decoration={brace,amplitude=0.5em},decorate,ultra thick,gray]
         (7.7,4.6) -- (7.7,-0.1);
\draw [decoration={brace,amplitude=0.5em},decorate,ultra thick,gray]
         (7.7,-0.4) -- (7.7,-5.1);
\draw[] (8,1.8) node[right]{$\boldsymbol{\hat{T}^*}$\textbf{-even}};
\draw[] (8,-2.2) node[right]{$\boldsymbol{\hat{T}^*}$\textbf{-odd}};
\end{tikzpicture}
\caption{Character table of the Shubnikov group ${D}^{\text{II}}_{4h}$.\label{ct_D4h}}
\end{table*}
As an example, the square lattice having the point group $D_{4h}$ is discussed. The group is generated by the elements $\left\{C_{4z},C_{2y},I\right\}$, where $C_{4z}$ denotes a four-fold rotation about the $z$-axis, $C_{2y}$ a two-fold rotation about the $y$-axis and $I$ the inversion. In total, $D_{4h}$ has 16 elements. Consequently, the corresponding Shubnikov group of the second kind ${D}^{\text{II}}_{4h}$ has 32 elements and is constructed according to Eq. \eqref{schub}. The character table of ${D}^{\text{II}}_{4h}$ is shown in Table \ref{ct_D4h}. For the irreducible representations the Mulliken notation is used \cite{mulliken1956}. Additionally, they are labeled with a superscript indicating an even (+) or odd (-) behavior with respect to time-permutation $\hat{T}^*$ according to Eq. \eqref{trchi1} and Eq. \eqref{trchi2}.

For the spin singlet gaps, the allowed irreducible representations occurring for a certain angular momentum $l$ can be determined by decomposing the representations of the orbital part only. In the following $D^l$ denote the irreducible representations of $SO(3)$, $D_x^l$ ($x=g,u$) the irreducible representations of $O(3)=\left\{E,I \right\}\times SO(3)$ and $D_{x,\pm}^l$ ($x=g,u$) the irreducible representations of $\left\{E,\hat{T}^* \right\}\times O(3)$. One obtains
\begin{align}
 \text{$s$-wave}:\,D^0_{g,+} &\simeq \text{A}_{\text{1g}}^+, \label{d4h_res1}\\
 \text{$p$-wave}:\,D^1_{u,-} &\simeq \text{A}_{\text{2u}}^- \oplus \text{E}_{\text{u}}^-,\label{d4h_res2} \\
 \text{$d$-wave}:\,D^2_{g,+} &\simeq \text{A}_{\text{1g}}^+ \oplus \text{B}_{\text{1g}}^+ \oplus \text{B}_{\text{2g}}^+ \oplus \text{E}_{\text{g}}^+.\label{d4h_res3}
\end{align}
Analogously, for the spin triplet gaps the allowed irreducible representations are found by decomposing the direct product belonging to the orbital part with $D^1_{g,-}$, representing the transformation properties of the spin triplet state,
\begin{align}
 \text{$s$-wave}:\,D^0_{g,+}  \otimes D^1_{g,-} &\simeq \text{A}_{\text{2g}}^- \oplus \text{E}_{\text{g}}^-, \label{d4h_res4}\\
 \text{$p$-wave}:\,D^1_{u,-} \otimes D^1_{g,-} &\simeq \text{A}_{\text{2u}}^+ \oplus \text{B}_{\text{2u}}^+ \oplus \text{B}_{\text{1u}}^+\oplus 2 \text{A}_{\text{1u}}^+ \oplus 2 \text{E}_{\text{u}}^+,\label{d4h_res5} \\
\text{$d$-wave}:\,D^2_{g,+} \otimes D^1_{g,-} &\simeq \text{A}_{\text{1g}}^- \oplus 2\text{A}_{\text{2g}}^- \oplus 2\text{B}_{\text{1g}}^- \oplus 2\text{B}_{\text{2g}}^- \oplus 4\text{E}_{\text{g}}^-.\label{d4h_res6}
\end{align}
\begin{table}[!h]
\begin{tabular}{lll}
\hline
\hline
\multicolumn{3}{l}{even-frequency}\\
\hline
$s$-wave: & $\text{A}^{+}_{\text{1g}}$ & $\Psi \simeq \text{const},\,k_x^2+k_y^2+k_z^2$\\
$p$-wave: & $\text{A}^{+}_{\text{1u}}$ & $\boldsymbol{d}\simeq k_x\boldsymbol{e}_x+k_y\boldsymbol{e}_y+k_z\boldsymbol{e}_z$\\
& $\text{A}^{+}_{\text{1u}}$ & $\boldsymbol{d}\simeq 2k_z\boldsymbol{e}_z-k_x\boldsymbol{e}_x-k_y\boldsymbol{e}_y$\\
& $\text{A}^{+}_{\text{2u}}$ & $\boldsymbol{d}\simeq k_y\boldsymbol{e}_x-k_x\boldsymbol{e}_y$\\
& $\text{B}^{+}_{\text{1u}}$ & $\boldsymbol{d}\simeq k_x\boldsymbol{e}_x-k_y\boldsymbol{e}_y$\\
& $\text{B}^{+}_{\text{2u}}$ & $\boldsymbol{d}\simeq k_y\boldsymbol{e}_x+k_x\boldsymbol{e}_y$\\
& $\text{E}^{+}_{\text{u}}$ & $\boldsymbol{d}\simeq k_x\boldsymbol{e}_z$\\
&  & $\boldsymbol{d}\simeq k_y\boldsymbol{e}_z$\\
& $\text{E}^{+}_{\text{u}}$ & $\boldsymbol{d}\simeq k_z\boldsymbol{e}_x$\\
&  & $\boldsymbol{d}\simeq k_z\boldsymbol{e}_y$\\
$d$-wave: & $\text{A}^{+}_{\text{1g}}$ & $\Psi \simeq 2k_z^2-k_x^2-k_y^2$\\
& $\text{B}^{+}_{\text{1g}}$ & $\Psi \simeq (k_x^2-k_y^2)$\\
& $\text{B}^{+}_{\text{2g}}$ & $\Psi \simeq k_x k_y$\\
& $\text{E}^{+}_{\text{g}}$ & $\Psi \simeq k_x k_z$\\
& & $\Psi \simeq k_y k_z$\\
\hline
\multicolumn{3}{l}{odd-frequency}\\
\hline
$s$-wave: &  $\text{A}^{-}_{\text{2g}}$ & $\boldsymbol{d}\simeq (k_x^2+k_y^2+k_z^2)\boldsymbol{e}_z$\\
 &  $\text{E}^{-}_{\text{g}}$ & $\boldsymbol{d}\simeq (k_x^2+k_y^2+k_z^2)\boldsymbol{e}_x$\\
 &    & $\boldsymbol{d}\simeq (k_x^2+k_y^2+k_z^2)\boldsymbol{e}_y$\\
$p$-wave: &  $\text{A}^{-}_{\text{2u}}$ & $\Psi \simeq k_z$ \\
  &  $\text{E}^{-}_{\text{u}}$ & $\Psi \simeq k_x$ \\
  &   & $\Psi \simeq k_y$ \\
$d$-wave: &  $\text{A}^{-}_{\text{1g}}$ & $\boldsymbol{d}\simeq k_y k_z \boldsymbol{e}_x - k_x k_z \boldsymbol{e}_y$ \\
 &  $\text{A}^{-}_{\text{2g}}$ & $\boldsymbol{d}\simeq k_x k_z \boldsymbol{e}_x + k_y k_z \boldsymbol{e}_y$ \\
 &  $\text{A}^{-}_{\text{2g}}$ & $\boldsymbol{d}\simeq (2 k_z^2 - k_x^2 - k_y^2) \boldsymbol{e}_z$ \\

 &  $\text{B}^{-}_{\text{1g}}$ & $\boldsymbol{d}\simeq k_y k_z \boldsymbol{e}_x + k_x k_z \boldsymbol{e}_y$ \\
 &  $\text{B}^{-}_{\text{1g}}$ & $\boldsymbol{d}\simeq k_x k_y \boldsymbol{e}_z$ \\
 &  $\text{B}^{-}_{\text{2g}}$ & $\boldsymbol{d}\simeq k_x k_z \boldsymbol{e}_x - k_y k_z \boldsymbol{e}_y$ \\
 &  $\text{B}^{-}_{\text{2g}}$ & $\boldsymbol{d}\simeq (k_x^2- k_y^2) \boldsymbol{e}_z$ \\

 &  $\text{E}^{-}_{\text{g}}$ & $\boldsymbol{d}\simeq k_x k_y \boldsymbol{e}_x$ \\
 &   & $\boldsymbol{d}\simeq k_x k_y \boldsymbol{e}_y$ \\

 &  $\text{E}^{-}_{\text{g}}$ & $\boldsymbol{d}\simeq k_z k_y \boldsymbol{e}_z$ \\
 &   & $\boldsymbol{d}\simeq k_z k_x \boldsymbol{e}_z$ \\

 &  $\text{E}^{-}_{\text{g}}$ & $\boldsymbol{d}\simeq (2 k_z^2 - k_x^2 - k_y^2) \boldsymbol{e}_x$ \\
 & & $\boldsymbol{d}\simeq (k_x^2 - k_y^2) \boldsymbol{e}_x$ \\
 &  $\text{E}^{-}_{\text{g}}$ & $\boldsymbol{d}\simeq (2 k_z^2 - k_x^2 - k_y^2) \boldsymbol{e}_y$ \\
 & & $\boldsymbol{d}\simeq (k_x^2 - k_y^2) \boldsymbol{e}_y$ \\
\hline
\hline
\end{tabular}
\caption{Even- and odd-frequency gap symmetries for the square lattice (${D}^{\text{II}}_{4h}$), considering $s$-, $p$- and $d$-wave superconductivity.\label{gap_sym}}
\end{table}
The obtained terms in \eqref{d4h_res1}-\eqref{d4h_res6} are in agreement with $\hat{P}\hat{S}\hat{T}^* = -1$ from Eq. \eqref{ist}. They reflect the cases:
\begin{itemize}
\item spin singlet, even parity, even time: Eq. \eqref{d4h_res1} and Eq. \eqref{d4h_res3}
\item spin singlet, odd parity, odd time: Eq. \eqref{d4h_res2}
\item spin triplet, odd parity, even time: Eq. \eqref{d4h_res5}
\item spin-triplet, even parity, odd time: Eq. \eqref{d4h_res4} and Eq. \eqref{d4h_res6}
\end{itemize}
Specific terms for gap symmetries are given in Table \ref{ct_D4h} and discussed subsequently.

\subsubsection{$s$-wave spin triplet}

\begin{figure}[b!]
\includegraphics[scale=0.8]{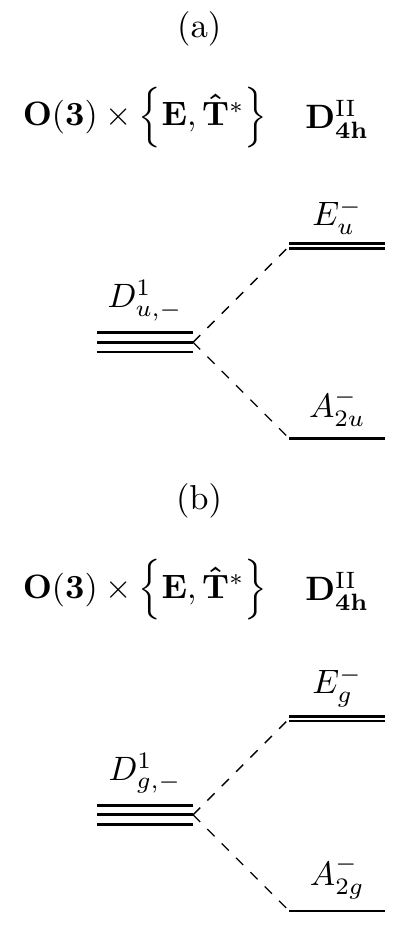}
\caption[fig]{\label{fig:trip:s:d4h} (Color online) Splitting of pairing states for a pairing potential with $D_{4h}^{\text{II}}$ symmetry. (a) $p$-wave spin-singlet and (b) $s$-wave spin-triplet.}
\end{figure}

As a first example, we consider the $s$-wave superconductivity. Whereas the conventional BCS theory describes a $s$-wave spin singlet pairing, even under $\hat{T}^*$, it is possible to construct a $s$-wave spin triplet that is odd under $\hat{T}^*$ Eq. \eqref{d4h_res4}. Under full rotational symmetry, a spin triplet transforms as the three-dimensional representation $D_{g,-}^1$. However, for the square lattice, the triplet state splits into $\text{A}_{\text{2g}}^-$ and $\text{E}_{\text{g}}^-$ as illustrated in Figure \ref{fig:trip:s:d4h}. Since the $z$-axis is chosen as principal axis, two linearly independent solutions belonging to $\text{E}_{\text{g}}^-$ are transforming as $\vk^2\boldsymbol{e}_x$ and $\vk^2\boldsymbol{e}_y$. Solutions belonging to $\text{A}_{\text{2g}}^-$ transform as $\vk^2\boldsymbol{e}_z$. The resulting gap functions are given by
\begin{align}
  \hat{\Delta}^{\text{E}_{\text{g}}^-}_1(\vk) &= -\vk^2\hat{\sigma}_z,\label{d4h:s:trip:1}\\
  \hat{\Delta}^{\text{E}_{\text{g}}^-}_2(\vk) &= i \vk^2 \hat{\sigma}_0,\label{d4h:s:trip:2}
\end{align}
and
\begin{equation}
  \hat{\Delta}^{\text{A}_{\text{2g}}^-}_1(\vk) = \vk^2\hat{\sigma}_x.\label{d4h:s:trip:3}
\end{equation}
As expected, all the three matrices are symmetric and thus even under spin interchange. They are even under parity since they contain $\vk^2$. But, they are odd with respect to the time-permutation introduced in Eq. \eqref{tr_gap}.
\subsubsection{$p$-wave spin singlet}
Another unconventional odd-frequency pairing is given by the $p$-wave spin singlet. Here, the three-dimensional odd-parity representation $D_{u,-}^1$ splits into the irreducible representations $\text{A}_{\text{2u}}^-$ and $\text{E}_{\text{u}}^-$. The gap transforms as $k_x$ and $k_y$ for $\text{E}_{\text{u}}^-$ and as $k_z$ for $\text{A}_{\text{2u}}^-$. The resulting superconducting gaps behave as
\begin{align}
  \hat{\Delta}^{\text{E}_{\text{u}}^-}_1(\vk) &= i k_x\hat{\sigma}_y,\\
  \hat{\Delta}^{\text{E}_{\text{u}}^-}_2(\vk) &= i k_y \hat{\sigma}_y,
\end{align}
and
\begin{equation}
  \hat{\Delta}^{\text{A}_{\text{2u}}^-}_1(\vk) = i k_z \hat{\sigma}_y.
\end{equation}
Clearly, the three matrices are anti-symmetric and odd under spin, odd under parity and also odd under time-permutation $\hat{T}^*$ according to Eq. \eqref{tr_gap}.

\section{\Ow pairing in bulk: mechanisms and properties}\label{sec:bulk}

The approach to induction of the \ow pairing generically falls into one of two categories. One approach is to induce the bulk \ow component due to some interaction. Another approach is to use the conversion of \ew pairs to \ow pairs in heterostructures and junctions where one uses the preestablished \ew state as a source of pairs that later are converted into \ow pairs. The latter approach, pioneered by Bergeret and collaborators, will be discussed in the subsequent chapter. Here, we focus on the possible intrinsic instabilities that drive \ow states.

\subsection{Microscopic mechanism}\label{sec:microscopic}

 The general framework for the symmetries of the \ow states was already covered in Sec. \ref{subsec:linearizedgap}.  We now will discuss possible specific mechanisms that might generate \ow states. In conventional BCS theory, it is electron-phonon coupling which provides the glue that binds electrons together in Cooper pairs. As a first attempt at identifying a microscopic mechanism for bulk \ow superconductivity, it is natural to consider the same type of interaction. Balatsky and Abrahams \cite{balatsky_prb_92} showed early on that an electron-electron interaction mediated by phonons could in principle lead to an \ow superconducting gap if the $\vk$-dependence of the phonon-mediated effective interaction $V_{\vk\vk'}$ was strong enough. To be more specific, the microscopic Eliashberg equations produce a matrix Green function of the form
\begin{align}
\hat{G}(\vk,\omega_n) = \frac{\i\omega_n Z_\vk(\omega_n)\tau_0 + W(\vk,\omega_n)\tau_1}{\omega_n^2 Z_k^2(\omega_n) + |W(\vk,\omega_n)|^2 + \epsilon_\vk^2}.
\end{align}
Here, $\tau_i$ are Pauli matrices in Nambu space, $\omega_n$ is the Matsubara frequency, $\vk$ is momentum, $\epsilon_\vk$ is the normal-state dispersion, and the one-loop self energies in the superconducting and normal channels are:
\begin{align}\label{eq:W}
W(\vk,\omega_n) &= -T_\text{temp}\sum_{n',\vk'} \frac{V_{\vk\vk'}(\omega_n-\omega_{n'}) W(\vk',\omega_{n'})}{\omega_{n'}^2 Z_{\vk'}^2(\omega_{n'}) + \epsilon_{\vk'}^2 + |W(\vk',\omega_{n'})|^2},\notag\\
\frac{1-Z_\vk(\omega_n)}{(\i\omega_n)^{-1}} &= T_\text{temp}\sum_{n',\vk'} \frac{V_{\vk\vk'}(\omega_n-\omega_{n'}) \i\omega_{n'}Z_{\vk'}(\omega_{n'})}{\omega_{n'}^2 Z_{\vk'}^2(\omega_{n'}) + \epsilon_{\vk'}^2 + |W(\vk',\omega_{n'})|^2}
\end{align}
Here, $T_\text{temp}$ is the temperature. The gap $\Delta$ used determined in, say tunneling spectra, is related to $W(\vk,\omega_n)$ and $Z_{\vk}(\omega_n)$ through $\Delta=W/Z$.
The effective interaction is written $V_{\vk\vk'}(\omega_n-\omega_{n'})$. Impurities have been neglected in the above equations for simplicity. Defining $\Omega=\omega_n-\omega_{n'}$ as a bosonic Matsubara frequency, an interaction mediated by phonons of the type
\begin{align}
V_{\vk\vk'}(\Omega) = \frac{2\alpha^2}{\pi} \int d\omega \frac{A_{\vk\vk'}(\omega)\omega}{\omega^2 + \Omega^2}
\end{align}
was shown in \cite{balatsky_prb_92} to produce an \ow gap under the assumption that the interaction has sufficiently strong $\vk$-dependence. Here, $\alpha$ is a measure of the coupling strength while $A$ is the spectral density.  In fact, the phonons do not contribute to the \ow pairing kernel of the expression for $W(\vk,\omega_n)$ in Eq. (\ref{eq:W}) if they are described in the Einstein approximation with a $\vk$-independent spectral density $A(\omega)$. \\

A crucial assumption in \cite{balatsky_prb_92} was that the renormalization of $Z_\vk$ in Eq. (\ref{eq:W}) caused by the interaction with phonons was neglected and $Z=1$ was instead set. The resulting odd-pairing kernel (odd in the quantities $\vk,\vk',\omega_n,\omega_n'$) can then be produced from the odd part of an interaction mediated by acoustic phonons with
\begin{align}
V_{\vk\vk'}(\Omega) = \alpha^2 \frac{c^2(\vk-\vk')^2}{c^2(\vk-\vk')^2+\Omega^2}.
\end{align}
This leads to a linearized gap equation
\begin{align}
\Delta(\vk,\omega_n) &= (4\alpha^2T_\text{temp}/c^2) \sum_{n',\vk'} \frac{\vk\cdot\vk'\omega_n\omega_n'}{(\vk^2+\vk'^2)^2 -4(\vk\cdot\vk')^2}\notag\\
&\times\frac{\Delta(\vk',\omega_{n'})}{\omega_{n'}^2+\epsilon_\vk^2}.
\end{align}

However, the effect of disregarding the renormalization turned out to be crucial. It was shown in a subsequent paper by Abrahams \etal~\cite{abrahams_prb_93} that a stable \ow singlet pairing state was unlikely to occur for a spin-independent effective potential coming \eg from a phonon interaction. The reason for this was precisely renormalization effects which would reduce the dressed coupling below a threshold value required to produce \ow superconductivity, irrespective of how strong the bare coupling was (this was originally pointed out by J. R. Schrieffer). It was instead argued in \cite{abrahams_prb_93} that if spin-dependent terms are added to the interaction, coming for instance from antiferromagnetic fluctuations that are present in \eg high-$T_c$ superconductors or other strongly correlated systems, this difficulty could be overcome. Specifically, they considered a general spin- and frequency-dependent electron-electron coupling
\begin{align}
g(\alpha k; \beta k'; \gamma p; \delta p') = g_c(k-p)\delta_{\alpha\beta}\delta_{\gamma\delta} + g_s(k-p)\sigma^i_{\alpha\beta} \sigma^i_{\gamma\delta}
\end{align}
where $\alpha,\beta,\ldots$ are spin indices while $k,p,\ldots$ are four-vectors and $\sigma^i$ are the Pauli matrices. Moreover, $g_c$ is the density-coupling while $g_s$ is the spin-dependent coupling. In such a scenario, the Eliashberg equations in the spin singlet $l$-wave channel become ($T_\text{temp}$ is temperature):
\begin{align}
\Delta_l(\omega_n) &= -\pi T_\text{temp}\sum_{n'} [g^l_c(\omega_n-\omega_{n'}) - 3g^l_s(\omega_n-\omega_{n'})]\notag\\
&\times \frac{\Delta_l(\omega_{n'})}{|Z(\omega_n)||\omega_{n'}|},\notag\\
Z(\omega_n) &= 1 - \pi T_\text{temp} \sum_{n'} [g^0_c(\omega_n-\omega_{n'}) + 3g^0_s(\omega_n-\omega_{n'})]\notag\\
&\times\frac{\omega_{n'}}{\omega_n|\omega_{n'}|}.
\end{align}
The key observation here is the different sign with which the spin-dependent coupling $g_s$ enters in the above equations. The sign difference provides the possibility of density and spin couplings adding in the pairing channel simultaneously as they oppose each other in the normal self-energy channel, so that $Z\sim 1$ or even $Z<1$ could be satisfied.

\begin{figure}[h!]
\includegraphics[scale=0.8]{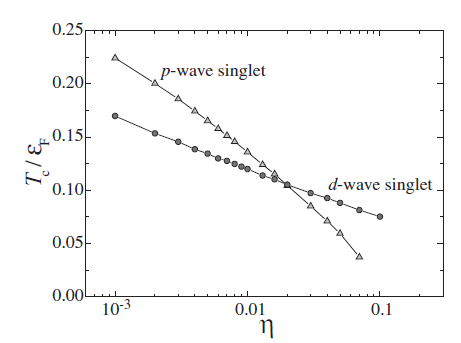}
\caption[fig]{\label{fig:fuseya} (Color online) Transition temperature $T_c$ for $p$- and $d$-wave spin-singlet pairing as a function of $\eta$. Figure adapted from \cite{fuseya_jpsj_03}.}
\end{figure}

Precisely an interaction mediated by spin fluctuations was later considered by Fuseya \etal~\cite{fuseya_jpsj_03} as a possible scenario for realizing \ow $p$-wave singlet pairing near the quantum critical point ($T_\text{temp}\to 0$ boundary between antiferromagnetic and superconducting phases) in CeCu$_2$Si$_2$.
The effective interaction considered mediated by spin fluctuations was taken to have the form
\begin{align}
V(\vq,\i\omega_m) = g^2\chi(\vq,\omega_m) = \frac{g^2N_F}{(\eta + A\boldsymbolr^2 + C|\omega_m|}
\end{align}
where $g$ is the coupling constant, $N_F$ the DOS at the Fermi level, $\eta$ is a measure of an inverse correlation length in the presence of magnetic correlations, $C$ is a constant, and $\boldsymbolr^2 = 4+2(\cos q_x + \cos q_y)$ in two dimensions. Such a pairing interaction had been used previously by Monthoux and Lonzarich \cite{monthoux_prb_99} to discuss strong coupling effects on superconducting order induced by critical antiferromagnetic fluctuations.
The linearized gap equation in the weak-coupling approximation serves
as the starting point for determining the favored superconducting state:
\begin{align}
\Delta(\vk,\i\omega_n) = -T_\text{temp}\sum_{\vk',\omega_{n'}} \frac{V(\vk-\vk',\i\omega_n-\i\omega_{n'})}{\xi_{\vk'}^2 + |\omega_{n'}|^2}\Delta(\vk',\i\omega_{n}),
\end{align}
where $\xi_\vk$ is the quasiparticle energy measured from the chemical potential. Following
\cite{fuseya_jpsj_03}, the pairing interaction can be further decomposed as
\begin{align}
V(\vk-\vk',\i\omega_n) = \sum_l V_l(\i\omega_n)\phi_l^*(\vk)\phi_l(\vk'),
\end{align}
where $\phi_l(\vk)$ are basis functions of irreducible representations of
the point group of the system and we defined
\begin{align}
V_l(\i\omega_n) = \sum_{\vk,\vk'}\phi_l(\vk) V(\vk-\vk',\i\omega_n)\phi_l^*(\vk').
\end{align}
The linearized gap equation may also be written out for each partial-wave component as
\begin{align}
\lambda(T)\Delta_l(\i\omega_n) = -T_\text{temp}\sum_{\vk',\omega_{n'}} \frac{V_l(\i\omega_n-\i\omega_{n'})}{\xi_{\vk'}^2 + |\omega_{n'}|^2}\Delta_l(\i\omega_{n'})
\end{align}
where $\Delta(\vk,\i\omega_n) = \sum_l \Delta_l(\i\omega_n)\phi_l(\vk)$. For spin-singlet pairing, the gap function
has to satisfy
\begin{align}
\Delta_d(\vk\,\i\omega_n) = \Delta_d(-\vk,\i\omega_n) = \Delta_d(\vk,-\i\omega_n)
\end{align}
for the $d$-wave orbital symmetry and
\begin{align}
\Delta_p(\vk,\i\omega_n) = -\Delta_p(-\vk,\i\omega_n) = -\Delta_p(\vk,-\i\omega_n)
\end{align}
for the $p$-wave case. Here, the eigenvalue $\lambda(T_\text{temp})$ determines the transition temperature via the condition $\lambda(T_c)=1$. By solving the linearized gap equation in the weak-coupling approximation
numerically with 512 Matsubara frequencies, the transition temperature $T_c$ could be determined for various pairing states. The transition temperature for the $p$-wave singlet and $d$-wave singlet state as a function of $\eta$ is shown in Fig. \ref{fig:fuseya}, and demonstrates that the \ow superconducting bulk state is indeed favorable for $\eta\simeq 0.02$ and smaller.

Kusunose \etal~\cite{kusunose_jpsj_11a} considered further aspects of bulk \ow superconductivity in strong-coupling
electron-phonon systems within the context of the Holstein-Hubbard model. The authors found numerical evidence
of an \ow state being realized, but cautioned that self-energy and vertex corrections were not included in their treatment,
which could potentially influence the conclusion. Shigeta \etal~\cite{shigeta_prb_09} also considered a possible bulk \ow pairing state on a triangular lattice, which we cover in more detail in Sec. \ref{sec:materials}. Shigeta \etal~have also theoretically examined a possible bulk \ow superconducting state appearing in the presence of a staggered field \cite{shigeta_prb_12}, where the latter suppresses the in-plane spin susceptibility and enhances the charge susceptibility, in addition to lattice models relevant for quasi-1D organic superconductors \cite{shigeta_jpsj_13}.



\subsection{The order parameter}\label{sec:op}

The question concerning the very existence of the order parameter for the \ow pairing deserves a special discussion. If the bulk \ow state can be realized, there has to be a set of attributes associated with the phase: an order parameter, wavefunction of the ground state,  a phase stiffness $\rho$, free energy difference between normal and ordered state $F_s - F_n$, and a Josephson energy associated with the phase difference across a Josephson junction. Moreover, if a quantum mechanical system with a broken symmetry satisfies a many-body Schr{\"o}dinger equation (which is first order in the time-derivative operator), there should exist some form of equal time order encoded in the corresponding wavefunction solving that equation.

On the other hand, one can take the view of  \ow state as a {\it dynamic order}.  Thus one might ask why the inherently dynamic order would have any of the attributes above developed in a stationary state or equilibrium ground state. In practice, much literature on \ow pairing, particularly in the context of hybrid structures, is using the Green function approach and hence deals with time dependent functions that can vanish at equal times. In this way, the question regarding the nature of the order parameter and wavefunction of the \ow state is muted. Technically one can proceed with \ow states without even asking the question concerning the existence of a steady equal-time order parameter. Nevertheless, if the Berezinskii state is a quantum phase of matter, there should exist a proper wavefunction, order parameter, and other ingredients that one expects when discussing such a phase. For completeness, we will lay out what has been discussed to date regarding this matter.

One approach to address the question about the order parameter in the \ow state is to ask what the equal time correlations are that control the pairing state. In other words we are looking for the {\em time independent} operators whose expectation value would represent the condensate that exists in the \ow state. One proposal, and the only one we know to date, was made in Refs. \cite{abrahams_prb_95,dahal_njp_09}. The proposal is to treat the \ow pairing anomalous correlator $F(t)$ at small times and use the time derivative as a definition for the equal time order parameter. Indeed, if $F(t) = \langle \mathcal{T}_t c(t)c(0)\rangle \sim K t$, where $K$ is a constant, is an odd function of time one can assume that at small time expansions (real time at temperature $T_\text{temp}= 0$ or Matsubara time for finite $T_\text{temp}$) is
\begin{equation}\label{EQ:OP1}
  \partial_t F(t) = K.
\end{equation}
For the purpose of qualitative discussion we use simplified notation and do not write all the other indices that are implied. To define the order parameter for the \ow state one has to use equations of motion for the fermion operator under the assumption of some Hamiltonian. On general grounds, using the equations of motion for $i \partial_t c(t) = [H,c(t)]$ one obtains a contribution in the commutator that arise from the kinetic energy terms. This contribution is irrelevant - instead, the interesting terms that yield a non-trivial result come from the interaction terms in the full Hamiltonian. For example, for the spin-fermion model, the interaction term.

\begin{align}
H_{int} = J \sum_{\boldsymbolr_n} S^i (\boldsymbolr_n) c^{\dagger}_{\alpha} (\boldsymbolr_n)\sigma ^i_{\alpha \beta} c_{\beta}(\boldsymbolr_n)
\end{align}
where $J$ sets the energy scale of the spin-fermion coupling and $S^i(\vecr)$ are spin operators, yields \cite{abrahams_prb_95}

\begin{equation}\label{EQ:OP2}
K \sim \langle S^i (\boldsymbolr_n) c_{\alpha} (\boldsymbolr_n)\sigma ^i_{\alpha \beta} c_{\beta}(\boldsymbolr_n)\rangle.
\end{equation}
The composite condensate $K$ represent the equal time condensate that has all the quantum numbers of the initial \ow state (the initial $F$ correlator). Taking a commutator with the Hamiltonian of any operator does not change the quantum numbers like spin $S$ and net charge $2e$. Hence the operator $K$ will have same spin and charge 2e expectation values as the initial  correlator $F$ of the \ow pair. However, by taking the time derivative we got rid of the time dependence and hence can talk about equal time correlations. We thus see that in order to discuss equal time order parameter of the \ow state one has to involve {\em composite pairs } represented by $K$. In the next section, we discuss this point in more detail.

\subsection{Composite pairing and relation to hidden orders}\label{sec:composite}

\begin{figure}[h!]
\begin{center}
  \includegraphics[width=0.4\textwidth]{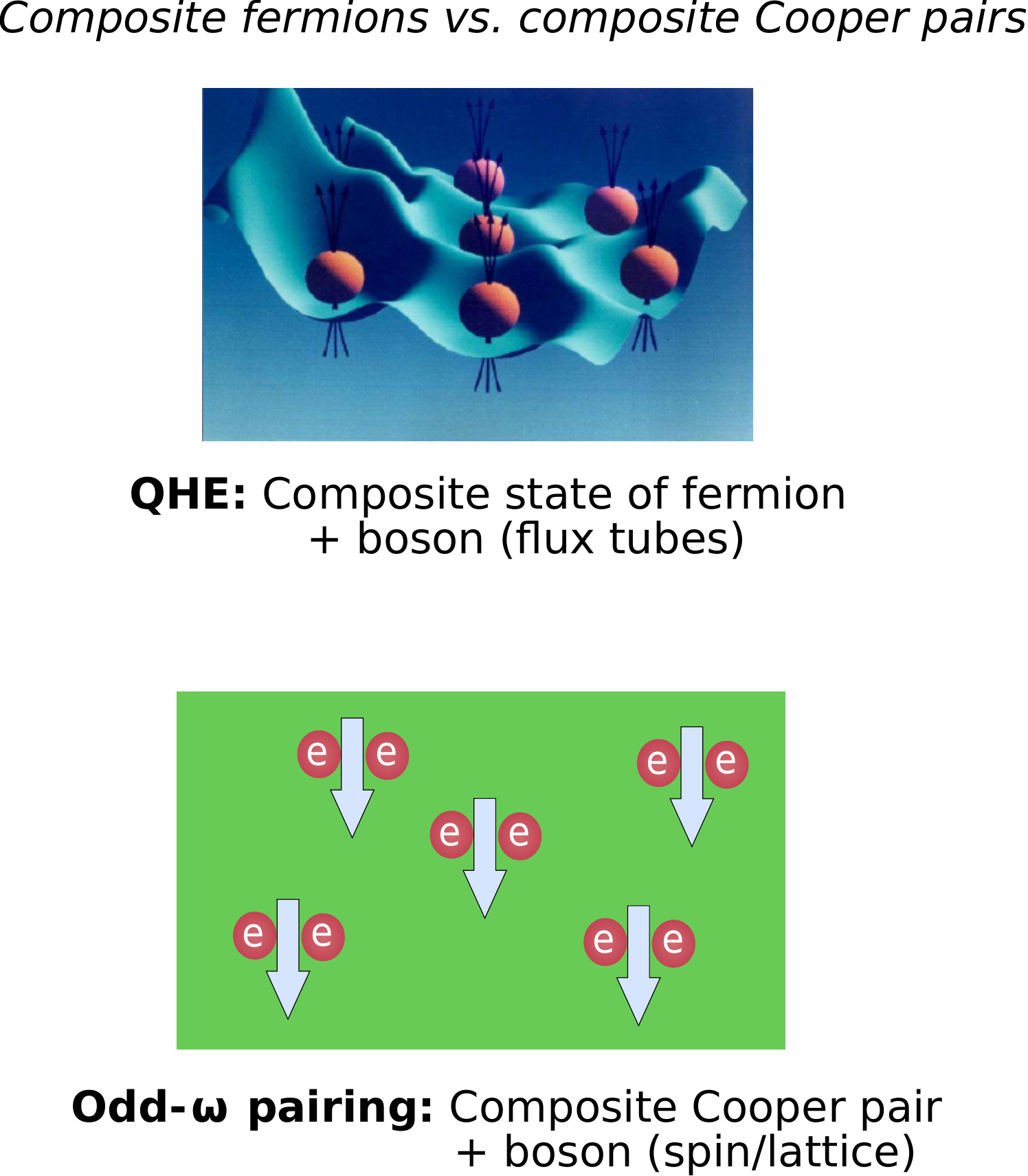}
  \caption{(color online) Illustration of the composite Cooper pairs as a condensate that is responsible for the \ow state. The upper panel illustrates the nature of a composite fermion  = fermion + boson [flux tubes as was shown to exist in the Quantum Hall Effect (QHE)]. The lower panel illustrates composite Cooper pairs = Cooper pair + boson (spin or lattice) that condenses in the \ow state. }
  \label{fig:CO1}
 \end{center}
\end{figure}

We can now illustrate the order parameter of the \ow Berezinskii state as a composite pair in Fig. (\ref{fig:CO1}). Namely, if one has a control of interactions to the degree where one can suppress the BCS pairing, i.e. the Cooper pairs alone do not condense, one can have a  {\em higher order} condensate forming where composite Cooper pair are formed. This is what the order parameter of the Berezinskii state seems to be telling us. We illustrate the nature of the composite order for singlet and triplet states. To be clear, we are giving here the symmetry analysis and list of possible composite states. At the moment, we are not aware of microscopic models that can prove the existence of these composite orders, although attempts to bring in higher order condensate were considered \cite{abrahams_prb_95, dahal_njp_09}.

\textit{Spin singlet composite.} A composite spin singlet \ow state could form as a result of binding a spin triplet Cooper pair with a spin-1 boson using the singlet part of the fusion decomposition $1_\text{Boson spin}\bigotimes 1_\text{Cooper pair spin} = 0 + 1 + 2$. In effect, the $S_\text{spin}=1$ Cooper pair combined with the spin-1 boson can form a net spin singlet that will have a charge $2e$:
\begin{equation}\label{EQ:CO1}
 K_\text{singlet} \sim \langle c_{\alpha}(\boldsymbolr)\sigma^i_{\alpha \beta}c_{\beta}(\boldsymbolr') S^i(\boldsymbolr) f_\text{odd}(\boldsymbolr,\boldsymbolr')\rangle
\end{equation}
To satisfy the odd orbital parity one would need to have $K$ be odd in $\boldsymbolr, \boldsymbolr'$. To accomplish this we have a weight function $f_\text{odd}$ being odd in $\boldsymbolr,\vecr'$. Thus, the defined $K$ has all the correct quantum numbers with respect to spin rotation (spin singlet $S_\text{spin} = 0$), inversion ($P_\text{parity} = -1$). Hence it has same quantum numbers as the \ow spin singlet pair. This is why the {\em time independent}  $K$ is the natural order parameter for the \ow state. A similar logic applies to spin triplet \ow Berezinskii state:

\textit{Spin triplet composite.}   The fusion rule would be that we can take a spin-1 Cooper pair and fuse it with the spin-0 boson:

\begin{equation}\label{EQ:CO2}
K_\text{triplet} \sim \langle c_{\alpha}(\boldsymbolr)\sigma^i_{\alpha \beta}c_{\beta}(\boldsymbolr')[ \phi(\boldsymbolr) - \phi(\boldsymbolr')] f_\text{even}(\boldsymbolr,\boldsymbolr')\rangle
   \end{equation}
with the neutral boson field $\phi$, \eg phonon displacement fields, and weight function $f_\text{even}$ being even functions of the spatial coordinate to ensure the $p$-wave nature of the neutral boson. Then, the composite pair field will have even parity $P_\text{parity}$ and have net spin of zero. An alternative possibility would be to take spin singlet $S_\text{spin}= 0$ pair and fuse it with the spin $S_\text{spin}= 1$ boson to form an $s$-wave composite boson.

Hence in both of these composite cases we have a net $2e$ condensate that has an opposite parity to the BCS case. These composite condensates are the order parameters that condense in the Berezinskii states.

\begin{figure}[h!]
\begin{center}
  \centering
  \includegraphics[width=0.5\textwidth]{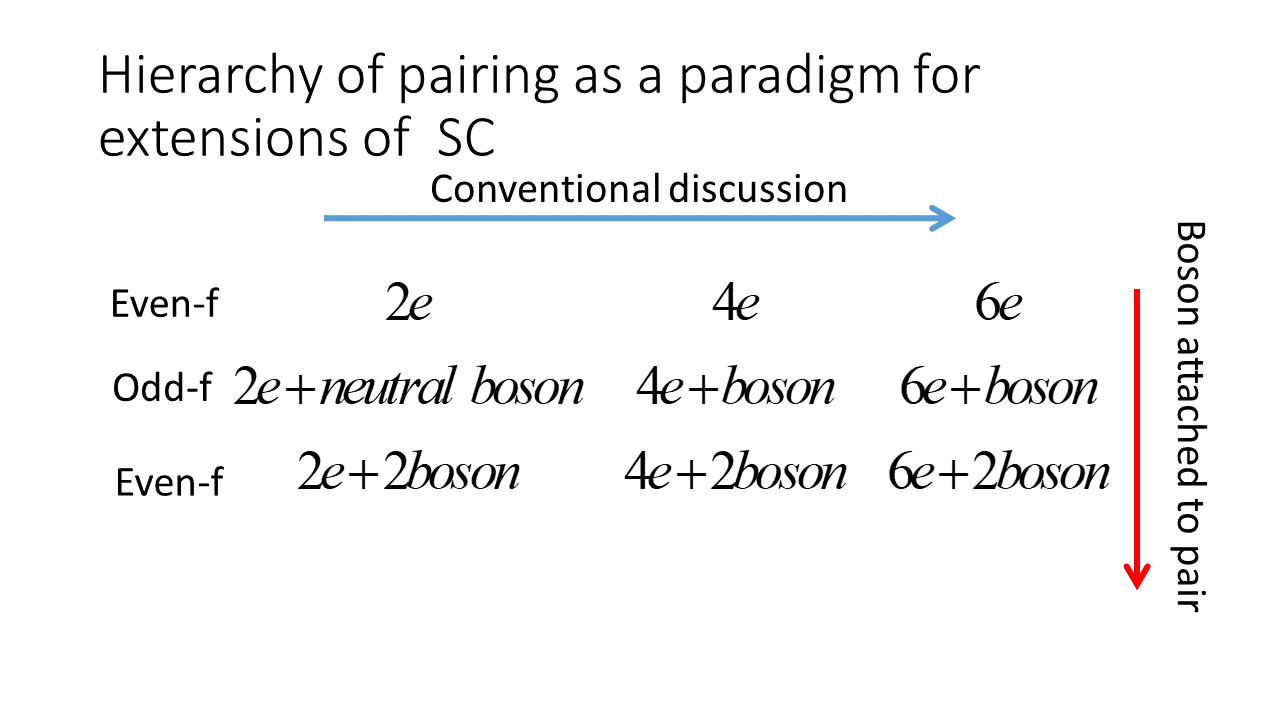}
  \caption{(Color online) A hierarchy of superconducting states beyond simple pairs is shown. The horizonal axis represents the increased even number of fermions that one condense such as $2e, 4e, 6e$ condensates. Alternative extensions based on the composite states are shown along the vertical axis. If we start with the conventional paired states, we can extend it to the $2e + 1$ boson state. This would correspond to the order parameter as a first derivative of the \ow amplitude and thus this line describes Berezinskii composite pairs. The next step in this process would be a paired state with $2e$ pair and two bosons that would correspond to second derivative and therefore to \ew pairing. The third line would correspond again to the \ow state with  three bosons attached to a pair, and so forth. }
  \label{fig:CO2}
 \end{center}
\end{figure}

A picture emerges where \ow states represent an extension of the conventional classification of superconducting states, i.e states that describe a condensate of fermion pairs. In addition to pairs, fermions can condense in the $4e$ and $6e$ channels. This way one can begin to develop a hierarchy of the superconducting states, as shown in Fig \ref{fig:CO2}. This discussion about composite orders implies another way to extend the hierarchy of the pairing states. We  include composite Cooper pairs in addition to the conventional extension of pairing to states with higher-charge condensates. Assume that one starts with fermionic particles in which case the lowest order condensate that is allowed to form is a two-fermion condensate. This establishes Cooper pairs as a key component of the pairing occurring in BCS states. Higher order charge condensates should be also allowed, like $4e$ and $6e$ condensates, but these are expected to be more fragile. Complementary to this extension there is another path to extend the family of paired states. One can imagine a state where neither Cooper pairs are formed and condense nor boson degrees of freedom condense (no BEC for spins for example). Yet conditions  exist that lead to  formation of the composite condensates where composite bosons made from Cooper pairs fused together with the neutral boson condense. The form of these condensates is captured in  Eq. (\ref{EQ:CO1},\ref{EQ:CO2}):

 \begin{equation}\label{EQ:CO3}
   \text{Composite pair} = \text{Cooper pair} \bigotimes \text{neutral boson}
 \end{equation}

We sum up the  proposed {\em hierarchy} of "higher order pairing"  in Fig. \ref{fig:CO2}. The composite pairing discussed here also can also  be viewed as an example of the {\em hidden order}  where neither conventional Cooper pairs nor a conventional Bose-field condenses separately, but only in the composite form.   Spectroscopy of these {\em composite hidden} orders would be more complicated. Therefore, we expect these {\em composite orders} will offer explanation to at least some of the {\em hidden and resonating orders}  that are ubiquitously observed in correlated quantum materials.
The extension of the pairing states to the realm of composite orders  would need to be explored further before we can see how plausible this proposal is.

\subsection{Dynamic induction of \ow state in superconductors}\label{sec:dynamic}

We here discuss the emergent understanding of the induction of a Berezinskii \ow  state in the time domain.
One view on the dynamic correlations revealed in \ow states is that \ow state as a dynamic order.  An \ow state  realizes strongly retarded order where in fact there are no equal time pairing correlations. This view is supported by the fact that a possible order parameter for \ow state is a time derivative of the pair correlation function $F$. An interesting question that arises is whether or not it is possible to induce the \ow state in the time domain using external fields as a drive.

The general argument about \ow induction in time domain is natural. We start with pair amplitudes that are only even in relative time. Upon turning on a time dependent drive, the pair amplitudes are modified by the drive field. What used to be a perfectly symmetric function upon reversal of relative time, $ t \rightarrow-t$,  now is no longer a function of a single time, but rather a function of two times. Symbolically and to lowest order in the drive potential $U(t)$, the parity properties of the function
\begin{align}
F(t_1,t_2) = F_0(t_1 - t_2) + \int dt' G_0(t_1- t') U(t') F_0(t'-t_2)
\end{align}
now depends on the drive field. Here, $G_0$ and $F_0$ are the unperturbed normal and anomalous Green functions. Hence, there are \ew and \ow components generated immediately in a driven superconductor. For this to happen, according to the SPOT constraint, we would need to also to break at least one more index. In the case of a one band material, one could break translational symmetry at the interface. In the case of a multiband superconductor, one would induce odd-interband index pairing that would also be odd in $T^*$. Both cases have been addressed for a driven superconducting state \cite{triola_prb_16, triola_prb_17}.  We thus can expect the induction of the \ew and \ow components and cross coupling of the even and odd channels in the case of the driven system. As mentioned in the introduction, one can take a view that once we have \ew pairs that are available in equilibrium, a time dependent drive will convert a fraction of \ew pairs into \ow pairs and vice versa.

\begin{figure}[h!]
 \begin{center}
  \centering
  \includegraphics[width=0.4\textwidth]{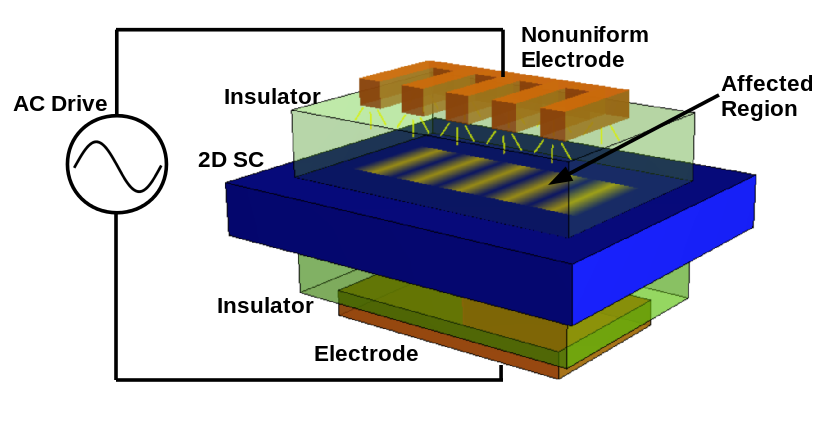}
  \caption{(Color online) Schematic of a driven superconducting system with a 2D superconducting region lying between two insulating slabs each capped by a conducting electrode configured in such a way as to generate an electric field. The AC voltage acts as a time-dependent drive. Such a device could be realized by sandwiching a thin film superconductor, like Pb and other superconductors, between two insulating wafers. Adapted from \cite{triola_prb_16}. }
  \label{fig:schematic}
 \end{center}
\end{figure}

We now will lay out mathematical arguments in support of this claim. One can induce the new components of the pair amplitude just like one induces new \ow components via scattering at interfaces in hybrid structures. We start with the general structure of any multiband superconducting state subject to the external electrostatic potential drive $U(t)$. We follow the above references where one can find a detailed description of the effect. A schematic overview of the possible driven system is shown in Fig. (\ref{fig:schematic}).

Following Triola \etal, we start with the multiband SC Hamiltonian
allowing for both interband and intraband pairing:
\begin{equation}
\begin{aligned}
H_{\text{sc}} &= \sum_{\textbf{k},\sigma}\left(\xi_{a,\textbf{k}} \psi^\dagger_{\sigma,a,\textbf{k}} \psi_{\sigma,a,\textbf{k}} + \xi_{b,\textbf{k}} \psi^\dagger_{\sigma,b,\textbf{k}} \psi_{\sigma,b,\textbf{k}} \right) \\
&+ \sum_{\alpha,\beta,\textbf{k}} \Delta_{\alpha\beta} \psi^\dagger_{\uparrow,\alpha,-\textbf{k}}\psi^\dagger_{\downarrow,\beta,\textbf{k}} + \text{h.c.} \\
&+ \sum_{\textbf{k},\sigma}\Gamma \psi^\dagger_{\sigma,a,\textbf{k}} \psi_{\sigma,b,\textbf{k}} + \text{h.c.}
\end{aligned}
\label{eq:H_sc}
\end{equation}
where $\xi_{\alpha,\textbf{k}}=\tfrac{k^2}{2m_\alpha}-\mu_\alpha$ is the quasiparticle dispersion in band $\alpha$ with effective mass $m_\alpha$ measured from the chemical potential $\mu_\alpha$, $\psi^\dagger_{\sigma,\alpha,\textbf{k}}$ ($\psi_{\sigma,\alpha,\textbf{k}}$) creates (annihilates) a quasiparticle with spin $\sigma$ in band $\alpha$ with momentum $\textbf{k}$, $\Delta_{\alpha\beta}\equiv \lambda \int \tfrac{d^dk}{(2\pi)^d}\langle \psi_{\uparrow,\alpha,-\textbf{k}}\psi_{\downarrow,\beta,\textbf{k}} \rangle$ is the superconducting gap, where $d$ is the dimensionality of the system, and we allow for the possibility of interband scattering with amplitude $\Gamma$.

With these conventions we write the time-dependent drive as:
\begin{equation}
H_t =\sum_{\textbf{k},\sigma,\alpha,\beta} U_{\alpha\beta}(t)\psi^\dagger_{\sigma,\alpha,\textbf{k}} \psi_{\sigma,\beta,\textbf{k}}.
\label{eq:H_drive}
\end{equation}
The bath and mixing terms take the form:
\begin{equation}
\begin{aligned}
H_{\text{bath}} &=\sum_{n,\sigma,\alpha,\textbf{k}}\left( \epsilon_{n} -\mu_\text{bath}\right) c^\dagger_{n;\sigma\alpha\textbf{k}} c_{n;\sigma\alpha\textbf{k}} \\
H_{\text{mix}} &=\sum_{\textbf{k},n,\sigma,\alpha}\eta_{n} c^\dagger_{n;\sigma\alpha\textbf{k}} \psi_{\sigma,\alpha,\textbf{k}} + \text{h.c.}
\end{aligned}
\label{eq:H_bath}
\end{equation}
where $\epsilon_{n}$ describes the energy levels of the Fermionic bath, $\mu_\text{bath}$ is the chemical potential of the bath, $c^\dagger_{n;\sigma\alpha\textbf{k}}$ ($c_{n;\sigma\alpha\textbf{k}}$) creates (annihilates) a fermionic mode with degrees of freedom indexed by $n$, $\sigma$, $\alpha$, and $\textbf{k}$, and $\eta_{n}$ specifies the amplitude of the coupling between the superconductor and the bath. The Dyson equation for the Keldysh Green functions is found to be:
\begin{align}
\hat{\mathcal{G}}(\textbf{k};t_1,t_2) &= \hat{\mathcal{G}}_0(\textbf{k};t_1-t_2) + \int_{-\infty}^\infty dt \hat{\mathcal{G}}_0(\textbf{k};t_1-t) \notag\\
&\times \left(
\begin{array}{cc}
\hat{U}(t) & 0 \\
0 & -\hat{U}(t)^*
\end{array}\right)\otimes \hat{\rho}_0 \hat{\mathcal{G}}(\textbf{k};t,t_2)
\label{eq:dyson_keldysh}
\end{align}
where $\hat{\rho}_0$ is the 2$\times$2 identity in Keldysh space, and $\hat{\mathcal{G}}_0(\textbf{k};t_1-t_2)$ is the Green function describing the unperturbed system in a Keldysh basis:
\begin{equation}
\hat{\mathcal{G}}_0(\textbf{k};t_1-t_2)=\left(\begin{array}{cc}
\hat{\mathcal{G}}_0^{\text{R}}(\textbf{k};t_1-t_2) & \hat{\mathcal{G}}_0^{\text{K}}(\textbf{k};t_1-t_2) \\
0 & \hat{\mathcal{G}}_0^{\text{A}}(\textbf{k};t_1-t_2)
\end{array} \right)
\end{equation}
where $\hat{\mathcal{G}}_0^{\text{R}}(\textbf{k};t_1-t_2)$, $\hat{\mathcal{G}}_0^{\text{A}}(\textbf{k};t_1-t_2)$, and $\hat{\mathcal{G}}_0^{\text{K}}(\textbf{k};t_1-t_2)$ are the retarded, advanced, and Keldysh Green functions, respectively.


 Iterating in powers of the drive via Eq (\ref{eq:dyson_keldysh}), one finds the  Green function to linear order in the drive. Fourier transforming with respect to the relative ($t_1-t_2$) and average ($(t_1+t_2)/2$) times Triola \etal~obtained the linear order corrections in frequency space:
\begin{widetext}
\begin{equation}
\hat{\mathcal{G}}(\textbf{k};\omega,\Omega) = 2\pi\delta(\Omega)\hat{\mathcal{G}}_0(\textbf{k};\omega) +\hat{\mathcal{G}}_0(\textbf{k};\omega+\tfrac{\Omega}{2}) \left(
\begin{array}{cc}
\hat{U}(\Omega) & 0 \\
0 & -\hat{U}(-\Omega)^*
\end{array}\right)\otimes \hat{\rho}_0 \hat{\mathcal{G}}_0(\textbf{k};\omega-\tfrac{\Omega}{2}).
\label{eq:linear_w}
\end{equation}
The terms to linear order in the drive are given by:
\begin{equation}
\begin{aligned}
\delta \hat{F}^{\text{R}}(\textbf{k};\omega,\Omega)&=\hat{G}_0^{\text{R}}(\textbf{k};\omega+\tfrac{\Omega}{2})\hat{U}(\Omega)\hat{F}_0^{\text{R}}(\textbf{k};\omega-\tfrac{\Omega}{2}) -\hat{F}_0^{\text{R}}(\textbf{k};\omega+\tfrac{\Omega}{2})\hat{U}^*(-\Omega)\hat{\overline{G}}_0^{\text{R}}(\textbf{k};\omega-\tfrac{\Omega}{2}) \\
\delta \hat{F}^{\text{A}}(\textbf{k};\omega,\Omega)&=\hat{G}_0^{\text{A}}(\textbf{k};\omega+\tfrac{\Omega}{2})\hat{U}(\Omega)\hat{F}_0^{\text{A}}(\textbf{k};\omega-\tfrac{\Omega}{2}) -\hat{F}_0^{\text{A}}(\textbf{k};\omega+\tfrac{\Omega}{2})\hat{U}^*(-\Omega)\hat{\overline{G}}_0^{\text{A}}(\textbf{k};\omega-\tfrac{\Omega}{2}) \\
\delta \hat{F}^{\text{K}}(\textbf{k};\omega,\Omega)&=\hat{G}_0^{\text{R}}(\textbf{k};\omega+\tfrac{\Omega}{2})\hat{U}(\Omega)\hat{F}_0^{\text{K}}(\textbf{k};\omega-\tfrac{\Omega}{2}) -\hat{F}_0^{\text{R}}(\textbf{k};\omega+\tfrac{\Omega}{2})\hat{U}^*(-\Omega)\hat{\overline{G}}_0^{\text{K}}(\textbf{k};\omega-\tfrac{\Omega}{2}) \\
&+\hat{G}_0^{\text{K}}(\textbf{k};\omega+\tfrac{\Omega}{2})\hat{U}(\Omega)\hat{F}_0^{\text{A}}(\textbf{k};\omega-\tfrac{\Omega}{2}) -\hat{F}_0^{\text{K}}(\textbf{k};\omega+\tfrac{\Omega}{2})\hat{U}^*(-\Omega)\hat{\overline{G}}_0^{\text{A}}(\textbf{k};\omega-\tfrac{\Omega}{2}).
\end{aligned}
\label{eq:dFRAK}
\end{equation}
\end{widetext}

To demonstrate the emergence of the even-$\omega$ and odd-$\omega$ terms one can focus on the retarded components of the anomalous Green functions in Eq (\ref{eq:dFRAK}). In general, these corrections, $\delta\hat{F}^{\text{R}}(\textbf{k};\omega,\Omega)$, could possess terms that are even in $\omega$ and terms that are odd in $\omega$. After explicitly separating even and odd frequency parts one can find  generically   even to even, even to odd, odd to even and odd to odd contributions of the pair amplitude upon turning on the drive.  The most relevant for our discussion are the terms that convert \ew pairs to \ow pairs:

 \begin{equation}
\begin{aligned}
\delta F_{\text{e}\rightarrow\text{o}}(\textbf{k};\omega,\Omega)&=\left[ \hat{G}^{\text{R}}_0\left(\textbf{k};\omega+\tfrac{\Omega}{2}\right) \hat{U}(\Omega),\hat{F}^{(\text{e})}\left(\textbf{k};\omega-\tfrac{\Omega}{2}\right) \right]_{-} \\
&-\left[ \hat{G}^{\text{R}}_0\left(\textbf{k};-\omega+\tfrac{\Omega}{2}\right) \hat{U}(\Omega),\hat{F}^{(\text{e})}\left(\textbf{k};\omega+\tfrac{\Omega}{2}\right) \right]_{-}, \\
\delta F_{\text{o}\rightarrow\text{e}}(\textbf{k};\omega,\Omega)&=\left[ \hat{G}^{\text{R}}_0\left(\textbf{k};\omega+\tfrac{\Omega}{2}\right) \hat{U}(\Omega),\hat{F}^{(\text{o})}\left(\textbf{k};\omega-\tfrac{\Omega}{2}\right) \right]_{-} \\
&-\left[ \hat{G}^{\text{R}}_0\left(\textbf{k};-\omega+\tfrac{\Omega}{2}\right) \hat{U}(\Omega),\hat{F}^{(\text{o})}\left(\textbf{k};\omega+\tfrac{\Omega}{2}\right) \right]_{-},
\end{aligned}
\label{eq:parity_reverse}
\end{equation}
where, for convenience, we have defined the bracket:
\begin{align}\label{eq:bracket}
[\hat{g}(\omega_1)\hat{u}(\omega_2),\hat{f}(\omega_3) ]_{\pm}&\equiv \frac{1}{2}\Big( \hat{g}(\omega_1)\hat{u}(\omega_2)\hat{f}(\omega_3) \\
& \pm\hat{f}(\omega_3)\hat{u}(-\omega_2)^*\hat{g}(\omega_1)^*\Big).
\end{align}

\begin{figure}[h!]
 \begin{center}
  \centering
  \includegraphics[width=0.45\textwidth]{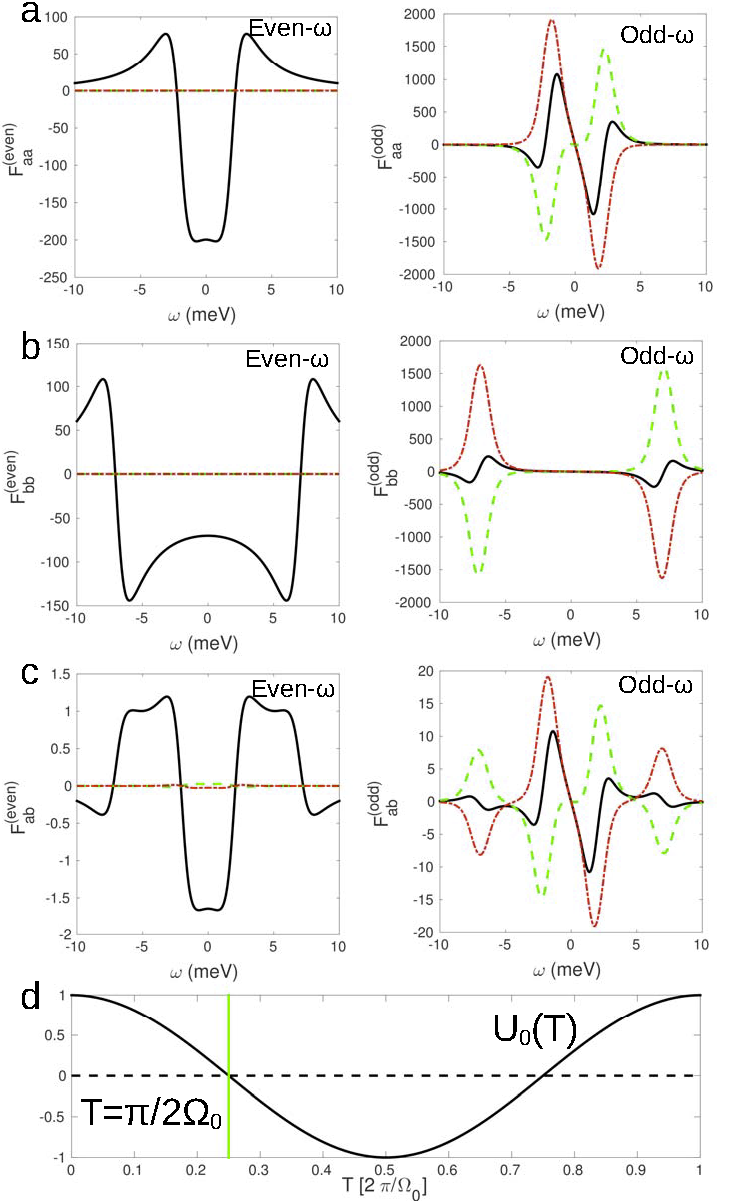}
  \caption{In the left (right) column we plot the even-$\omega$ (odd-$\omega$) terms of the real part of the Wigner transform   of the anomalous part of the Green function, $\langle \hat{F}^{\text{R}}(\omega,T=\pi/2\Omega_0)\rangle$, in black (solid), where we have taken the average value of $\hat{F}^{\text{R}}(\textbf{k};\omega,T=\pi/2\Omega_0)$ at $|\textbf{k}|=k^{(a)}_\text{F}$ and $|\textbf{k}|=k^{(b)}_\text{F}$. In each case we have also plotted the parity-preserving terms (green-dashed) and parity-reversing terms (red-dash-dotted). (a) the diagonal component for band-$a$, (b) the diagonal component for band-$b$, (c) the interband component. (d) The components of the drive, plotted in the time domain over a full period, the green vertical line denotes the time, $T_\text{cm} \equiv T=\pi/2\Omega_0$, at which all plots in this figure are evaluated. The parameters used to describe the driven multiband superconductor in this case are: effective masses, $m_a=0.5$ \AA$^{-2}$/eV and $m_b=1$ \AA$^{-2}$/eV; chemical potentials, $\mu_a=\mu_b=2$eV; $s$-wave gaps, $\Delta_{aa}=2$meV, $\Delta_{bb}=7$meV, $\Delta_{ab}=\Delta_{ba}=0$, consistent with MgB$_2$\cite{choi_nature_02}; interband scattering, $\Gamma=10$ meV; dissipation described by $\eta=1$meV; and a drive $U(t) = U_0 cos(\Omega_0 t)$ with $U_0=10$meV, and $\Omega_0=1$meV (242 GHz). Adapted from \cite{triola_prb_16}.     }
  \label{fig:even_odd_u0_10meV_T_halfpi}
 \end{center}
\end{figure}

\begin{figure}[h!]
 \begin{center}
  \centering
  \includegraphics[width=0.45\textwidth]{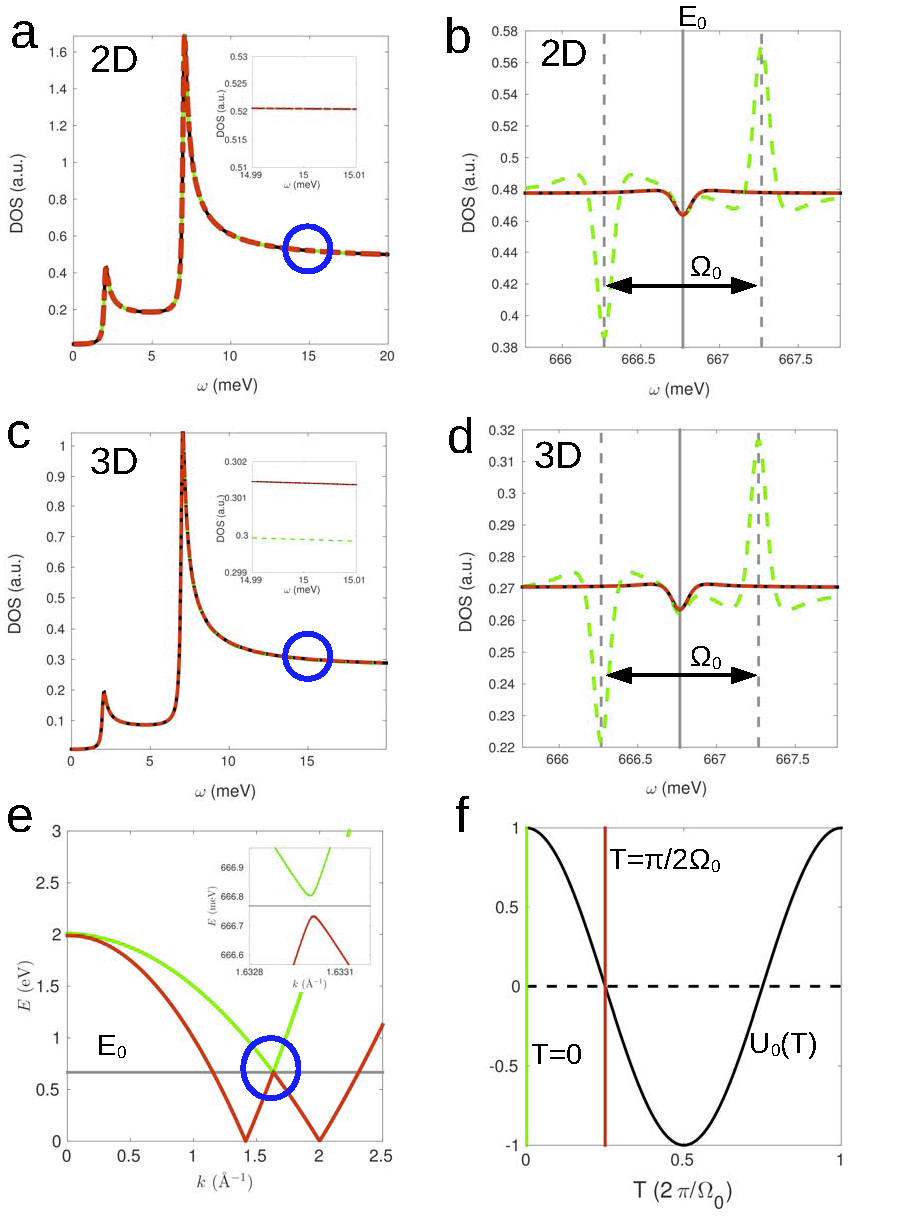}
  \caption{In (a) and (b), the 2D DOS computed using: effective masses, $m_a=0.5$ \AA$^{-2}$/eV and $m_b=1$ \AA$^{-2}$/eV; chemical potentials, $\mu_a=\mu_b=2$eV; $s$-wave gaps, $\Delta_{aa}=2$meV, $\Delta_{bb}=7$meV, $\Delta_{ab}=\Delta_{ba}=0$, consistent with MgB$_2$\cite{choi_nature_02}; interband scattering, $\Gamma=10$ meV; dissipation described by $\eta=0.1$meV; and a drive with $U_0=10$meV, and $\Omega_0=1$meV (242 GHz). In both panels we show the case for no drive in black (solid), and the cases with the drive at times $T_\text{cm} \equiv T=0$ and $T=\pi/2\Omega_0$ in green (dashed) and red (dash-dotted), respectively. In (a) we focus on the states near the Fermi surface, in (b) we focus on the range of energies near the crossing of the two bands at which we find the driven DOS at $T=0$ possesses two peaks shifted from the avoided crossing at $E_0$ by, $\pm\Omega_0/2$. In (c) and (d), the 3D DOS plotted for the same parameters as in (a) and (b). Notice that the main difference is that in 3D the driven DOS at $T=0$ is slightly suppressed relative to the undriven DOS (see inset). In (e) we plot the spectrum of the two band superconductor given by $\epsilon_{\pm}(\textbf{k})$. The horizontal grey line denotes the avoided crossing (see inset) at $E_0$, due to the finite interband scattering, $\Gamma$. In (f) we show the drive  plotted in the time domain over a full period, the green vertical line denotes the beginning of the period at $T=0$ where the drive has maximum amplitude, while the red line denotes $T=\pi/2\Omega_0$ where the drive amplitude is zero. Adapted from \cite{triola_prb_16}.
         }
  \label{dos_plots}
 \end{center}
\end{figure}

The induced \ow components are plotted in Fig. \ref{fig:even_odd_u0_10meV_T_halfpi}. The effect of the dynamically induced components can be observed in the density of  states as satellite features induced by Stokes satellites due to external potential pumping. We would like to stress the general nature of the proposed phenomena. The induction of the \ow component in time driven systems is a quite general phenomena and will not depend on the specifics of the mechanism and experimental setup. The general rule to anticipate the induction of the new components is  guided only by the Berezinskii classification and rule that $SPOT = -1$. Conventional  pairs with $S = -1, P = +1, O = +1, T = +1$ can be converted into odd-in-time pairs with $S = -1, P = +1, O = -1, T = -1$ while $SPOT = -1$ remains intact. As we go forward, we will see that this is a general rule that applies to other cases, \eg the induction of \ow and \ew pairing correlations in Majorana systems.

\subsection{Sign of the Meissner effect}\label{sec:meissner}

The Meissner effect is the most fundamental property of the superconducting state
as it incorporates both the zero resistance property of a
superconductor as well as the flux expulsion due to screening currents.
The diamagnetic currents blocking external magnetic fields remain constant with
time and hence do not decay. A
superconductor is thus not primarily defined by the existence of charge currents
flowing without resistance, a property which is shared by many other physical systems
such as the edge states of the quantum Hall state or field-induced persistent currents in resistive conductors. The Meissner effect is a direct consequence of the Higgs mechanism
that takes place in a superconductor which spontaneously breaks U(1) gauge symmetry:
the superconducting ground-state is independent on the phase $\varphi$ of the
order parameter $\Delta=|\Delta|\e{\i\varphi}$, but a particular ground-state is
characterized by a certain value of $\varphi$. When this symmetry is spontaneously
broken, the Higgs mechanism renders the gauge field (photon) in the superconductor
massive and causes it to have a finite range, leading to the Meissner effect.
The sign of the Meissner effect in conventional BCS superconductors is thus negative:
the currents are diamagnetic in nature and attempt to screen any external flux.

Taking into account the fundamental role played by the Meissner effect in
superconductivity, there was clearly reason for concern when Abrahams \etal~pointed
out that \ow Berezinskii bulk superconductors appeared to have a sign problem with the
Meissner effect \cite{abrahams_prb_95}. This issue had also previously been remarked on by A. Garg \cite{abrahams_private} The Meissner effect calculated to
lowest order provided an opposite sign to the BCS case, providing a superfluid
density which was negative. This result seemed to suggest that a bulk \ow
superconducting state had to be thermodynamically unstable.

The work by Coleman, Miranda, and Tsvelik \cite {coleman_prl_93, coleman_prb_94,
coleman_prl_95} who studied
 \ow-pairing in a Kondo lattice and heavy fermion compounds, however, did not
have any problem with a negative superfluid density. Their idea was
 built on the interesting twist that \ow superconductivity was driven by an
anomalous composite, staggered three-body scattering amplitude which turned out to provide a stable
superconducting phase with a diamagnetic Meissner response. A similar resolution
was indeed proposed in \cite{abrahams_prb_95}, who suggested that a stable
Meissner state could be achieved by introducing a composite condensate where
there existed a joint condensation of Cooper pairs and density fluctuations.

The problem nevertheless remained that within the standard framework
with a two-body interaction where only Cooper pairs would condense, the
\ow bulk state appeared to be thermodynamically unstable. Heid \cite{heid_zphysb_95}
summarized the stability analysis problem related to \ow superconductivity in
the following manner. Consider first the case of weak-coupling superconductivity
with a continuous (second-order) phase transition, in which case the change $\delta\Omega_\text{pot}$ in
the thermodynamical potential $\Omega_\text{pot}$ due to a two-body interaction reads
\cite{abrikosov_book_75}:
\begin{align}\label{eq:free_energy}
\delta\Omega_\text{pot} \propto -\frac{1}{\beta} \sum_{\omega_n,\vq} \frac{\Delta(\omega_n,\vq)\Delta^+(\omega_n,\vq)}{\omega_n^2+\xi_{\vq}^2}
\end{align}
where we have used the notation of \cite{solenov_prb_09}. Above, $\xi_{\vq}$
is the quasiparticle normal-state dispersion, $\omega_n$ is the Matsubara frequency,
whereas the gap functions $\Delta(\omega_n,\vq)$ are connected to the anomalous
Green functions $F(\omega_n,\vq)$ in terms of the self-consistency equation:
\begin{align}\label{eq:selfconsistent}
\Delta(\omega_n,\vq) = \sum_{\omega_n',\vq'} D(\omega_n-\omega_n',\vq-\vq')F(\omega_n',\vq')
\end{align}
$\beta$ is inverse temperature and $D$ is the irreducible interaction between quasiparticles, i.e.
the pairing glue of the Cooper pairs, the latter assumed to be real and both even in
$\omega_n$ and $\vq$. We underline that there is no contradiction between choosing a pairing interaction
that is even in $\omega_n$ and obtaining an \ow superconducting state: the self-consistency equation
allows for both even and odd-frequency solutions of $\Delta(\omega_n,\vq)$ even if $D$ is even with
respect to $\omega_n$, as can be verified by direct inspection. The anomalous Green functions are here defined as
\begin{align}
F(\omega_n,\vq) = \int^\beta_0 d\tau \e{\i\omega_n\tau}\langle \mathcal{T}_\tau\{c_{\vq}(\tau) c_{-\vq}(0)\} \rangle,\notag\\
F^+(\omega_n,\vq) = \int^\beta_0 d\tau \e{\i\omega_n\tau}\langle \mathcal{T}_\tau\{c_{-\vq}^\dag(\tau) c^\dag_{\vq}(0)\} \rangle.
\end{align}
 The relation between $F^+$ and $\Delta^+$ is
identical to Eq. (\ref{eq:selfconsistent}). The sign of $\delta\Omega$, which determines whether or not the bulk \ow state is
thermodynamically stable, is determined by establishing the relation between
$\Delta(\omega_n,\vq)$ and $\Delta^+(\omega_n,\vq)$, since it is this
combination that determines $\delta\Omega_\text{pot}$ in Eq. (\ref{eq:free_energy}).
To do so, one needs to compute the averages $\langle \mathcal{T}_\tau\{c_{\vq}(\tau) c_{-\vq}(0)\} \rangle$
and $\langle \mathcal{T}_\tau\{c_{-\vq}^\dag(\tau) c^\dag_{\vq}(0)\} \rangle$ which are nonzero if taken with respect to a state with broken U(1) symmetry (absence
of particle number conservation for single-particle excitations). Assume that
there exists an appropriate symmetry-breaking mean field Hamiltonian $H_\text{MF}$ for this
purpose. In this case, one obtains
\begin{align}\label{eq:f_even}
F(\tau,\vq) &= \frac{1}{Z} \text{Tr}\{ \e{-\beta H_\text{MF}} \mathcal{T}_\tau \e{\tau H_\text{MF}} c_{\vq} \e{-\tau H_\text{MF}} c_{-\vq} \},\notag\\
F^+(\tau,\vq) &= \frac{1}{Z} \text{Tr}\{ \e{-\beta H_\text{MF}} \mathcal{T}_\tau \e{\tau H_\text{MF}} c^\dag_{-\vq} \e{-\tau H_\text{MF}} c^\dag_{\vq} \},
\end{align}
where $Z=\text{Tr}\{\e{-\beta H_\text{MF}}\}$ is the partition function. Inspecting
Eq. (\ref{eq:f_even}) shows that the two Green functions are related via
\begin{align}\label{eq:fcross}
F^+(\tau,\vq) = [F(\tau,\vq)]^*
\end{align}
Because of this property, one can verify from Eq. (\ref{eq:selfconsistent})
that the product $\Delta(\omega_n,\vq)\Delta^+(\omega_n,\vq)$ is negative
definite and thus producing $\delta\Omega_\text{pot}>0$. Since the free energy is larger
in the \ow superconducting state than the disordered state, one concludes
that the \ow superconducting phase is thermodynamically unstable. Accompanying this
conclusion is the property of a negative superfluid density or Meissner kernel $\mathcal{K}$
that relates the supercurrent
$\boldsymbolj$ and vector potential $\boldsymbolA$ via $\boldsymbolj = -\mathcal{K}(\vk)\boldsymbolA$.

The problem with the above reasoning was discussed in detail by Belitz and Kirkpatrick \cite{belitz_prb_99} who explained that the reality properties of the gap function (its real and imaginary parts), beyond what is possible to manipulate via global gauge transformations, were crucial in order to obtain a thermodynamically stable \ow-state. Later, Solenov \etal~\cite{solenov_prb_09} argued that the reality properties of the gap function that caused the sign problem in the Meissner effect relied on the existence of a mean field Hamiltonian $H_\text{MF}$ that
can describe \ow superconductivity. They further conjectured that an effective Hamiltonian formulation cannot capture
the strong retardation effects which are inherent to \ow pairing correlations. Instead,
one can describe these by an effective action $\mathcal{S}$ which is non-local in time. The
latter approach was utilized in Ref. \cite{solenov_prb_09} with the outcome
that Eq. (\ref{eq:fcross}) for an \ow superconductor is modified to
\begin{align}\label{eq:fcross_odd}
F^+(\tau,\vq) = -[F(\tau,\vq)]^*,
\end{align}
i.e. with an extra minus sign compared to the \ew case described by Eq. (\ref{eq:fcross}). This was a different, but physically equivalent, way of arriving at the same conclusion as \cite{belitz_prb_99}. This additional
sign restores the thermodynamic stability of the \ow superconducting state, since
the product $\Delta(\omega_n,\vq)\Delta^+(\omega_n,\vq)$ now becomes \textit{positive definite}
so that $\delta\Omega_\text{pot}<0$. Moreover, one can explicitly verify that the Meissner kernel now yields
a diamagnetic response corresponding to a positive superfluid density. The kernel $\mathcal{K}$
is defined as \cite{abrikosov_book_75}
\begin{align}
&\mathcal{K}(\vk) = \frac{Ne^2}{m} + \frac{2e^2}{m^2\beta} \sum_{\omega_n}
\int \frac{d\boldsymbolp}{(2\pi)^3}\boldsymbolp^2\times  \notag\\
&[G(\omega_n,\boldsymbolp_+)G(\omega_n,\boldsymbolp_-)
+ F(\omega_n,\boldsymbolp_+)F^+(\omega_n,\boldsymbolp_-)].
\end{align}
We defined $\boldsymbolp_\pm = \boldsymbolp \pm \boldsymbolk/2$ and the Green functions for an \ow
superconductor are, making sure to utilize the correct equation (\ref{eq:fcross_odd}) instead
of (\ref{eq:fcross}):
\begin{align}\label{eq:oddwgreenfunctions}
G(\omega_n,\boldsymbolq) &= \frac{\i\omega_n+\xi_{\vq}}{\omega_n^2+\xi_{\vq}^2 + 2|\Delta(\omega_n,\vq)|^2},\notag\\
F(\omega_n,\boldsymbolq) &= \frac{2\Delta(\omega_n,\vq)}{\omega_n^2+\xi_{\vq}^2 + 2|\Delta(\omega_n,\vq)|^2},\notag\\
F^+(\omega_n,\boldsymbolq) &= \frac{2[\Delta(\omega_n,\vq)]^*}{\omega_n^2+\xi_{\vq}^2 + 2|\Delta(\omega_n,\vq)|^2}.
\end{align}
The factor 2 appearing in front of $|\Delta(\omega_n,\vq)|^2$ has no special meaning: it can readily be absorbed into the definition of the order parameter by incorporating a factor $\frac{1}{2}$ into the pairing interaction, as is often done. The Meissner kernel diverges and is regularized by subtracting its value for
$\Delta=0$, so that the new $\mathcal{K}(\vk)$ equals zero in the normal phase
as it should. In the long wavelength limit $\vk\to 0$ and assuming a $\vq$-
independent gap ($s$-wave pairing), one obtains
\begin{align}
\mathcal{K}(\vk \to 0) = \frac{\pi Ne^2}{m\beta} \sum_{\omega_n} \frac{2|\Delta(\omega_n)|^2}{[\omega_n^2 + 2|\Delta(\omega_n)|^2]^{3/2}}
\end{align}
This equation is clearly positive definite, whereas an incorrect result (negative definite $\mathcal{K}$)
would have been obtained if we had used Eq. (\ref{eq:fcross}) to obtain the Green functions for the \ow superconducting case. Consequently, a second-order transition to a spatially
homogeneous, odd-frequency superconducting state is in principle allowed, in contrast
to the conclusion of Ref. \cite{heid_zphysb_95}.

The technical derivation of this result provided in Ref. \cite{solenov_prb_09}
was further refined and expanded upon in Ref. \cite{kusunose_jpsj_11b} where
the importance of choosing the appropriate mean field solution that minimizes
the effective free energy was pointed out. Note that in the above treatment of the
thermodynamic potential and Meissner kernel, spinless fermions were assumed for
simplicity the entire way
so that in the \ew case the gap function would have an odd-parity symmetry
(such as $p$-wave) whereas in the \ow case the gap function would have an even-parity
symmetry (such as $s$-wave).

\begin{figure}[h!]
\includegraphics[scale=0.6]{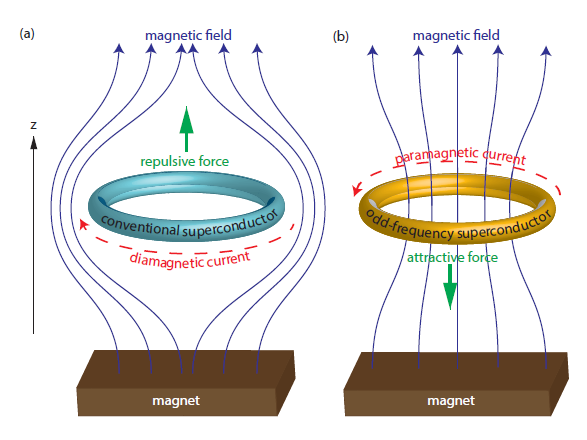}
\caption[fig]{\label{fig:levitation} (Color online) Meissner respones of (a) a conventional superconducting ring and (b) an \ow superconducting ring. In the event of a paramagnetic supercurrent response, the \ow superconductor experiences an attractive force to the underlying magnet, causing superconducting antilevitation. Figure adapted from \cite{lee_arxiv_16}.}
\end{figure}

Fominov and co-workers \cite{fominov_prb_15} studied the possible coexistence of \ow states with both a diamagnetic and paramagnetic response. As shown above, a bulk \ow superconducting state with a conventional diamagnetic Meissner response is possible under the assumption that there exists a microscopic mechanism (pairing interaction $D$) that creates this type of superconductivity. In contrast, the \ow superconducting state induced in \eg diffusive S/F structures has a paramagnetic Meissner response \cite{yokoyama_prl_11, mironov_prl_12, dibernardo_prx_15}. An interesting issue is thus to consider if these two types of superconducting correlations can coexist. It was demonstrated in \cite{fominov_prb_15} that such a coexistence would lead to unphysical properties such as complex superfluid densities and Josephson couplings. A paramagnetic Meissner response from an odd-frequency superconductor, whether bulk or artificially generated via the proximity effect, would provide superconducting anti-levitation as shown Fig. \ref{fig:levitation}. \\

We underline that by introducing a composite order parameter it was shown by Abrahams \etal~\cite{abrahams_prb_95} that it is possible to write down a mean-field Hamiltonian describing a thermodynamically stable \ow Berezinskii state. This finding is not necessarily inconsistent with the arguments put forward by \cite{solenov_prb_09} and \cite{fominov_prb_15}, because in those papers the condensate (and corresponding anomalous Green function) consist of two fermions whereas the condensate described by a mean-field Hamiltonian in \cite{abrahams_prb_95} is composed of two fermions and a bosonic fluctuation.

Paramagnetic Meissner effects have been discussed in previous literature in the context of high-$T_c$ superconductors \cite{kostic_prb_96, higashitani_jpsj_97, shan_prb_05,zhuravel_prl_13}. In this case, the presence of Andreev surface-bound states can also provide a paramagnetic contribution to the shielding supercurrent. However, this contribution is unable to render the total Meissner response paramagnetic in large superconducting samples \cite{suzuki_prb_14}. Moreover, it has been shown \cite{fauchere_prl_99} that repulsive interactions in the normal metal of an SN bilayer could induce a midgap bound state (residing at the Fermi level) at the interface. In turn, this led to a paramagnetic Meissner response. The common aspect of both these scenarios is thus the appearance of surface-states, which strongly suggests an intimate link between these and the paramagnetic Meissner response. In Sec. \ref{sec:ABS}, we shall indeed show that midgap-surface states in superconductors are always accompanied by \ow pairing which explains the unconventional shielding response whenever such states are present.\\

We finally mention that metastable paramagnetic Meissner effects have been shown to originate from effects which are not related to unconventional superconductivity, but rather to flux capturing at the surface of small superconductors \cite{geim_nature_98}. Care must thus be exerted when interpreting the physical origin of paramagnetic Meissner measurements.

\subsection{Vortex cores}\label{sec:vortex}

When translational symmetry is absent, one expects additional superconducting correlation
components with different symmetry properties than the leading instability channel to
be generated. For instance, as will be discussed in detail in Sec. \ref{sec:hetero}, interfaces between
superconductors and non-superconducting materials break translational symmetry and
thus serve as a source for \ow pairing. However, there are other ways to break
translational symmetry apart from creating hybrid structures. A conventional BCS $s$-wave
superconductors will also break translational symmetry in its bulk when vortices appear.
Applying a magnetic field $H$ that exceeds the lower critical field $H_{c1}$ of a type II
superconductor leads to the formation of vortices, which have a normal core of size
$\xi_S$ and a flux core of size $\lambda$ where $\lambda > \xi_S$. In the clean limit where
the impurity scattering time is long, low-energy bound states $E<\Delta$ are generated inside
the normal core of the vortex \cite{caroli_pl_64}, assisted by the pair potential $\Delta$ vanishing in the center
of the vortex. This leads to an enhancement of the zero-energy density of states locally
in the vortex core, an effect which has been observed via scanning tunneling microscope (STM)
measurements \cite{hess_prl_89, gygi_prb_91, fischer_rmp_07}.

These so called Caroli-de Gennes-Matricon states are in fact a manifestion of \ow superconductivity, as shown by Yokoyama \etal~\cite{yokoyama_prb_08}. More specifically,
they showed that for a vortex with vorticity $m$ in a superconductor,
the pairing function of the Cooper pair at the vortex center
has the opposite symmetry with respect to frequency compared to that of the bulk if $m$
is an odd integer. For a
conventional vortex with $m= 1$, corresponding to a phase-winding of $2\pi$ around the
vortex core, the zero-energy
local DOS would thus be enhanced at the center of the vortex
core in an \ew superconductor due to the generation of \ow Cooper pairs. At the center
of a vortex core in a conventional ballistic $s$-wave superconductors, \ow $p$-wave pairing
would thus arise. Conversely, if the vorticity $m$ is an even integer, the Cooper pairs
at the vortex core would have the same pairing symmetry with respect to frequency as
the leading instability of the bulk.

The above conclusions were obtained based on a quasiclassical approach which allows one
to distinguish between the \ew and \ow superconducting correlations. This is a powerful theory to use as long as one is interested in physical quantities that change slowly compared to the Fermi wavelength, for instance on the scale of the superconducting coherence length $\xi$. The essence of the method \cite{serene_physrep_83,rammer_rmp_86,belzig_sm_99} is to integrate out the high energy degrees of freedom corresponding to the rapid, small-scale oscillations in the Green function describing particle and hole propagators. One is left with the low energy behavior near the Fermi level, which is suitable for describing systems where the Fermi energy $E_F$ is much larger than any other energy scale.\\

\begin{figure}[h!]
\includegraphics[scale=0.8]{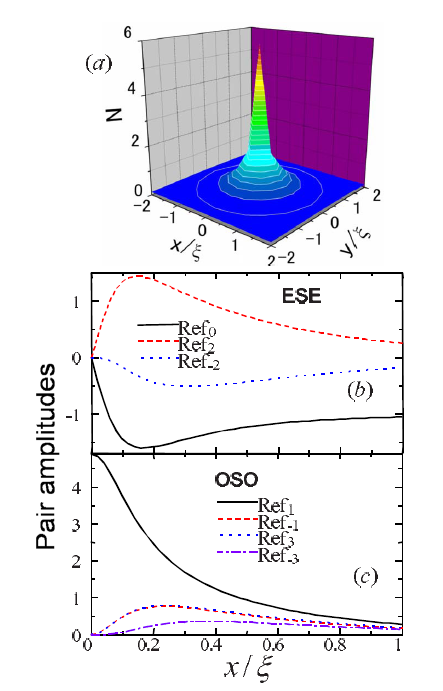}
\caption[fig]{\label{fig:vortex} (Color online) Results for the DOS and Cooper pair symmetry
near the vortex core of a conventional $s$-wave BCS superconductors. (a) Normalized
local DOS around the vortex at $E=0$. The center of
the vortex is situated at $x=y= 0$. Spatial dependencies of (b) \ew singlet (ESE)
and (c) \ow singlet (OSO) correlations at $E= 0$. $f_j$ corresponds to the different angular momentum components of the anomalous Green functionf $f = \sum_n f_n\e{\i n\theta}$, and all have a spin-singlet symmetry. Figure adapted from \cite{yokoyama_prb_08}.}
\end{figure}

To describe the electronic structure of the vortex core in a single Abrikosov vortex in a
ballistic superconductor, the Ricatti-parametrized Eilenberger equation was used in
\cite{yokoyama_prb_08}. Considering the Eilenberger equation along a quasiparticle
trajectory $\boldsymbolr(x) = \boldsymbolr_0 + x\hat{\boldsymbol{v}}_F$ where $\hat{\boldsymbol{v}}_F$
is the Fermi velocity unit vector reduces the problem to solving two decoupled
differential equations for the quantities $a(x)$ and $b(x)$:
\begin{align}
\hbar v_F\partial_xa((x) + [2\omega_n + \Delta^\dag a(x)]a(x) - \Delta=0,\notag\\
\hbar v_F\partial_xb(x) - [2\omega_n + \Delta b(x)]b(x) + \Delta^\dag = 0.
\end{align}
Above, $\omega_n$ is the Matsubara frequency whereas $\Delta^\dag$ is defined as
$\Delta^\dag =\Delta^*$ for an \ew superconductor and $\Delta^\dag=-\Delta^*$ for an
\ow superconductor.
With the solutions for $a$ and $b$, one then obtains both the anomalous Green function describing
the symmetry of the Cooper pair correlations $f=-2a/(1+ab)$ and the local DOS at position
$\boldsymbolr_0$ and energy $E$ normalized
to its value in the normal state:
\begin{align}
N(\boldsymbolr_0,E) = \int^{2\pi}_0 \frac{d\theta}{2\pi} \text{Re}\Big[ \frac{1-ab}{1+ab} \Big] _{\i\omega_n \to E+\i \delta}
\end{align}
where $\delta$ represents inelastic scattering usually taken as $\delta \ll \Delta_0$ and $\theta$ denotes the quasiparticle trajectory according to $\boldsymbol{v}_F = v_F(\cos\theta\hat{\boldsymbol{x}} + \sin\theta\hat{\boldsymbol{x}})$.
Focusing on the experimentally most relevant case of a bulk \ew BCS superconductor,
one can choose the following form of the pair potential in order to incorporate the effect
of a vortex:
\begin{align}
\Delta(\boldsymbolr,\theta) = \Delta_0 F(r) \e{\i m\phi},
\end{align}
where $F(r) = \tanh(r/\xi_S)$ describes the spatial profile of the gap while the phase-winding
associated with a vortex core of vorticity $m$ is described by $\e{\i m\phi}$ where
$\e{\i\phi} \equiv (x+\i y)/\sqrt{x^2+y^2}$. Solving the above equations gives the normalized
local DOS at $E=0$ shown in Fig. \ref{fig:vortex}(a) and the spatial dependences of the \ew
superconducting correlations at $E=0$ in Fig. \ref{fig:vortex}(b) and the \ow correlations in Fig.
\ref{fig:vortex}(c) \cite{yokoyama_prb_08}.

The DOS near the vortex core features the characteristic zero-energy peak, which is well-known, but Figs. \ref{fig:vortex}(b) and (c) show a more surprising result: only \ow Cooper pairs (the $f_1$ component to be specific) exist at the vortex core. Moving away from the core, all components are suppressed except the one corresponding to the bulk order parameter, namely the $s$-wave \ew function $f_0$. The zero-energy state in a superconducting vortex is thus a direct signature of \ow correlations. Moreover, the fact that it is the odd-parity component $f_1$ that exists at the vortex core is consistent with the experimental fact that the zero-energy peak is highly sensitive to disorder \cite{renner_prl_91}, which inevitably would suppress $p$-wave pairing and thus $f_1$. To connect this observation with the claim that all known examples obey simple design principles, we note that this set up converts  \ew $S = -1, P =+1, O = +1, T = +1 $ pairs into \ow pairs with $S = -1, P = -1, T = -1, O = +1$ where $P$ is now parity of the amplitude inside the  vortex core. It was further shown in \cite{yokoyama_prb_08} that if one instead considered a bulk \ow superconductor with a conventional vortex of vorticity $m=1$, only \ew pairing existed at the core, causing a suppression of the DOS at $E=0$.

The relation between \ow pairing and vortex core states in more exotic chiral $p$-wave superconductors was studied in
\cite{daino_prb_12}. In contrast to most previous works regarding \ow pairing at the time, the authors went beyond the quasiclassical
regime $\Delta \ll E_F$ and considered the quantum limit where $\Delta \sim E_F$. Zero-energy states appearing in
half-quantum vortex cores of chiral $p$-wave superconductors are Majorana bound-states \cite{read_prb_00, ivanov_prl_01} and it was shown in \cite{daino_prb_12}
how these states are related to emergent \ow superconductivity in the vortex core. The two were found to be strongly correlated: when zero-energy Majorana states
were present, the \ow triplet anomalous Green function had precisely the same spatial structure as the local density of states revealing the Majorana modes.
However, for finite energy bound states in the vortex-core of a chiral $p$-wave superconductor, the correspondence between \ow pairing and the density of states
depends on the vortex winding relative the chirality of the order parameter \cite{daino_prb_12}. Further aspects of \ow Cooper pairs near vortices in chiral $p$-wave superconductors were studied in \cite{tanuma_prl_09, tanaka_prb_16}. Yokoyama \etal~determined how \ow pairing arises in the vortex lattice that is present in the Fulde-Ferrell-Larkin-Ovchinnikov vortex state \cite{yokoyama_jpsj_10}. Finally, Bj{\"o}rnson \etal~\cite{bjornson_prb_15} studied the relation between \ow pairing and Majorana states bound to vortex cores in semiconductor/superconductor heterostructures.


\subsection{Multiband systems}\label{sec:multiband}

 In the single-band case, an order parameter with a $s$-wave and spin-singlet symmetry must necessarily be an \ew superconductor, and so forth (see Table. \ref{tab:symmetries}). In the multiband case, this is no longer the case. The reason for this is that the transformation of the Cooper pair wavefunction under an exchange of \textit{band-indices} $O$  also comes into play as part of the SPOT = -1  constraint. In this subsection, we also treat multichannel and multiorbital models since they, similarly to the multiband case, also are characterized by the fermion operators acquiring an extra quantum number index which becomes part of the Pauli principle requirement.

Following \cite{blackschaffer_prb_13b}, and as discussed previously in this review, it is convenient to introduce the generalized parity operators below which have the following effect on the two electrons that comprise the Cooper pair:
\begin{itemize}
\item Spin parity $S$: exchanges the spin-coordinates.
\item Spatial parity $P$: exchanges the positions.
\item Orbital parity $O$: exchanges the band indices.
\item Time parity $T$: exchanges the time-coordinates.
\end{itemize}
In the single-band case, the Pauli-principle dictates $PST = -1$. In the multiband case, one instead has $SPOT = -1$. In this way, it is possible to generate for instance \ew $s$-wave triplet superconducting correlations, which is not permitted in the single-band case. Formally, the operators act as follows on the general superconducting anomalous Green function defined in Eq. (\ref{eq:fundamentalf}):
\begin{align}
S f_{\alpha\beta, ab}(\boldsymbolr,t)S^{-1} &= f_{\beta\alpha, ab}(\boldsymbolr,t),\notag\\
P f_{\alpha\beta, ab}(\boldsymbolr,t)P^{-1} &= f_{\alpha\beta, ab}(-\boldsymbolr,t),\notag\\
O f_{\alpha\beta, ab}(\boldsymbolr,t)O^{-1} &= f_{\alpha\beta, ba}(\boldsymbolr,t),\notag\\
T f_{\alpha\beta, ab}(\boldsymbolr,t)T^{-1} &= f_{\alpha\beta, ab}(\boldsymbolr,-t).
\end{align}
Here, $\boldsymbolr=\boldsymbolr_1-\boldsymbolr_2$ and $t=t_1-t_2$ are the relative space- and time-coordinates.

It was shown in \cite{blackschaffer_prb_13b} that \ow pairing should appear ubiquitously in the multiband case. The authors started with a generic two-band superconductor model as an example of the simplest case:
\begin{align}
H &= \sum_{\vk\sigma} \epsilon_{a,\vk} a_{\vk\sigma}^\dag a_{\vk\sigma} + \epsilon_{b,\vk} b_{\vk\sigma}^\dag b_{\vk\sigma} \notag\\
&+\sum_{\vk\sigma} (\Gamma_\vk a_{\vk\sigma}^\dag b_{\vk\sigma} + \text{h.c.})
+\sum_{\vk} (\Delta_{a,\vk} a_{\vk\uparrow}^\dag a_{-\vk\downarrow}^\dag \notag\\
&+ \Delta_{b,\vk} b_{\vk\uparrow}^\dag b_{-\vk\downarrow}^\dag + \text{h.c.}).
\end{align}
Here, $a_{\vk\sigma}^\dag$ is the creation operator for an electron in band $a$ with momentum $\vk$ and spin $\sigma$, and equivalently for band $b$, $\Gamma_\vk$ is the hybridization between the bands, and $\epsilon_{a(b),\vk}$ is the band dispersion. The hybridization $\Gamma_\vk$ will in general have a finite value in realistic systems, for instance if the superconducting pairing occurs in a basis of atomic or molecular orbitals where the kinetic energy is not fully diagonal, as proposed for the iron-pnicitide superconductors \cite{moreo_prb_09}. It will also occur in the presence of disorder-induced interband scattering \cite{komendova_prb_15}. By diagonalizing the kinetic energy into two new bands $c$ and $d$, a set of intraband ($\Delta_c$ and $\Delta_d$) and interband ($\Delta_{cd}$) superconducting order parameters appear. Focusing on the $s$-wave singlet pairing amplitude denoted $F^\pm(t)$, one finds a contribution which is even $(+)$ in the band indices and one that is odd $(-)$:
\begin{align}
F^\pm(t) \equiv \frac{1}{2N_\vk} \sum_\vk \mathcal{T}_t \langle c_{-\vk\downarrow}(t) d_{\vk\uparrow}(0) \pm d_{-\vk\downarrow}(t)c_{\vk\uparrow}(0)\rangle,
\end{align}
where $c_{\vk\sigma}$ and $d_{\vk\sigma}$ are fermion operators for the previously defined bands $c$ and $d$ while $N_\vk$ is the number of points in the first Brillouin zone. Moreover, $F^\pm(t)$ can be even or odd in the relative time coordinate $t$. Since the \ow amplitude must vanish at $t=0$, it is natural to define the singlet $s$-wave amplitude with $O=+1$ as $F_\text{\ew} \equiv F^+(t\to0)$, but it is not immediately clear how the \ow amplitude should be defined as it vanishes at equal-times. However, it is in fact still possible to define an equal-time order parameter for the \ow amplitude in the same way as Eq. (\ref{EQ:OP1})] by considering the \textit{time derivative} at equal times:
\begin{align}
F_\text{\ow} \equiv \frac{\partial F^-(t)}{\partial t}\Bigg|_{t\to0}
\end{align}
as the \ow pairing amplitude is necessarily accompanied by the $P=-1$ symmetry for a singlet $s$-wave order parameter. It was found in \cite{blackschaffer_prb_13b} that the \ow amplitude would in general be finite, whether intraband pairing is present or not. In the special case of exclusive interband pairing in the diagonal kinetic energy basis $(\Delta_c=\Delta_d=0)$, one finds the analytical expression
\begin{align}
F_\text{ow} = \frac{\i}{2 N_\vk} \sum_\vk \frac{\Delta[\eta \text{sinh}(\frac{\epsilon_c-\epsilon_d}{2k_BT}) + (\epsilon_c-\epsilon_d)\text{sinh}(\frac{\eta}{2k_BT})]}{\eta[\text{cosh}(\frac{\epsilon_c-\epsilon_d}{2k_BT}) + \text{cosh}(\frac{\eta}{2k_BT})]}.
\end{align}
where $\eta = \sqrt{(\epsilon_c+\epsilon_d)^2+4|\Delta|^2}$ and $\Delta\equiv\Delta_{cd}$. This shows that \ow odd-interband pairing (meaning $O=-1)$ is always present in a superconductor that has even-interband interaction between the electrons as long as the bands are non-identical, $\epsilon_c\neq\epsilon_d$, which is ensured when $\Gamma_\vk\neq0$. More generally, \ow pairing exists if there is finite intraband pairing $\Delta_c$ and $\Delta_d$ so long as an interband pairing of the \ew type is present.

The induction of \ow superconductivity hybridization (single-quasiparticle scattering) between two superconducting bands in a multiband superconductor was also studied in \cite{komendova_prb_15}, where an interesting signature in the density of states was identified. The \ow correlations were shown to cause hybridization gaps located at higher energies than the superconducting gaps which could constitute an experimentally measurable signatures of odd-frequency pairing in multiband superconductors.

The multiband case was further explored in \cite{asano_prb_15}, including also the case of spin-orbit interactions. The authors showed that band hybridization not only generates \ow correlations, but in general also gives rise to \ew Cooper pairs whose symmetry is distinct from that of the original order parameter itself. This result also extends to the multilayer case \cite{parhizgar_prb_14} where the layer index plays the role of the band. \Ow-pairing arising in the bulk of the two-band superconductor MgB$_2$ has also been discussed \cite{aperis_prb_15}, but we cover this scenario in more detail in the next section. Recently, Komendova and Black-Schaffer~\cite{komendova_prl_17} predicted the existence of bulk odd-frequency superconductivity in a multi-orbital model of Sr$_2$RuO$_4$ as a result of hybridization between different orbitals in the normal state, suggesting an intrinsic Kerr effect as the experimental probe.

\begin{figure*}[t!]
\includegraphics[scale=0.7]{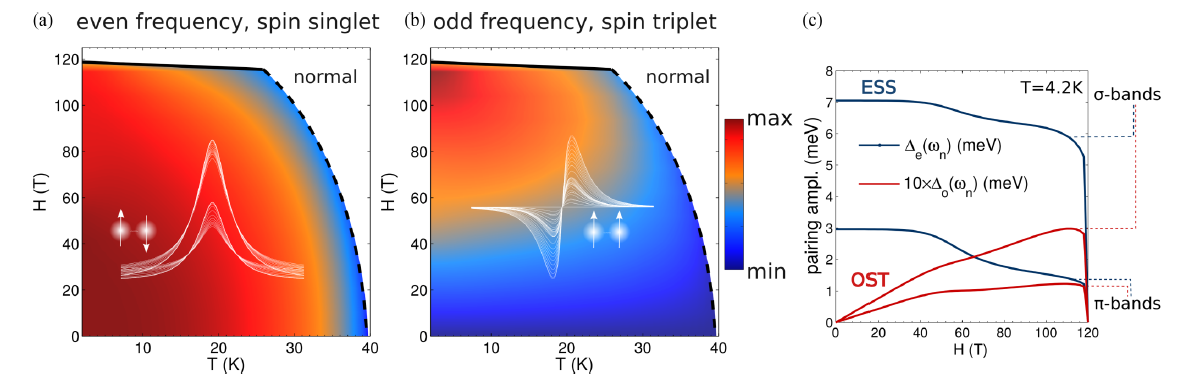}
\caption[fig]{\label{fig:mgb2} (Color online) (a) The $H-T_\text{temp}$ phase diagram for the \ew superconducting order parameter in MgB$_2$. Dashed (solid) lines indicate a second (first) order phase transition. (b) $H-T_\text{temp}$ phase diagram for the \ow superconducting order parameter. The insets in both (a) and (b) show the Matsubara frequency dependence of the order parameters for different magnetic field values. The color bar max/min values are 7 mev/0 meV for the \ew amplitude and 0.3/0.0 meV for the \ow amplitude. (c) The band-resolved field dependence of the \ew and \ow order parameters at low temperature. The lines correspond to the maximum values in Matsubara space of the momentum averaged superconducting order parameters on each band, which is equivalent to the peaks in the insets of (a) and (b). Adapted from \cite{aperis_prb_15}.}
\end{figure*}

The possibility of bulk \ow superconductivity realized in multichannel Kondo systems \cite{cox_aip_98} has also been studied in several works ever since the pioneering work of Emery and Kivelson \cite{emery_prb_92} who showed that an exact solution of the anisotropic two-channel Kondo problem in the continuum limit was permissible under specific conditions. In turn, this implied that an \ow pairing instability might also appear in the lattice case. A large number of work have since then investigated the two-channel Kondo and Anderson lattice models, the latter taking into account the $f$-electron charge degrees of freedom. Jarrell \etal \cite{jarrell_prl_96} examined the two-channel Kondo lattice model with quantum Monte Carlo simulations in the limit of infinite dimensions and found a superconducting transition to an odd-frequency channel. Anders studied composite triplet pairing in the two-channel Anderson lattice model \cite{anders_epjb_02} and found that an \ow superconducting phase developed out of a non-Fermi liquid phase. The order parameter in this case was comprised of a local spin or orbital degree of freedom bound to triplet Cooper pairs with an isotropic and a nearest-neighbor form factor. The scenario of \ow composite pairing in the context of heavy-fermion superconductors was further examined by Flint \etal~\cite{flint_prl_10, flint_prb_11}. Using dynamical mean-field theory combined with continuous-time quantum Monte Carlo simulations, Hoshino and Kuramoto found an \ow-superconducting pairing instability which was equivalent to a staggered composite-pair amplitude with even frequencies \cite{hoshino_prl_14}. A mean-field description of \ow superconductivity with a staggered ordering vector and its implication for the Meissner effect was provided in \cite{hoshino_prb_16}.

Interestingly, order parameters with an \ow symmetry have recently been studied beyond superconductivity in multiorbital systems. In particular, a new type of composite-ordered state in multi-orbital Hubbard systems, the so-called
spontaneous orbital selective Mott state, which may be regarded as a state with a nonzero odd-frequency orbital moment, was studied in \cite{hoshino_prl_17}.

\subsection{Josephson and tunneling effects}\label{sec:josephson}

Here we discuss a number of effects one should expect when investigating the Josephson effect in the context of \ow pairing. When two superconductors are coupled in a tunneling junction, a Josephson effect is permitted: a supercurrent flow driven by the U(1) phase-difference $\varphi$ between the superconducting order parameters. The precise nature of such a Josephson coupling depends on the symmetries of the order parameters in the two superconductors. The lowest order term in the hopping matrix element gives rise to a $\sin\varphi$ dependence  when there is no orthogonality between the symmetries of the order parameters in the spin, parity, frequency, or band channels. For instance, considering an $s$-wave singlet superconductor such as Al and a $p$-wave triplet superconductor such as UGe$_2$, the lowest order Josephson coupling would vanish due to the orthogonality in both spin and parity channel between the superconductors. It should be noted that such a strict orthogonality is only relevant when spin-orbit interactions can be neglected, since the latter generates a mixture of parity components. Below, we first describe the Josephson effect when at least one bulk \ow superconductor is present and then give an exposition of how Josephson-induced intralead \ow correlations appear even for conventional \ew superconductors.

\subsubsection{Josephson effect between \ow and \ew frequency states}

Consider the case of a Josephson effect in a junction where one of the component is \ow. According to the above argument, one might expect that the Josephson effect between an \ow and \ew superconductor should vanish to lowest order, so that the first non-trivial contribution to the supercurrent would be $\sin 2\varphi$, corresponding to tunneling of "pairs of Cooper pairs" with charge $4e$ \cite{abrahams_prb_95}.
However, it was realized more than a decade later \cite{tanaka_prl_07b} that, contrary to what has previously been believed, a first harmonic coupling was in fact possible between \ew and \ow superconductors in the form of $\cos\varphi$ rather than $\sin\varphi$. The physics behind this phenomenon can be understood by considering role of the interface separating the superconductors, which breaks translational symmetry \cite{tanaka_prl_07b, eschrig_jltp_07}. As a result, additional parity components in the superconducting order parameter are generated near the interface region where the superconducting correlations vary spatially. This means that near in the \ew superconductor, an \ow component with opposite parity symmetry of the \ew component is generated near the interface region. Similarly, in the \ow superconductor, an \ew component is generated close to the interface, and a Josephson coupling now becomes possible. Its peculiar $\pi/2$ shift, manifested as a $\cos\varphi$ current-phase relation, means that the Josephson coupling breaks time-reversal symmetry as a consequence of the frequency-symmetries of the superconductors being different.

The lowest order Josephson coupling was also found to be restored in a diffusive junction, where only $s$-wave pairing can survive due to impurity scattering so that no parity mixing exists, consisting of an \ow and \ew superconductor separated by a ferromagnet (F). Due to the magnetic exchange field in F, \ow and \ew components would mix due to their differing spin symmetries \cite{linder_prb_08a} and restore the Josephson coupling. The Josephson coupling between different types of superconductors with various symmetries in spin- and frequency-space have also been studied in \cite{fominov_prb_15, hoshino_prb_16}.

We mention briefly here that dissipative transport in the form of quasiparticle tunneling and Andreev reflection is also different for \ow superconductors compared to the usual BCS case. Fominov \cite{fominov_jetp_07} studied the conductance of a diffusive junction consisting of a normal metal in contact with an $s$-wave triplet \ow superconductor, with the motivation to suggest a simple experimental setup that would still be sensitive to the \ow dependence of the superconducting state. The fundamental process of Andreev reflection in N/S bilayers was indeed found to be sensitive to the \ow symmetry of the order parameter. An effective low-energy behavior $f^\text{R} = \Delta(E)/\sqrt{[\Delta(E)]^2 -E^2}$ with constant $a$ and $\Delta(E) = E/(1+ a^{-2})$ was chosen as a model for an \ow superconductor in \cite{fominov_jetp_07}, where it was established that the conductance of the junction could exceed the normal-state value even in the tunneling limit, in stark contrast to conventional \ew superconductors, in spite of the vanishing Andreev reflection amplitude at $E \to 0$ in the \ow case. The conductance of ballistic junctions N/S junctions with \ow superconductors having different parity symmetries was studied in \cite{linder_prb_08b}, where an enhanced conductance at low bias voltages compared to the conventional spin-singlet \ew case was also found.

Most of the works
giving predictions for experimentally verifiable properties of the \ow state so far have focused on an indirect property, such as
the spin-polarization of the \ow triplet state imposed by the Pauli principle in dirty systems. However, such a spin-polarization is
not unique for the \ow state and a true smoking gun signature should arguably instead be related to the \textit{time-dependence}
of the order parameter.  The lowest order Josephson
coupling between an even- and odd-frequency superconductor in an SIS tunneling junction vanishes \cite{abrahams_prb_95} [although an
inverse proximity effect can restore it \cite{tanaka_prl_07b}] for symmetry reasons, both for the DC and AC effect. However one could envision that
by applying either an AC voltage or alternatively causing the tunneling matrix elements to be time-dependent by using \eg capacitors, the AC Josephson effect between an even- and odd-frequency superconductor should be restored.

Coupling to the \ow order parameter with an explicitly time-dependent perturbation and in this way inducing an otherwise absent Josephson effect would help to reveal  the existence of this superconducting state.



\subsubsection{Josephson effect induced Berezinskii components}

\begin{figure}[h!]
	\includegraphics[scale=0.38]{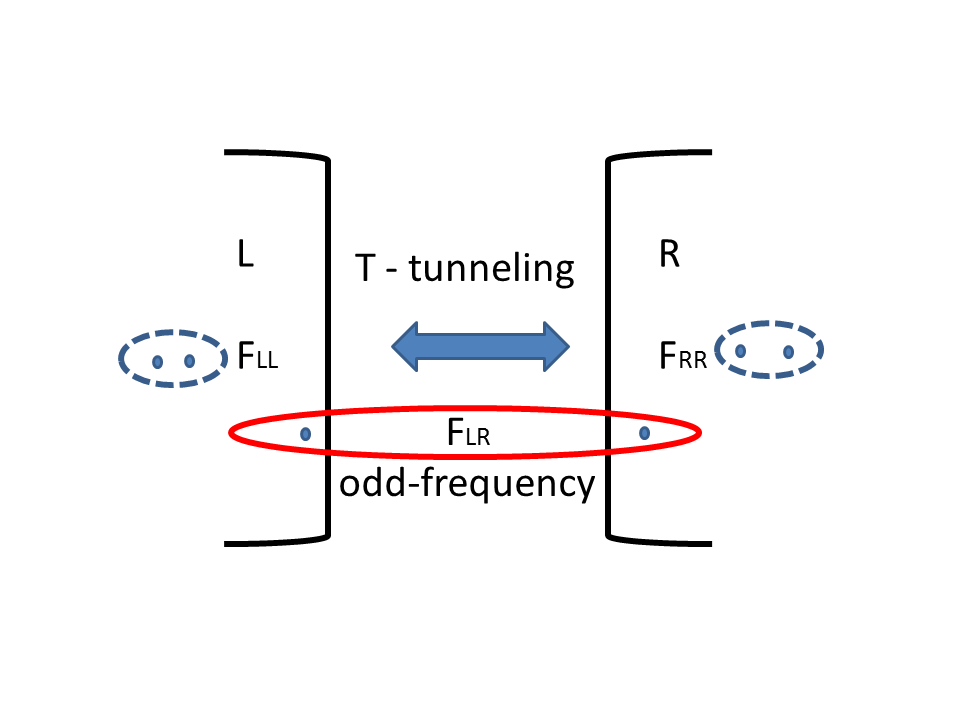}
	\caption[fig]{\label{JJFig1} (Color online) Schematics of the conventional Josephson junction is shown. Interlead tunneling induces diagonal pairing amplitudes $F_{LL}, F_{RR} \sim T_\text{tun}^2$,  $T_\text{tun}$ being the tunneling matrix element. We also indicate the presence of \ow inter-lead pairing amplitude $F_{LR} \sim T_\text{tun} (\Delta_L - \Delta_R) \omega$ that is \ow,  odd under $L \to R$ permutation, while preserving the product $SPOT = -1$. In addition to the conventional Cooper pairs present in each of the leads tunneling induces the interlead superconducting correlations. Traditional textbook analysis predicts the corrections to the intralead pairing and explains the Josephson effect as an induction of the $T_\text{tun}^2\text{Re}\{\Delta^*_R \Delta_L\}$ term in free energy.  The interlead pairing amplitude is much larger at small tunneling amplitudes $T_\text{tun}$ \cite{balatsky_unpub_17}.}
\end{figure}

As discussed, Berezinskii pairing components are generated and modified in the presence of interfaces.   We now illustrate how \ow component is generated in a Josephson junction (JJ) effect between two conventional superconductors, as shown in Fig. \ref{JJFig1} \cite{balatsky_unpub_17}.  We start by considering point-like tunneling between the leads of BCS superconductors. Tunneling between the left ($L$) and right ($R$) leads is given by tunneling matrix element $T_\text{tun}$. There are native pairing correlations which are diagonal in the junction index (intralead pairing), $F_{LL}, F_{RR}$. Josephson pointed out the coherent pair tunneling between superconducting leads. Yet in the discussion of the effect all the attention is devoted to the {\em tunneling of the pairs}. At no point in time the real, non-virtual pair breakup is allowed. The new observation is that there are also \textit{interlead} correlations $F_{LR}$ {\em present} in conventional Josephson effect. It is these $LR$ correlations that are found to be \ow on the very general grounds: it naturally follows from the $SPOT$ classification. The $L$ and $R$ leads of the JJ represent effectively a new discrete index that can be viewed as a band index: band $L =$ left lead and band $R =$ right lead. Using the junction index $L,R$ as an effective orbital index, pairing correlations can be even and odd in this index.

 Consider for simplicity only the spin singlet component of the pairing correlations, $S = -1$. For any allowed pairing due to  Berezinskii classification,  the remaining product $POT = 1$ where again $P$ interchanges the spatial coordinates in the pair, $O$ is the lead ($=$ band) permutation operator, and $T$ interchanges the time coordinates. Two possible pairing states may be generated due to tunneling  in the conventional Josephson effect: intralead singlet \ew

\begin{equation}
F_{LL}, F_{RR},\; (\ S = -1, \ P = \ T = \ O = +1)
\end{equation}

 and interlead, \ow singlet correlations

 \begin{equation}
 F_{LR} = - F_{RL},\; (S = -1, \ P = +1, \  T = \ O \ = -1)
 \end{equation}

 While keeping $POT  = 1$, one thus immediately realizes that the \ow, odd junction (orbital index) pairing is allowed. Previous  literature focused on the intralead ($LL, RR$) corrections due to tunneling. These corrections are of the order $T_\text{tun}^2$. The \ow interlead correction is {\em linear} in $T_\text{tun}$ and hence is largest in the case of weak tunneling. The intralead corrections due to tunneling are well studied. The Josephson phase coupling between the superconductors emerge as a result of Cooper pair tunneling and the effect is {\em even order} in the tunneling matrix element $T_\text{tun}$. To lowest order they are quadratic  $\sim T_\text{tun}^2$  for a low transparency barrier.  An \ow interlead amplitude instead emerges to {\em odd order}, to keep the pair amplitude odd under $ L \leftrightarrow R$ permutation,  and thus is linear in $T_\text{tun}$. This separation of the even- and odd- in $T_\text{tun}$ components is general and will hold for a barrier of any transparency.  In this sense, the \ow component is more robust than the \ew in the Josephson junction as it emerges even in lower order in $T_\text{tun}$. We now outline the proof, following \cite{balatsky_unpub_17}.

Consider the JJ Hamiltonian with
\begin{equation}
H = H^L_{BCS} + H^R_{BCS} + T_\text{tun} c^{\dag L}_{s}(r = 0)c^R_{s}(r = 0) + h.c.
\label{Eq:JJ1}
\end{equation}
where $H^{LR}_{BCS}$ is the BCS like Hamiltonian for $L,R$ leads taken independently, $s$ being the spin index. Each lead will have respective dispersions of quasiparticles $\varepsilon_{L,R}(k)$ and the respective gaps $\Delta_{L,R}$. We assume that tunneling is spin independent, is occuring at one point $r =0$, and we consider effects to lowest order in $T_\text{tun}$. Higher order terms have also been calculated and checked: as is intuitively reasonable, they will modify the scale of the effect but not the symmetry. Hence, for the easiest illustration we keep the analysis confined to lowest order in $T_\text{tun}$.

One can introduce a normal and anomalous correlation function $G$ and $F$. Each of these correlators will have the lead index and one can expect Green functions of the following type:  $G_{LL}, G_{RR}, G_{LR}, F_{LL}, F_{RR}, F_{LR}$, (leaving aside obvious indices). Let us define:

\begin{align}
G_{ij, ss'}({\bf k},\tau) = -\langle \mathcal{T}_{\tau} c^{\dag}_{is}({\bf k},\tau)c_{js'}(-{\bf k},0)\rangle
\end{align}
and
\begin{align}
F_{ij, ss'}({\bf k},\tau) = -\langle {\mathcal{T}_\tau} c_{is}({\bf k},\tau)c_{js'}(-{\bf k},0) \epsilon_{ss'} \rangle
\end{align}
where $i,j = L,R$ and $\epsilon_{ss'}$ is the projector to spin singlet pairs one considers here. Using standard methods it can be shown that

\begin{align}
F_{LR, ss'}(r=0,i\omega_n) &= T_\text{tun}\sum_{{\bf k}, {\bf k'}}[G^0_{LL}({\bf k}, i\omega_n)F^0_{RR, ss'}({\bf k'},i\omega_n)+ \notag\\
&+F^0_{LL,ss'}({\bf k},i\omega_n)G^0_{RR}({\bf k'},-i\omega_n)].
\label{Eq:FLR1}
\end{align}

The summation over ${\bf k,k'}$ in Eq.(\ref{Eq:FLR1}) is carried out independently and hence one deals with the quasiclassical Eilenberger functions. Simple algebra yields

\begin{align}
F_{LR,ss'}(r = 0, i\omega_n) = (\pi N_0)^2 T_\text{tun} \epsilon_{ss'} \frac{i\omega_n(\Delta_L - \Delta_R)}{D_LD_R}
\label{Eq:FLR2}
\end{align}
with $D_{L,R} = \sqrt{ \omega^2_n + |\Delta_{L,R}|^2}$. We indeed see that induced inter-lead component is singlet, \ow, odd in the lead index and is {\em linear} in  tunneling matrix element $T_\text{tun}$.

Several observations are in order here. Firstly, the induction of the \ow interlead SC amplitude occurs even in the case of a conventional Josephson effect between conventional superconductors.  This unexpected finding supports our claim about the ubiquity of the \ow states in the presence of the underlying \ew states.  An \ow interlead component is in fact expected to emerge immediately in any JJ.  The physical picture is similar to the induction of the \ow component in the multiband superconductors due to conversion of conventional pairs into \ow pairs. In this particular case, the \ow component is induced as a result of the intralead pairing correlations that leak into to the opposite lead and generate \ow interlead correlations. A possible reason for why these pairing correlations have not been discussed previously is due to the dynamic nature of the interlead pairing.

Secondly, $F_{LR}$ represents the tunneling induced entanglement between two leads. As the leads are coupled, we can view them as a degenerate two level system. Hence, it is natural to expect that Rabi-like oscillations are induced by a phase difference between the leads. Indeed, from Eq. (\ref{Eq:FLR2}) we can estimate the real time behavior of the $F_{LR}$.  For the case of identical leads with $\Delta_{L,R} = \Delta \exp(\i\phi_{LR})$ one can easily find the time dependence of $F_{LR}$. In the zero-temperature limit, one obtains
\begin{align}
F_{LR,ss'}(r = 0, t) &= i\epsilon_{ss'} 4\pi^3 N^2_F T_\text{tun} \notag\\
&\times\Delta \exp (i\Theta) \sin \varphi \sin(\Delta t)
\label{Eq:FLR3}
\end{align}
with $\Theta = (\varphi_L + \varphi_R)/2$, $\varphi = (\varphi_L - \varphi_R)/2$, and $N_F$ being the DOS at the Fermi level. The coherent Rabi-like oscillations of the interlead pair amplitude with the frequency $\Omega = \Delta$ reinforces the notion of a connection of \ow states to time crystals (tX). Indeed, some would argue that even the dc Josephson effect with the oscillating Josephson current can be viewed as tX; the system spontaneously violates translational symmetry in time as only a dc voltage is applied. In the case of \ow oscillations, we see that the interlead correlations develop a time dependent correlation without any voltage. Therefore, the system spontaneously violates time translation due to oscillations in the off diagonal pairing amplitude. Oscillations are present only as long as the phase difference is maintained, $F_{LR} = 0$ for $\phi_L  =\phi_R$. As long as the finite phase difference across the junction the system is maintained the junction is in the non-equilibrium steady state. As such one concludes that the Berezinskii state can only exist for finite phase difference across the junction. 
The connection of the \ow state and any other dynamical order including tX is a fascinating idea that will likely be explored more in the future.

Finally, the standard results for the free energy as a function of the phase difference and Josephson current are not modified to linear order in $T_\text{tun}$ and the presence of the \ow interlead component does not change the established results.  Hence, one would need to have a nonlocal observable to reveal the interlead \ow component. A physical observable that could reveal the presence of odd-frequency interlead pairing is the nonlocal spin susceptibility, which is predicted to be finite at low temperatures for a fully gapped $s$-wave superconductor, and proportional to second power of Josephson current \cite{balatsky_unpub_17}. Both predictions are quite striking: a non-exponential susceptibility for a fully gapped system would clearly point to a non-BCS states. The current dependence is a consequence of the Eq.(\ref{Eq:FLR3}).



We also mention that the AC Josephson effect for odd-frequency superconductors have not been considered so far in the literature.   The AC Josephson effect could potentially probe the dynamic nature of \ow correlations and offer a direct signature of the \ow Berezinskii superconductivity.

\subsection{Candidate materials}\label{sec:materials}

Even in the absence of a bulk \ow pairing state, \ow superconductivity arises at the interface to other materials or vacuum under quite general circumstances. This will be discussed in detail in the next chapter, and \ow pairing also arises at surface of superfluids such as $^3$He \cite{mizushima_prb_14, higashitani_prb_12}. However, can \ow pairing exist as a pairing instability in the bulk of a material? This question has historically been a controversial one, as suggested by our previous discussion regarding the stability of the \ow pairing state and the sign of the Meissner effect. While several works have shown that a diamagnetic bulk \ow pairing state is in principle possible \cite{belitz_prb_99, solenov_prb_09, kusunose_jpsj_11b}, it should be noted that \cite{fominov_prb_15} concluded oppositely. As of today, there is no clear consensus on the microscopic mechanism that would underlie this phenomenon. Nevertheless, several works have in recent years attemped to establish a model that would yield an \ow pairing instability, both primary and subdominant, with direct relevance to existing materials.

To investigate this issue, an appropriate framework to use is the one due to Eliashberg where the frequency-dependence of the pairing interaction and gap function are fully taken into account. Aperis \etal~\cite{aperis_prb_15} used the anisotropic Eliashberg framework to study pairing in the two-band superconductor MgB$_2$ which is known to have two Fermi surfaces of $\pi$ and $\sigma$ character, respectively. On its own, MgB$_2$ does not show any signs of \ow pairing. Using \textit{ab initio} calculations it was shown \cite{aperis_prb_15} that an applied magnetic field would generate a considerable \ow order parameter in the bulk of MgB$_2$. Confirming the highly anisotropic $s$-wave two-gap structure of MgB$_2$ with $\Delta_\pi=2.8$ meV and $\Delta_\sigma=7$ meV in the absence of a
magnetic field, it was shown in \cite{aperis_prb_15} that an \ow triplet state appeared
and coexisted with a conventional \ew pairing state in the $H$-$T_\text{temp}$ phase diagram where $H$ is an external magnetic field (see Fig. \ref{fig:mgb2}), when neglecting orbital effects. As an experimental signature of the emergence \ow
bulk pairing state $\Delta_\text{\ow}(\boldsymbolk,\omega)$, the authors computed the spin-resolved electronic density of states
\begin{align}
\frac{N_\sigma(\omega)}{N_F} = \frac{1}{2}\text{Re}\Bigg\{ \langle \frac{|\omega + \sigma \tilde{H}(\boldsymbolk,\omega)}{\sqrt{[\omega+\sigma \tilde{H}(\boldsymbolk,\omega)]^2 - [\Delta_\sigma(\boldsymbolk,\omega)]^2}} \rangle_\boldsymbolk\Bigg\}
\end{align}
where we defined the total order parameter:
\begin{align}
\Delta_\sigma(\vk,\omega) \equiv \Delta_\text{\ew}(\boldsymbolk,\omega) + \sigma\Delta_\text{\ow}(\boldsymbolk,\omega).
\end{align}
Moreover, $\langle \ldots \rangle_\vk$ denotes Fermi surface averaging, $\tilde{H}$ is a renormalized magnetic field including self-energy effects, and $n_F$ is the density of states at the Fermi level in the non-superconducting state. Using self-consistent \textit{ab initio} calculations, the magnetic field evolution of the tunneling spectra showed clear subgap features. Due to the imaginary part of the \ow order parameter being finite, $\text{Im}\{\Delta_\text{\ow}(\vk,\omega)\} \neq 0$, a finite density of states arises at $\omega=0$ which would be absent if $\Delta_\text{\ow}(\vk,\omega)=0$. The physical origin of the imaginary part is damping processes of quasiparticle excitations caused by the magnetic field, which broadens the quasiparticle lifetime \cite{aperis_prb_15}.
These results reinforce the broader possibilities of inducing \ow pairing states in multiband superconductors \cite{triola_prb_17}. As mentioned previously, bulk \ow superconductivity has also recently been predicted \cite{komendova_prl_17} in a multiorbital model of Sr$_2$RuO$_4$ when taking into account orbital hybridization.

A bulk \ow superconducting state had also been proposed earlier \cite{fuseya_jpsj_03} for CeCu$_2$Si$_2$ in order to explain unusual experimental features, such as gapless superconductivity coexisting with antiferromagnetism \cite{kawasaki_prl_03} even in very clean samples. The existence of \ow pairing in heavy fermion superconductors in fact dated back to the early work by Coleman and co-workers \cite{coleman_prl_93}.
The key idea of Fuseya \etal~was that an \ow $p$-wave singlet superconducting pairing state could be realized close to the quantum critical point and/or in the coexistent superconducting and antiferromagnetic state. This state was shown \cite{fuseya_jpsj_03} to
arise to critical spin fluctuations, granted that two conditions were fulfilled. First, the pair scattering interaction was required to host a sharp peak as a function of frequency with a width smaller than the thermal energy. Secondly, the dominant process for pair scattering with the antiferromagnetic ordering vector $\boldsymbolQ$ would have to be weakened by the nodes in a competing \ew $d$-wave singlet state. The authors argued that it could be  reasonable to assume that these criteria were fulfilled in CuCu$_2$Si$_2$. Spin fluctuations and nesting also played a key part in the work by Johannes \etal~\cite{johannes_prl_04}, who proposed that the most compatible superconducting pairing state with the nesting structure of Na$_x$CoO$_2 \cdot y$H$_2$O featured an \ow $s$-wave triplet symmetry.

A possible bulk \ow pairing state on a quasi one-dimensional triangular lattice was proposed in \cite{shigeta_prb_09}. Starting with the single-band Hubbard model on an anisotropic triangular lattice
\begin{align}
H = \sum_{\langle i,j\rangle,\sigma} (t_{ij}c_{i\sigma}^\dag c_{j\sigma} + \text{h.c.}) + \sum_i U n_{i\uparrow}n_{i\downarrow}\; (n_{i\sigma} = c_{i\sigma}^\dag c_{i\sigma}),
\end{align}
the authors computed the Green function in the case of half-filling in both the random-phase approximation and the FLEX approximation. By linearizing the Eliashberg equation in the singlet (triplet channel):
\begin{align}
\lambda\Delta(\omega_n,\boldsymbolk) &= - \frac{T_\text{temp}}{N} \sum_{m,\vk'} V^{s(t)}(\omega_n-\omega_{m'},\boldsymbolk-\boldsymbolk') G(\omega_m,\boldsymbolk') \notag\\
&G(-\omega_m,-\boldsymbolk') \Delta(\omega_m,\boldsymbolk')
\end{align}
and inserting the effective pairing interactions
\begin{align}
V^s(\omega_m,\boldsymbolq) = U + \frac{3}{2}U^2\chi_s(\omega_n,\boldsymbolq) - \frac{1}{2}U^2\chi_c(\omega_m,\boldsymbolq),\notag\\
V^t(\omega_n,\boldsymbolq) = -\frac{1}{2}U^2\chi_x(\omega_m,\boldsymbolq) - \frac{1}{2}U^2\chi_c(\omega_m,\boldsymbolq)
\end{align}
the pairing state providing the highest critical temperature could be computed. Above, $T_\text{temp}$ is the temperature, $N = N_x\times N_y$ is the number of $\boldsymbolk$-point meshes on the lattice, $\chi_s$ and $\chi_c$ is the spin and charge susceptibility, while $G(\omega_m,\boldsymbolk)$ is the Green function determined by the Dyson equation:
\begin{align}
G^{-1}(\omega_n,\boldsymbolk) = G_0^{-1}(\omega_n,\boldsymbolk) - \Sigma(\omega_n,\boldsymbolk).
\end{align}
$G_0$ is the bare Green function while $\Sigma$ is the self-energy. In the regime where the hopping along one direction, say $t_x$, of the lattice dominated the other hopping terms, the authors found that the \ow singlet state provided the largest $T_c$ using an onsite interaction $U/t_x=1.6$.

\begin{figure}[h!]
\includegraphics[scale=0.5]{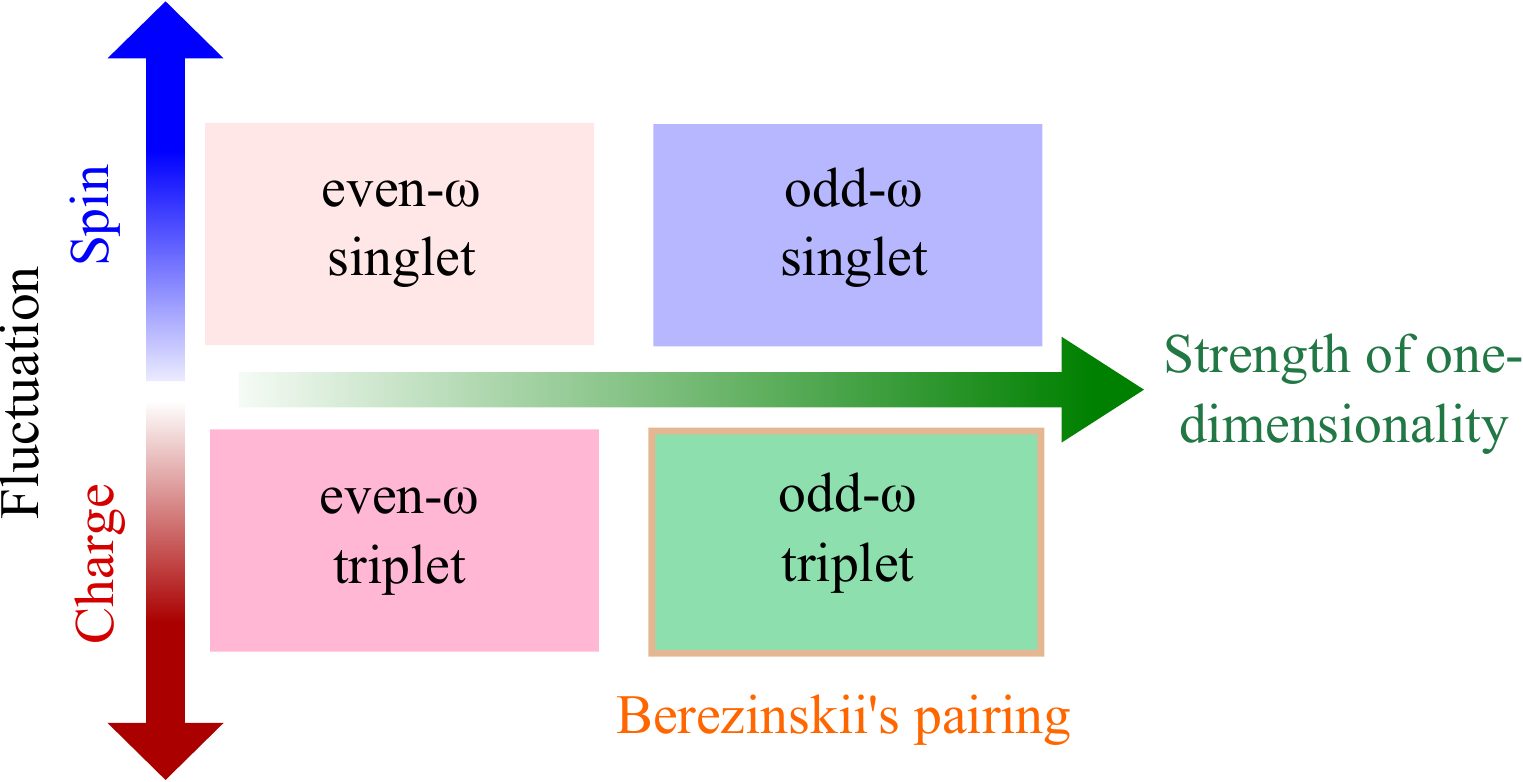}
\caption[fig]{\label{fig:onedimow} (Color online) Qualitative dependence of the most stable superconducting pairing symmetries on the degree of one-dimensionality and spin/charge fluctuations. Adapted from \cite{shigeta_prb_11}.}
\end{figure}

A further step toward identifying a clear mechanism for generating \ow superconductivity in a bulk material was taken in \cite{shigeta_prb_11}. The authors noted that in the context of quasi one-dimensional systems, such as the organic superconductor $($TMTSF$)_2X$, spin-triplet $f$-wave pairing could become favorable compared to singlet $d$-wave pairing when the charge fluctuations strongly exceeded the spin fluctuations. At the same time, a quasi one-dimensional geometry should favor on-site pairing ($s$-wave) of electrons to form Cooper pairs. Taking these two facts into account, it would thus appear that the geometrical constraint resulting from a quasi one-dimensional setup combined with strong charge fluctuations should provide the ideal scenario for realizing $s$-wave triplet pairing, which due to the Pauli principle must have an \ow symmetry. This is precisely the same type of pairing as in the original proposal by Berezinskii. Shigeta \etal~\cite{shigeta_prb_11} investigated this via the linearized Eliashberg framework described above using the extended Hubbard model on a quasi one-dimensional lattice, the latter in the sense that the hopping parameter $t_y$ in the $y$-direction was much smaller than $t_x$ in the $x$-direction. Their main result was that the \ow triplet state provided the highest $T_c$ when the charge fluctuations exceeded the spin fluctations. The favored superconducting state is schematically shown in Fig. \ref{fig:onedimow}.

\section{\Ow pairing in heterostructures}\label{sec:hetero}
Having reviewed the properties of \ow pairing in bulk superconductors, \eg where this type of superconductivity is the leading instability, we now turn our attention to a different type of situation. In hybrid structures with conventional BCS superconductors, where $s$-wave spin-singlet pairing is the leading instability, it turns out to be possible to induce \ow pairing under quite general circumstances, where Berezinskii component is induced as a result of scattering, consistent with the SPOT constraint and design rules. The \ow pairing in this way can either exist in the non-superconducting part of the heterostructure itself, by means of the proximity effect, or even be created as a subdominant pairing amplitude in the superconductor itself.

\subsection{Normal-superconductor}\label{sec:NS}

\begin{figure}[h!]
\includegraphics[scale=0.8]{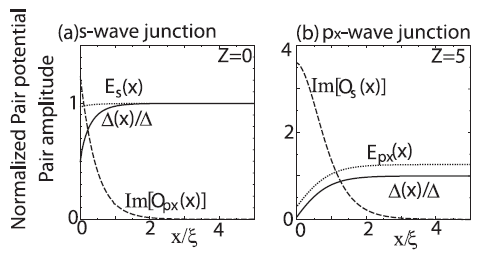}
\caption[fig]{\label{fig:SN} (Color online) Spatial dependence of the pair potential normalized against its bulk (solid line) and the \ew spin-singlet pair amplitudes $E_s(x)$ ($s$-wave channel) and $E_{p_x}(x)$ ($p$-wave channel) for an SN ballistic bilayer. The $x$-axis extends into the superconducting layer. The \ow pair amplitudes in the corresponding angular momentum channels are denoted $O_s(x)$ and $O_{p_x}(x)$. The parameter $Z$ quantifies the junction transparency, with $Z=0$ corresponding to a perfect interface and $Z\gg1$ corresponding to the tunneling limit. In (a), the superconductor is of the conventional $s$-wave BCS type whereas in (b) the superconductor is of the $p_x$ type. The ballistic superconducting coherence length is $\xi = v_F/\Delta$. Figure adapted from \cite{tanaka_prl_07b}.}
\end{figure}

It is interesting to note that the prediction of \ow pairing in the conceptually most simple heterostructure, a superconductor/normal metal bilayer, came later than its prediction in more complex heterostructures involving magnetic materials \cite{bergeret_prl_01}.
Tanaka \etal~\cite{tanaka_prl_07b, tanaka_prb_07a} and Eschrig \etal~\cite{eschrig_jltp_07} established in 2007 that magnetic ordering was in fact not required to generate \ow pairing in hybrid structure: any type of inhomogeneous superconducting state, such as a spatially inhomogeneous one due to the presence of an interface, must host \ow pairing. This means that even a ballistic S/N bilayer would allow for the existence of \ow pairing due to the broken translational symmetry. An $s$-wave \ew spin-singlet state would transform into a $p$-wave \ow spin-singlet state near the interface region, preserving its spin symmetry (see Fig. \ref{fig:SN}).

Following Ref. \cite{eschrig_jltp_07}, a solution of the Eilenberger equation in a balllistic S/N bilayer provides the following anomalous Green function $f_s$ in the N region, the subscript $s$ denoting that it is a spin-singlet correlation:
\begin{align}
f_s^{(l)} = T_\text{int} \frac{\pi\Delta}{|\omega_n|}[\text{sgn}(\omega_n)]^l Q_l(2|\omega_n|x/v_F),
\end{align}
where $Q_l$ is a purely real function whose details are not important for the present purpose, while $l$ denotes the angular momentum quantum number of the Cooper pair: $l=0$ for $s$-wave, $l=1$ for $p$-wave, and so on. Moreover, $T_\text{int}$ is the transparency of the interface while $v_F$ is the Fermi velocity. All the odd components in $l$ are clearly seen to have an \ow symmetry due to the factor $[\text{sgn}(\omega_n)]^l$.

The possibility to induce \ow pairing in a normal metal without the requirement of magnetic ordering had in fact been noted a few years earlier \cite{tanaka_prb_05a, asano_prl_07, tanaka_prl_07a}, but in these works the authors proposed to use a spin-triplet superconductor as the host. This meant that \ow triplet pairing was generated at the interface, which could survive even in diffusive normal metals where frequent impurity scattering would suppress any non $s$-wave amplitude (higher order angular momentum) due to the Fermi surface averaging.

An interesting consequence of the fact that \ow pairing can appear in a normal metal is that it is intimately linked to a phenomenon discovered in the 1960s, namely McMillan-Rowell oscillations \cite{rowell_prl_66, rowell_prl_73}. This effect consists of the density of states in a normal metal connected to a superconductor displaying a series of sharp subgap peaks, indicating the presence of resonant energy levels in the system. In \cite{tanaka_prb_07a}, the authors showed that the energies $\varepsilon$ where the  McMillan-Rowell peaks occurred coincided precisely with the points where \ow pairing amplitude $f_\text{\ow}(\varepsilon)$ would strongly dominate over the \ew pairing amplitude $f_\text{\ew}(\varepsilon)$, their ratio $f_\text{\ow}/f_\text{\ew}$ in fact formally diverging. The conclusion is that the McMillan-Rowell oscillations can be taken as direct evidence of \ow pairing.

To show this \cite{tanaka_prb_07a}, one may consider the case of a long $N$ region $L=5L_0$ where $L_0 = v_F/2\pi T_c$ is a measure of the superconducting coherence length ($T_c$ is the bulk superconducting critical temperature). Focusing for simplicity on the case of a fully transparent interface, the local DOS acquires a series of peaks arising due to electron-hole interference effects in the N region (precisely the McMillan-Rowell peaks). The amplitudes of the corresponding \ew and \ow components can be computed via quasiclassical theory by solving the Eilenberger equation which in the notation of \cite{tanaka_prb_07a} takes the form:
\begin{align}
\i v_{F,x} \hat{g}_{\pm} = \mp [\hat{H}_\pm, \hat{g}_\pm],
\end{align}
where we defined
\begin{align}
\hat{H}_\pm = \i\omega_n \hat{\tau}_3 + \i \Delta_\pm(x)\hat{\tau}_2.
\end{align}
and their ratio is found to depend on both energy and position in the N region. Here, $v_{F,x}$ is the component of the Fermi velocity in the direction normal to the SN interface, $\omega_n=2\pi T_\text{temp}(n+1/2)$ is the Matsubara frequency and $\Delta_\pm$(x) is the pair potential for left/right-going quasiparticles. Solving this equation for the Green function matrix $\hat{g}_\pm$ and applying suitable boundary conditions (we do not go into details on this matter here, as these are technically too comprehensive to fully account for here), one is able to identify an odd-frequency component $f_\text{\ow}$ and even-frequency component $f_\text{ew}$. Their ratio is:
\begin{align}\label{eq:ratiof}
\frac{|f_\text{\ow}|}{|f_\text{\ew}|} = \Big|\tan\Big( \frac{2E}{v_{F,x}}(L+x)\Big)\Big|.
\end{align}
At the edge of the normal region $(x=-L)$, the \ow component vanishes for all energies. In contrast, at the SN interface ($x=0$) it does not in general and Eq. (\ref{eq:ratiof}) then establishes a direct relation between the energy of the bound-states forming the resonances in the system and the ratio $f_\text{ow}/f_\text{\ew}$. To see this, consider the bound-state energy derived \cite{rowell_prl_66, rowell_prl_73} in the limit $L\gg L_0$ for a perfect interface transparency:
\begin{align}\label{eq:boundmcmillan}
E_n = \frac{\pi v_{F,x}}{2L}(n+ \frac{1}{2}),\; n=0,1,2,\ldots
\end{align}
Inserting Eq. (\ref{eq:boundmcmillan}) into Eq. (\ref{eq:ratiof}) one obtains
\begin{align}
\frac{|f_\text{\ow}|}{|f_\text{\ew}|} = |\tan(\pi/2 + \pi n)| \to \infty.
\end{align}
In effect, the ratio between \ow and \ew correlations diverges precisely at the subgap peak energies where the McMillan-Rowell resonances exist.

\Ow pairing in SN hybrid structures has also been investigated for the case of unconventional (non $s$-wave) superconductors \cite{tanaka_prl_07a, asano_prl_11, matsumoto_jpsj_13, lu_prb_16}. The general rule is that unless some spin-dependent interactions are present, either in the form of a magnetic exchange field in the normal region or due to spin-active scattering at the interface, the induced \ow pairing in SN structures will have the same symmetry in the spin-part of the Cooper pair correlation function as the host superconductor. Thus, for a normal metal/$p$-wave triplet superconductor (such as SrRu$_2$O$_4$) bilayer, the induced \ow correlations would have a spin-triplet symmetry and can thus survive even in a diffusive normal metal \cite{tanaka_prb_04} due to the orbital part being even. However, they are not necessarily restricted to one particular angular momentum channel: in general, higher order angular momentum pairing is also generated, such as $d$-wave in the above example, but with decreasing magnitude.

The first study of \ow pairing and its relation to zero-energy surface states in normal metal junctions involving unconventional superconductors such as $p$-wave (presumably relevant for SrRu$_2$O$_4$ and ferromagnetic superconductors such as UGe and UCoGe) were reported in by Tanaka and co-workers \cite{tanaka_prb_05a, tanaka_prb_05b, tanaka_prberratum_06}. Before discussing these findings, it is instructive to establish a more general understanding of the interplay between zero-energy states and how the proximity effect is manifested in normal metal/unconventional superconductor systems, including the $d$-wave case relevant for the high-$T_c$ cuprates \cite{yokoyama_prb_05}.

\begin{figure}[h!]
\includegraphics[scale=0.4]{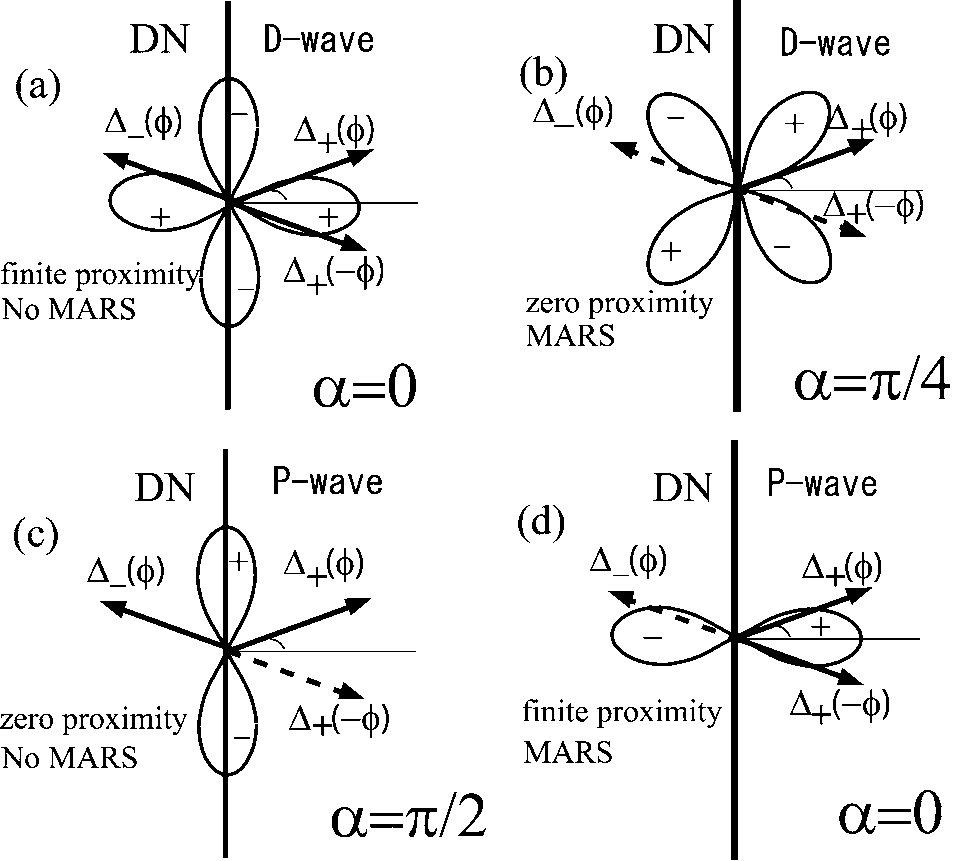}
\caption[fig]{\label{fig:marsproxy} (Color online) The arrows illustrate the trajectories of scattered quasiparticles at the interface between a diffusive normal metal and an unconventional superconductor with a $d$-wave symmetry [(a) and (b)] and a $p$-wave symmetry [(c) and (d)]. The angle $\alpha$ denotes the angle between the normal to the interface and the crystal axis in the $d$-wave case and the lobe direction in the $p$-wave case. The angle $\phi$ denotes the injection angle of quasiparticles as measured from the $x$-axis. Figure adapted from \cite{yokoyama_prb_05}. }
\end{figure}

Considering a diffusive normal metal, as is often the case experimentally, in contact with a $p$- or $d$-wave superconductor as shown in Fig. \ref{fig:marsproxy}. Due to the frequent impurity scattering in the normal part, the effective pair potential felt by quasiparticles near the interface is obtained by averaging over the backscattered half of the Fermi surface. Only when a finite average pair potential exists in this way, can there be a net proximity effect. This is seen to be the case for $p_x$-wave and $d_{x^2-y^2}$-wave pairing, whereas no proximity effect is present in a diffusive normal metal for the crystallographic orientations corresponding to $p_y$-wave and $d_{xy}$-wave pairing. On the other hand, the existence of zero-energy states [denoted MARS (midgap Andreev resonant state in the figure] is based on solely on the orientation of the $\vk$-dependent gap in the superconductor relative the interface. This can lead to interesting situations such as the absence of a proximity effect in spite of the presence of zero-energy states in the $d$-wave case, in contrast to the coexistence of a proximity effect and zero-energy states in the $p$-wave case.

With the above considerations in mind, we can understand why, for certain crystallographic orientations of the interface, \ow-pairing does not arise in diffusive normal metals in contact with $d$-wave superconductors despite the presence of zero-energy surface states. The reason is that the proximity effect (leakage of superconducting Cooper pairs) into the normal region is absent due to the net pair potential experienced upon scattering at the interface averages to zero. On the other hand, this problem is not present for $p_x$-wave pairing and in such a scenario it was shown \cite{tanaka_prb_05a} that \ow superconductivity is induced in the diffusive normal region despite the absence of any magnetism. We also note that more recent work has investigated the appearance of \ow-pairing in normal-superconductor systems when Rashba spin-orbit interactions are present \cite{reeg_prb_15, ebisu_ptep_16}, including an extension to bilayer-superconductor systems \cite{parhizgar_prb_14}. Finally, it has been shown \cite{higashitani_prb_14} that translational symmetry-breaking in non-uniform \ew superconductors also produces \ow pairing by a similar physical mechanism as in S/N heterostructures.


%


\subsection{Ferromagnet-superconductor}\label{sec:FS}

Hybrid structures consisting of ferromagnetic materials in contact with conventional $s$-wave superconductors have historically played the most important role with regard to proximity-induced \ow pairing, both theoretically and experimentally. The key breakthrough theoretically was obtained in 2001 with Bergeret \etal~\cite{bergeret_prl_01} demonstrating that when a diffusive ferromagnetic material with an inhomogeneous magnetic texture, such as a domain wall, was placed in contact with an $s$-wave superconductor, this would induce an \ow triplet component in the ferromagnet. This component would moreover be able to penetrate far into the magnetic region, beyond the range of the conventional \ew singlet
component for strong exchange fields $h \gg \Delta$. This phenomenon became known as the long-ranged proximity effect. This result was also obtained virtually simultaneously by Kadigrobov \etal~\cite{kadigrobov_epl_01}. The \ow dependence of the triplet component that arises in hybrid structures consisting of conventional BCS superconductors and ferromagnets is formally equivalent to the \ow correlations proposed in Ref. \cite{berezinskii_pisma_74}. However, an important difference is that no unusual pairing mechanism is required to obtain the \ow component in hybrid structures, presumably in contrast to the originally proposed \ow pairing by Berezinskii. The physics of \ow pairing in SF structures was reviewed twelve years ago \cite{bergeret_rmp_05}, but since then experimental progress in this field has been substantial. We therefore here focus on the most recent developments regarding \ow pairing in SF hybrid systems which in recent years have emerged as a promising building block for superconducting spintronics \cite{eschrig_rpp_15, linder_nphys_15}.

\subsubsection{Broken spin rotational symmetry}

The broken spin rotational symmetry lies at the heart of the appearance of \ow pairing in a S/F bilayer. As was mentioned in the introduction, the principle is to trade off change in $T$ parity for a change in parity of spin $S$ while keeping the SPOT parity intact.  In the same way as translational symmetry breaking produced higher-angular momentum pairing in the N/S case due to the interface region (see Sec. \ref{sec:NS}), i.e. causing a mixing of different \textit{parity} components of the superconducting anomalous Green function, the broken spin-rotational symmetry caused by the exchange field in a ferromagnet causes a mixing of different \textit{spin} components of the Cooper pairs, i.e. producing both singlets and triplets. In the diffusive limit, only $s$-wave correlations can survive due to the frequent impurity scattering causing an isotropization of all correlations in momentum space. According to the Pauli principle, an $s$-wave triplet component must thus have an frequency-symmetry which is odd under $\omega \to -\omega$. It is important to point out that magnetic inhomogenities are not a prerequisite for \ow pairing, but only for the long-ranged components of these pairing correlations. \Ow pairing indeed arises in an S/F bilayer even if the ferromagnet has a homogeneous exchange field, although in this case the \ow amplitude decays equally fast as the singlet \ew amplitude. To see this, one may compute the proximity-induced correlations in a simple S/F bilayer conveniently using the quasiclassical theory of superconductivity. We perform this calculation explicitly here as it also allows us to recover the S/N result treated in Sec. \ref{sec:NS}. In the diffusive limit, the Usadel equation \cite{usadel_prl_70} governs the behavior of the $4\times4$ Green function matrix $\hat{g}$ which contains both a normal $(2\times2)$ part $\underline{g}$ and an anomalous $(2\times2)$ part $\underline{f}$:
\begin{align}\label{eq:greenmat}
\hat{g} = \begin{pmatrix}
\underline{g}(E,\boldsymbolr) & \underline{f}(E,\boldsymbolr) \\
-\underline{f}^*(-E,\boldsymbolr) & -\underline{g}(-E,\boldsymbolr) \\
\end{pmatrix}.
\end{align}
The normal part describes the propagation of electrons and holes in addition to spin-flip processes. The anomalous part describes the presence of superconducting correlations in the system and is decomposed into the singlet $(f_s)$ and triplet $(f_{\uparrow\uparrow}, f_{\downarrow\downarrow}, f_t)$ components as follows:
\begin{align}
\underline{f}(E,\boldsymbolr) = \begin{pmatrix}
f_{\uparrow\uparrow}(E,\boldsymbolr) & f_{\uparrow\downarrow}(E,\boldsymbolr)\\
f_{\downarrow\uparrow}(E,\boldsymbolr) & f_{\downarrow\downarrow}(E,\boldsymbolr) \\
\end{pmatrix},
\end{align}
where $f_{\uparrow\downarrow}(E,\boldsymbolr) =  f_t(E,\boldsymbolr) +  f_s(E,\boldsymbolr)$ and $f_{\downarrow\uparrow}(E,\boldsymbolr) =  f_t(E,\boldsymbolr) - f_s(E,\boldsymbolr)$. We underline that the singlet component is \ew whereas the triplet components are \ow. In order to obtain analytically transparent results, we assume here that the superconducting proximity effect is weak. Such a scenario is valid either in the case of a temperature close to $T_c$ or if there is a high interface resistance between the superconducting and magnetic material, causing in both cases the induced superconducting correlations in the ferromagnet to be quantitatively weak.

The Green function matrix $\hat{g}$ satisfies in the diffusive limit the Usadel equation
\begin{align}\label{eq:usadel}
D\nabla(\hat{g}\nabla\hat{g}) + \i[E\hat{\rho}_3 + \hat{M} + \hat{\Delta},\hat{g}]=0.
\end{align}
Here, $D$ is the diffusion coefficient, $E$ is the quasiparticle energy, whereas the exchange field $\boldsymbolh$ of the ferromagnet and the order parameter $\Delta$ of the superconductor are described by the matrices
\begin{align}
\hat{M} &= \begin{pmatrix}
\boldsymbolh\cdot\underline{\boldsymbolsigma} & \underline{0} \\
\underline{0} & \boldsymbolh\cdot\underline{\boldsymbolsigma}^* \\
\end{pmatrix},\;
\hat{\Delta} = \begin{pmatrix}
\underline{0} & \Delta \i\underline{\sigma}_y \\
\Delta^*\i\underline{\sigma}_y & \underline{0 }\\
\end{pmatrix}.
\end{align}
In the weak proximity regime, one assumes that $\hat{g}$ only has a small deviation from its normal-state value $\hat{g} = \hat{\rho}_3$ where $\hat{\rho}_3 = \text{diag}(1,1,-1,-1)$. This means that $\hat{g} = \hat{\rho}_3 + \hat{f}$ where $\hat{f}$ is given by Eq. (\ref{eq:greenmat}) with $\underline{g}=0$. Inserting this form of $\hat{g}$ into Eq. (\ref{eq:usadel}) and linearizing the equation in $\hat{f}$, one obtains the following set of coupled equations
\begin{align}\label{eq:weak}
D\nabla^2 f_s + 2\i E f_s + 2\i \boldsymbolh\cdot\boldsymbolf = 0,\notag\\
D\nabla^2 \boldsymbolf + 2\i E \boldsymbolf + 2\i \boldsymbolh f_s = 0,
\end{align}
where we have defined the triplet anomalous Green function vector
\begin{align}
\boldsymbolf = [f_{\downarrow\downarrow} - f_{\uparrow\uparrow}, -\i(f_{\downarrow\downarrow}+f_{\uparrow\uparrow}), 2f_t]/2.
\end{align}
The quantity $\boldsymbolf$ is mathematically equivalent to the $\boldsymbold$-vector commonly used to analyze $p$-wave triplet superconductivity \eg in the context of SrRu$_2$O$_4$ \cite{maeno_rmp_03}.

The functions $f_s$ and $\boldsymbolf$ describe the singlet and triplet superconducting correlations induced in the ferromagnet, respectively. The penetration depth into the magnetic region for the different types of Cooper pairs can be illustrated most simply by considering a magnetic region with a homogeneous exchange field, taking along the $\hat{\boldsymbolz}$-direction for concreteness. Defining $f_\pm = f_t \pm f_s$, the general solution of Eq. (\ref{eq:weak}) reads
\begin{align}
f_\pm = A_\pm \e{\i k_\pm x} + B_\pm\e{-\i k_\pm x},\; k_\pm = \sqrt{\frac{2\i(E\pm h)}{D}},\notag\\
f_{\sigma\sigma} = C_{\sigma\sigma} \e{\i kx} + D_{\sigma\sigma}\e{-\i kx},\; k = \sqrt{\frac{2\i E}{D}}.
\end{align}
The value of the unknown coefficients $\{A_\pm, B_\pm, C_{\sigma\sigma}, D_{\sigma\sigma}\}$ are determined by the boundary conditions of the system \cite{kupriyanov_jetp_88, nazarov_sm_99, cottet_prb_09, cottet_prberratum_11, eschrig_nphys_15}. As there by now are a number of these available in the literature, it is instructive to briefly consider their regime of validity. Continuity of the Green function and its derivative corresponds to a perfectly transparent interface, which substantially simplifies analytical calculations but clearly corresponds to an idealized situation. The Kupriyanov-Lukichev boundary conditions \cite{kupriyanov_jetp_88} are commonly used and are valid for non-magnetic, low-transparency interfaces where the probability $\tau_n$ of tunneling for a given interface channel $n$ is low ($\tau_n \ll 1$). Nazarov \cite{nazarov_sm_99} derived a boundary condition valid for arbitrary transparency $\tau_n$ for a non-magnetic interface. In the presence of a tunneling $(\tau_n\ll 1$) magnetic interfaces, either realized via an explicit magnetic barrier separating a superconductor from a normal metal or simply a superconductor/ferromagnet bilayer, the boundary conditions due to Cottet \etal~\cite{cottet_prb_09, cottet_prberratum_11} are valid under the assumption of a weak magnetic polarization. Recently, Eschrig \etal presented the most general boundary conditions for magnetic interfaces to date \cite{eschrig_nphys_15}, valid for arbitrary polarization magnitude and thus applicable to half-metallic compounds as well.

\begin{figure}[h!]
\includegraphics[scale=0.6]{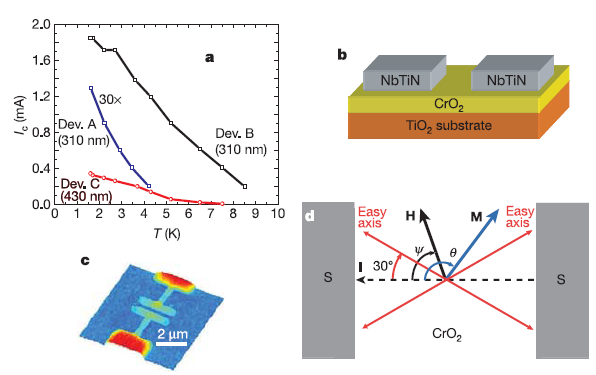}
\caption[fig]{\label{fig:keizer} (Color online) (a) Critical supercurrent as a function of temperature for different separation distances between the superconducting electrodes. (b) Schematic setup of the studied devices, consisting of a lateral Josephson junction with two superconducting electrodes deposited on the half-metal CrO$_2$. (c) Scanning electron micrograph of a typical final device. (d) Illustration of the alignment of the current direction with respect to the magnetization axes: $I$ is the current, $H$ is the applied magnetic field, and $M$ is the magnetization. Figure adapted from \cite{keizer_nature_06}.}
\end{figure}

As we are usually interested in energies close to the superconducting gap $E\sim\Delta_0$ in order to see \eg the signature of the correlations in the density of states, and magnetic exchange fields in ferromagnets typically satisfy $h\gg\Delta_0$, it is clear from the expression for $f_\pm$ that both $f_s$ and $f_t$ decay on a length scale $\xi_f = \sqrt{D/h}$. These Cooper pairs are then said to be short-ranged in the ferromagnet. Values of $\xi_f$ typically takes values from a few nm to (at most) a few tens of nm. On the other hand, the equal spin-pairing Cooper pairs $f_{\sigma\sigma}$ as seen relative the quantization axis $\hat{\boldsymbolz}$ decay on a length scale $\sqrt{D/E}$. As $E\to 0$, this length diverges (in practice, the correlations are limited by the temperature-dependent coherence length $\sqrt{D/T_\text{temp}}$). Therefore, it is clear that such pairs can, once created, penetrate a very long distance into a ferromagnet. The existence of such long-ranged pairs carrying a supercurrent is the commonly accepted explanation for the experiment of Keizer \etal~\cite{keizer_nature_06} where a supercurrent flowing between two superconducting electrodes through $\sim 1\mu$m of half-metallic CrO$_2$ was observed (see Fig. \ref{fig:keizer}). Such long-ranged supercurrent were later also observed by \cite{anwar_prb_10}. We emphasize again that the short-ranged component $f_t$ is \ow and present even in the absence of magnetic inhomogeneities or spin-orbit interactions. We note in passing that a proximity structure consisting of a ferromagnet and the spin-triplet superconductor Sr$_2$RuO$_4$ was recently considered experimentally \cite{anwar_natcom_16}, but no clear signs of \ow pairing were observed.

\begin{figure}[h!]
\includegraphics[scale=0.66]{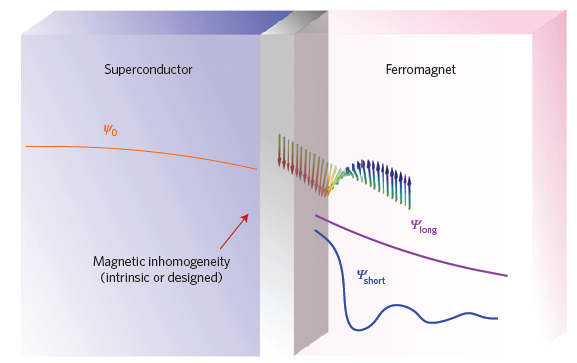}
\caption[fig]{\label{fig:inhomogeneous} (Color online) Starting out with a conventional $s$-wave \ew superconductor described by a wavefunction $\psi_0$, a proximity-coupling to a homogeneous diffusive ferromagnet creates short-ranged \ow Cooper pairs with a wavefunction $\psi_\text{short}$. These rapidly decay in an oscillatory manner inside the magnetic region. In the presence of a magnetic inhomogeneity at the interface, long-ranged \ow Cooper pairs $\psi_\text{long}$  which are spin-polarized (triplet) emerge which penetrate a much longer distance compared to $\psi_\text{short}$. Figure adapted from \cite{linder_nphys_15}. }
\end{figure}

The discovery that the previously hypothesized \ow pairing amplitude \cite{berezinskii_pisma_74} could now actually be experimentally realized in a relatively simple way triggered the interest among several research groups.  Various geometries and structures proposed to date to host Berezinskii \ow state represent different pathways to accomplish conversion of conventional pairs into \ow Berezinskii pairs consistent with the design rules we summarized in the introduction. A key ingredient in most of the proposals was to use magnetic inhomogeneities (see Fig. \ref{fig:inhomogeneous}) of some sort, either in the form of magnetic layers with misaligned magnetizations or magnetic layers featuring an intrinsic texture such as domain wall \cite{bergeret_prb_03}. The reason for this is that if the degree of magnetic inhomogeneity could be controlled, it would provide a mean to turn on and off the long-ranged \ow correlations.

Volkov \etal~\cite{volkov_prl_03} studied a Josephson setup with misaligned magnetic layers and showed that one could control not only the long-ranged proximity effect, but also trigger a transition between 0- and $\pi$-states via the relative magnetization orientation.

Eschrig \etal~\cite{eschrig_prl_03} studied an extreme case of a half-metallic Josephson geometry, where a fully polarized ferromagnet was sandwiched between two $s$-wave superconductors. As only one spin-band existed in the half-metallic region, it would be impossible for singlet Cooper pairs to exist there and any supercurrent carried between the superconductors would have to carried by triplet pairs. In the diffusive limit where the mean free path $l_\text{mfp}$ of the half-metal is much shorter than the superconducting coherence length $\xi_S$ and the length $L$ of the sample, $l_\text{mfp} \ll \{\xi_S, L\}$, an observation of a finite supercurrent could thus be taken as evidence of \ow pairing.
Eschrig \etal~proposed that when spin-flip processes existed at the interface between the superconductor and the half-metal, this would create the long-ranged pairs described by $f_{\uparrow\uparrow}$ (assuming the half-metal magnetization $\boldsymbolm\parallel\hat{\boldsymbolz}$), thus allowing for a finite supercurrent flow. The original proposal considered a ballistic half-metallic junction, where the triplets had an \ew $p$-wave amplitude, but this was later expanded on in Ref. \cite{eschrig_nphys_08} to account for the presence of impurity scattering and where the role of \ow pairing was explicitly discussed. The half-metallic case with spin-active interfaces was also studied in Refs. \cite{asano_prl_07, asano_prb_07,braude_prl_07}, who also pointed out that so-called $\varphi_0$ junction behavior [where the supercurrent-phase relation takes the form $I = I_c\sin(\varphi + \varphi_0)$] could arise for suitably oriented magnetic moments at the interface regions.

\subsubsection{Relation between \ow pairing and zero-energy states}

As mentioned, \ow pairing arises in diffusive structures as soon as the conduction electrons experience a magnetic exchange field, and thus would give rise to observable consequences even in the absence of inhomogeneities. A particular feature that traditionally has been taken as a hallmark property of \ow pairing is that it produces a zero-energy enhancement of the density of states, even exceeding the normal-state value, which is completely opposite to the conventional fully gapped density of states predicted by BCS theory in $s$-wave superconductors such as Nb and Al where no electronic states are available for subgap energies $E<\Delta_0$. To understand the enhancement of the zero-energy density of states, consider the normal Green function $G(\vp,\omega_n)$ of a superconductor which according to Eq. (\ref{eq:oddwgreenfunctions}) has the form (we absorb the factor 2 in front of $|\Delta|$ into the order parameter itself for convenience):
\begin{align}
G(\vp,\omega_n) = \frac{\i\omega_n + \xi_\vp}{\omega_n^2 + \xi_\vp^2 + |\Delta(\vp,\omega_n)|^2},
\end{align}
where $\xi_\vp$ is the kinetic energy, $\omega_n$ is the Matsubara frequency, and $\Delta(\vp,\i\omega_n)$ is the superconducting order parameter. Consider first the case of a BCS \ew superconductor. In this case, $\Delta(\vp,\omega_n) = \Delta$, i.e. it is independent on both momentum (since it is $s$-wave) and frequency. The poles of $G$ (the values of $\omega_n$ which causes the denominator of $G$ to become zero) correspond to the allowed quasiparticle energies and take the form:
\begin{align}
\i\omega_n = \sqrt{\xi_\vp^2+|\Delta|^2}
\end{align}
This is the usual quasiparticle energy for a superconductor, as can be seen after performing an analytical continuation $\i\omega_n \to E + \i 0^+$. Now, consider instead the case of an \ow superconductor (as realized in an S/F structure). In this case, we cannot neglect the frequency dependence of $\Delta$, so we set $\Delta(\vp,\omega_n) = \Delta(\omega_n)$ where now $\Delta(\omega_n) = -\Delta(-\omega_n)$ reflects the odd symmetry while it remains $s$-wave (independent on momentum). For the sake of illustrating the DOS enhancement effect in the simplest way possible, consider an order parameter of the form $\Delta(\omega_n) = \alpha\omega_n$ where $\alpha$ is a constant, which clearly is odd in frequency. This should be a reasonable choice for small frequencies $\omega_n$, since only the lowest order in frequency needs to be retained as $\omega_n\to 0$. The Green function now becomes:
\begin{align}
G(\vp,\omega_n) = \frac{\i\omega_n + \xi_\vp}{\omega_n^2 + \xi_\vp^2 + |\Delta(\omega_n)|^2} = \frac{\i\omega_n + \xi_\vp}{\omega_n^2(1+|\alpha|^2) + \xi_\vp^2}.
\end{align}
In other words, the Green function now looks like that of a \textit{non-superconducting} state $(\Delta=0)$, but with a \textit{renormalized mass}. To see this, observe that the poles of the Green function $G$ now occur at:
\begin{align}
\i\omega_n = \frac{\xi_\vp}{\sqrt{1+|\alpha|^2}}.
\end{align}
In a free electron model where $\xi_\vp = \vp^2/2m$, we see that this corresponds to a mass renormalization $m^* = m\sqrt{1+|\alpha|^2}$. One consequence of this is precisely to enhance the DOS \textit{above its normal-state value}, since the DOS scales as $m^{3/2}$ in a free electron model. This explains why odd-frequency superconductivity allows for gapless excitations and also increases the DOS above its normal-state value. The mass renormalization effect was first noted in \cite{balatsky_prb_92}. A detailed discussion on the restrictions on the exchange field $h$ which would allow clear observation of the zero-energy enhancement of the DOS in S/F structures was given in Ref. \cite{yokoyama_prb_07}.

\begin{figure}[h!]
\includegraphics[scale=0.6]{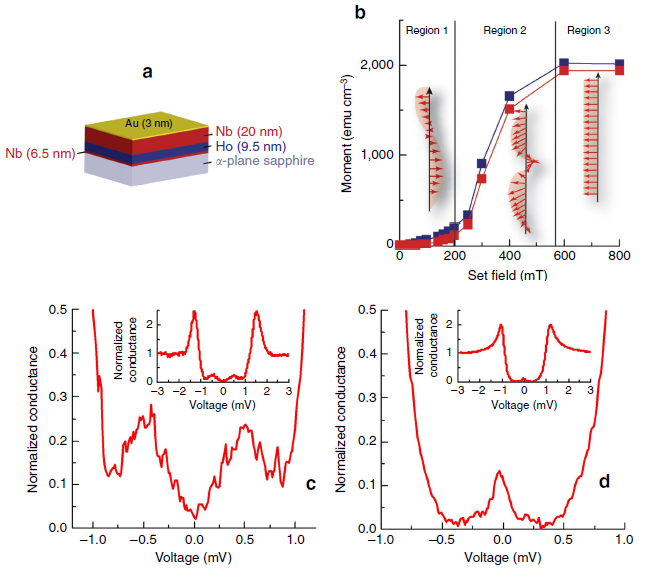}
\caption[fig]{\label{fig:stm} (Color online) (a) The sample structure on which the STM measurements were performed: an Au/Nb/Ho/Nb multilayer. (b) The magnetization of Ho at zero field (remanent magnetization $M_r$: red line) and with the set field $H$ switched on (blue line). The vertical (black) lines separate different magnetic phases of Ho: a bulk helix (region 1), coexisting helix and F component (region 2), and F state (region 3). (c) and (d) show typical subgap features obtained in the normalized conductance.} Figure adapted from \cite{dibernardo_natcom_15}.
\end{figure}


\subsubsection{Further proposals for \ow effects in S/F}

Nearly a decade after the prediction of \ow pairing in S/F structures, the field was enjoying much attention and several proposals were put forth in terms of how one would be able to apply external control over \ow pairing, dictating when it would appear or not, by utilizing for instance spin-active interfaces \cite{linder_prl_09}, multilayered magnetic structures \cite{houzet_prb_07}, or spin-pumping \cite{yokoyama_prb_09}. Several studies focused on the diffusive limit of transport, investigating the signatures of \ow pairing in the experimentally accessible DOS \cite{yokoyama_prb_07, cottet_prb_07, linder_prl_09, linder_prb_10, cottet_prl_11}, whereas Halterman \etal studied the manifestation of \ow pairing in the ballistic limit \cite{halterman_prl_07, halterman_prb_08}. Whereas \ow and \ew superconductivity in general coexists in S/F structures, it is possible to find ways to separate them spatially. One way would be to use very strong ferromagnets, such that any superconducting correlations existing deep inside such a magnetic region would necessarily have a spin-triplet symmetry in order to survive despite the strong local exchange field. This would additionally require some form of magnetic inhomogeneity, as discussed previously. Another way to isolate pure \ow superconductivity without requiring strong ferromagnets or magnetic inhomogeneities is to make use of magnetic insulators as interfaces \cite{linder_prl_09, linder_prb_10}. We now show this in more detail as an practical example of how to use the quasiclassical theory for superconducting proximity structures. Consider a normal metal/superconductor bilayer where the two materials are separated by a magnetic interface, \eg EuO or GdN (the latter particularly compatible crystallographically with the normal metal TiN and the superconductor NbN). Let us start by using the linearized Usadel equations presented earlier in this section, supplemented by the relevant boundary conditions. For this system, the latter should describe a tunneling interface with spin-dependent scattering, meaning that the Kupriyanov-Lukichev boundary conditions expanded to include spin-dependent phase-shifts can be used at $x=0$ (the superconducting interface):
\begin{align}
2L\frac{R_B}{R_N} \hat{g}\partial_x\hat{g} = [\hat{g}_S,\hat{g}_N] + \i \frac{G_\phi}{G_T}[\hat{\tau}_3,\hat{g}_N].
\end{align}
Here, $\hat{g}_{N(S)}$ is the Green function matrix in the N (S) region, $L$ is the length of the N region, $R_B$ ($R_N$) is the resistance of the barrier (normal region), $G_T$ is the barrier conductance, $\hat{\tau}_3 = \text{diag}(1,-1,1,-1)$, and $\hat{g}_S$ is the Green function in the superconducting region. The latter is taken as its bulk value for now, and we later show that the results do not change upon solving the problem self-consistently (accounting for the inverse proximity effect in the superconductor which alters $\hat{g}_S$). At the vacuum N interface, the boundary condition is simply $\partial_x\hat{g}=0$. The key term here is the $G_\phi$ which describes the spin-dependent phase-shifts of quasiparticles being reflected at the interface. Microscopically, $G_\phi$ is determined from \cite{cottet_prb_09} $G_\phi \propto G_q\sum_n d\phi_n$ where $G_q = e^2/h$ is the conductance quantum and $d\phi_n$ is the spin-dependent phase-shift occuring from reflection in interface transport channel $n$. It is defined from the reflection coefficient for spin-$\sigma$ via $r_\sigma = |r_\sigma|\e{\i\phi_n + \sigma d\phi_n}$ where $\phi_n$ is the spin-independent part of the scattering phase. The term $G_\phi$ will in general be present at any magnetic interface (whether one inserts an explicit magnetic insulator or considers an FS interface). Both its sign and magnitude will vary with the magnitude of the interface spin polarization and the precise shape of the spin-dependent scattering potential \cite{grein_prb_13}, and thus it is usually treated as a phenomenological parameter. We note in passing that $G_\phi$ is closely related to the so-called spin mixing conductance which is often used in spintronics \cite{cottet_prb_09}.\\

Solving the linearized Usadel equations (\ref{eq:weak}) with the above boundary conditions, provides the solution \cite{linder_prb_10}:
\begin{align}
f_\pm = \frac{\pm s[\e{\i k(x-2L)} + \e{-\i kx}]}{\i k \frac{R_B}{R_N}L(1-\e{-2\i kL}) + \Big(c \pm \i \frac{G_\phi}{G_T}\Big)(1+\e{-2\i kL})}.
\end{align}
We defined $k=\sqrt{2\i E/D}$ and $s=\text{sinh}(\Theta)$, $c=\text{cosh}(\Theta)$ with $\Theta=\text{atanh}(\Delta/(E+\i\delta)$ and $\delta$ describing the inelastic scattering energy scale $(\delta/\Delta \ll 1)$. Recall that $f_\pm = f_t \pm f_s$ where $f_t = f_\text{\ow}$ is the \ow anomalous Green function while $f_s = f_\text{\ew}$ is the \ew anomalous Green function. In the limiting case of a non-magnetic insulator $G_\phi \to 0$, it is seen that $f_+ = -f_-$, meaning that $f_t=0$. There is no \ow pairing in the system, as expected for a diffusive SN system. However, $f_t\neq 0$ when $G_\phi\neq0$.
The remarkable aspect of the above result is that precisely at the Fermi level $E=0$, where $k=c=0$ and $s=-\i$, one finds
\begin{align}
f_\pm = -G_T/G_\phi.
\end{align}
so long as $G_\phi\neq 0$. In other words, the conventional spin-singlet amplitude has been completely erased and \textit{pure \ow pairing} exists: $f_\pm = f_\text{\ow}$. In fact, even the non-linearized (full proximity effect) Usadel equation can be solved analytically at $E=0$, and one obtains the following result. For $|G_\phi|>G_T$:
\begin{align}
f_\text{\ew} = 0,\; f_\text{\ow} \propto G_T/\sqrt{G_\phi^2 - G_T^2},
\end{align}
whereas for $|G_\phi| < G_T$:
\begin{align}
f_\text{\ew} \propto G_T/\sqrt{G_T^2 - G_\phi^2},\; f_\text{\ow} = 0.
\end{align}
This conversion from pure \ew to pure \ow pairing taking place at $|G_\phi|=G_T$ is a robust effect, as the above results are independent on the interface resistance $R_B$ and the length $L$ of the normal metal, so long it remains below the inelastic scattering length. Moreover, the pure \ow correlations do not exist solely at the superconducting interface, but extend throughout the entire N region so that they can be probed even at the vacuum interface. The experimental signature of this effect can be obtained via STM measurements of the DOS, which acquires the form:
\begin{align}
\frac{N(E=0)}{N_0} = \text{Re}\Bigg\{ \frac{|G_\phi|}{\sqrt{G_\phi^2 - G_T^2}}\Bigg\}.
\end{align}
At zero energy, the DOS has a usual minigap when $|G_\phi|<G_T$ whereas it has a peak that strongly exceeds the normal-state value of the DOS $N_0$ when $|G_\phi|>G_T$. This conversion also takes place in the ballistic limit \cite{linder_prb_10}.

Regarding experimental studies, early work by Kontos \etal~\cite{kontos_prl_01} demonstrated signs of a very weak zero energy peak in SF bilayers (0.5\% of the normal-state value) which was inverted into a suppression at $E=0$ upon altering the F thickness. This was consistent with the predicted oscillatory behavior of the zero energy DOS \cite{buzdin_prb_00}, but was not understood as a signature of \ow pairing at the time. More recently, clear evidence of \ow pairing at SF interfaces was reported \cite{dibernardo_natcom_15} via STM-measurements of Nb superconducting films proximity coupled to epitaxial Ho. By driving Ho through a metamagnetic transition where the magnetization pattern changes from a helical antiferromagnetic pattern to a homogeneous magnetic state, signatures of \ow pairing in the form of substantial subgap peaks (up to 30\% of the normal-state value) were observed (see Fig. \ref{fig:stm}).  \\

\begin{figure}[t!]
\includegraphics[scale=0.8]{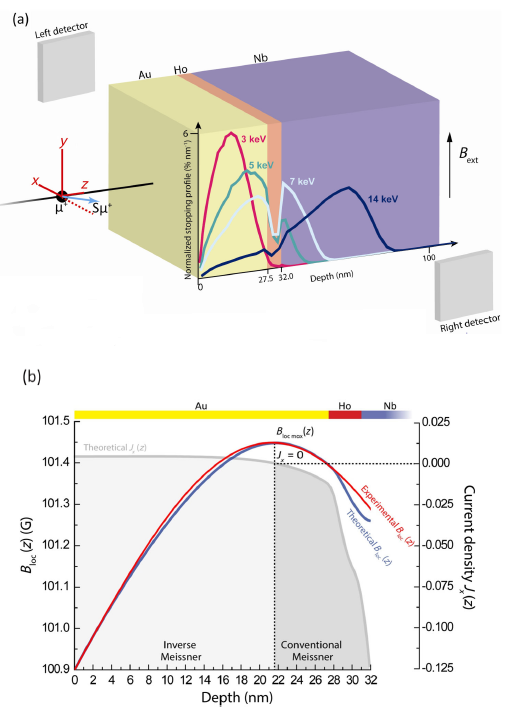}
\caption[fig]{\label{fig:meissnerexp} (Color online) (a) Setup used for observation of the paramagnetic Meissner effect due to \ow-triplets in Ref. \cite{dibernardo_prx_15}: an Au/Ho/Nb trilayer exposed to an external field $\boldsymbolB$. Low-energy muons injected in Au provided information about the local magnetization profile. (b) Experimental measurement and theoretical fit to the local magnetization signal $B_\text{loc}$ as well as the theoretically computed spatial distribution of the shielding current density $J_x$ throughout the system.}
\end{figure}


\subsubsection{Anomalous Meissner effect and spin-magnetization}

Other works discussed the anomalous paramagnetic Meissner effect occurring in proximity-coupled superconductor/ferromagnet layers precisely due to the presence of \ow pairing \cite{yokoyama_prl_11}, a fact which had been noted in earlier work \cite{bergeret_prb_01}. It was recently shown that the paramagnetic Meissner effect becomes highly anisotropic as a function of the field orientation angle $\theta$ in the presence of spin-orbit interactions \cite{espedal_prl_16} as a result of the dependence of the \ow triplet depairing energies on $\theta$.
The effect of a paramagnetic screening current on the induced magnetization in a hybrid structure can be illustrated with a simple quantitative analysis \cite{yokoyama_prl_11}. Consider an SN bilayer with a magnetic interface so that both \ow and \ew correlations can be generated inside the proximitized normal region, as discussed above. Assuming normalized units for brevity of notation, the Maxwell equation determining the magnetic response from a supercurrent can be written as:
\begin{align}
\frac{d^2\boldsymbolA}{dx^2} = -\boldsymbolJ = -J'(x)\boldsymbolA
\end{align}
where $\boldsymbolJ$ is the screening supercurrent density which here is computed via its linear-response to the applied field and resulting presence of a vector potential $\boldsymbolA$.  Moreover, $x$ is the coordinate perpendicular to the SN interface. The induced magnetization (normalized against the externally applied field $B$ reads:
\begin{align}
\boldsymbolM = \frac{d\boldsymbolA}{dx}-1.
\end{align}
This set of equations can be solved by supplying boundary conditions. A crude, but physically reasonable approximation, would be to assume that the superconductor shields completely the external magnetic field whereas the proximity effect is sufficiently weak at the vacuum edge of the normal region so that no screening-induced magnetization exists there. If the SN interface exists at $x=0$ while the vacuum edge resides at $x=1$ (the position coordinate has been normalized to the length of the N region), the boundary conditions take the form:
\begin{align}
\boldsymbolA(x=0) = 0,\; \frac{d\boldsymbolA}{dx}\Bigg|_{x=1} = 1.
\end{align}
For a conventional Meissner response due to \ew pairing, the induced supercurrent is negative: $J'(x) < 0$. Neglecting for simplicity the spatial dependence of the current magnitude $J'(x)$, we may write $J'(x) = -k^2$ where $k$ is a real number, which gives the following solution for the amplitude of the magnetization $\boldsymbolM$:
\begin{align}
M(x) =\frac{\text{cosh}(kx)}{\text{cosh}(k)} - 1.
\end{align}
Since $x\in[0,1]$, $M(x)$ is always negative and decays monotonically away from the vacuum edge as expected for a a conventional Meissner response. In contrast, if a positive supercurrent (anti-screening) is generated due to the presence of \ow Cooper pairs $(J'(x) = k^2 >0)$, one obtains instead:
\begin{align}
M(x) = \frac{\cos(kx)}{\cos(k)}- k.
\end{align}
The proximity-induced magnetization now displays an oscillatory behavior and can assume both positive and negative values. This means that the induction of a \ow pairing supercurrent does not necessarily have to give an inverse (paramagnetic) Meissner response, in the framework of the approximations made in this treatment. \\

An interesting experimental result was achieved in 2015 when Di Bernardo \etal~\cite{dibernardo_prx_15} measured a paramagnetic Meissner response in an Au/Ho/Nb trilayer. The Ho layer consisted of a conical magnetization pattern which created \ow triplet Cooper from singlet pairs leaking in from the superconducting Nb. In turn, these triplet pairs further penetrated into the normal Au region where the local magnetization was measured via low-energy muon spectroscopy (see Fig. \ref{fig:meissnerexp}). Whereas samples without the Ho layer previously had been shown to give a conventional Meissner effect, with a local magnetization induced oppositely to the external $\boldsymbolB$ field, the Au/Ho/Nb trilayer showed an \textit{increased magnetization} below the superconducting critical temperature. The enhancement of the local magnetization above the external field value was shown to be consistent with the presence of \ow pairing.

\begin{figure}[t!]
\includegraphics[scale=0.9]{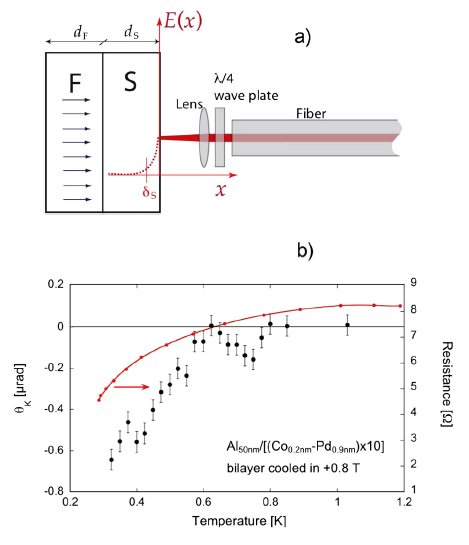}
\caption[fig]{\label{fig:xia}(Color online) (a) Schematic measurement setup used in \cite{xia_prl_09}: two perpendicularly linearly polarized lights emerging from the fiber become cirularly polarized and focus on the sample using a lens. The electric field $E$ penetrates a short distance $\ll d_S$ into the superconductor. (b) Kerr effect measurement of an Al/(Co-Pd) bilayer system with a 50 nm Al-sample. Figure adapted from \cite{xia_prl_09}.}
\end{figure}

A final aspect worth mentioning is how to detect \ow superconductivity indirectly via spin measurements. Due to the symmetry requirements dictated by the Pauli principle with respect to the Cooper pair correlation function at equal times, \ow pairing in the diffusive limit must have a spin-triplet symmetry. In principle, this means that measuring an induced magnetization due to a superconducting proximity effect could be taken as a signature of \ow Cooper pairs. This idea was explored in Ref. \cite{bergeret_prb_04} where an SF bilayer was considered and the magnetization induced in the superconducting part was computed. It was found that the magnetic moment carried by free electrons (non-localized) in the superconductor was oppositely directed to the magnetization in the F region and penetrated a distance of $\sim\xi$, indicating a spin screening effect. The physical origin was proposed to be that $S_z=0$ Cooper pairs which were spatially "shared" between the magnetic and superconducting layer, with one residing in each part (made possible due to the finite spatial extent $\sim\xi$ of the pairs). In this case, the electron with magnetic moment parallell with the magnetization in the F region would energetically favor to stay there, leading to the electron with opposite spin to reside in the superconductor and thus induce an opposite magnetic moment compared to F. Experimental measurements \cite{xia_prl_09} of the polar Kerr effect using a magnetometer on Pb/Ni and Al/(Co-Pd) bilayers provided supporting experimental evidence of such a scenario (see Fig. \ref{fig:xia}). Later work examined the proximity-induced magnetization in both superconducting and non-superconducting regions of magnetically textured systems, demonstrating that the sign and magnitude of $\delta M$ would change depending on parameters such as the spin-dependent phase shifts occuring at the SF interface \cite{linder_prb_09} and the superconducting phase difference in a Josephson junction geometry \cite{hikino_prb_15, hikino_jpsj_17}. It is then clear that \ow triplets can provide a magnetic signal both via their spins and their anomalous Meissner effect.

\subsection{Topological insulator- and quantum dot-superconductor}\label{sec:TQS}

\Ow superconductivity has also been predicted to appear in superconductor-topoplogical insulator heterostructures. Yokoyama \cite{yokoyama_prb_12} showed that attaching an $s$-wave superconductor to the surface of a 3D topological insulator (TI) would induce \ow triplet pairing in the presence of an exchange field. The various types of superconducting correlations induced among the Dirac electrons on the topological surface can be described via an anomalous Green function $2\times2$ matrix $\underline{f}_\text{TI}$ which in the absence of impurity scattering and in the low-doping limit $\mu\to 0$ takes the form \cite{yokoyama_prb_12}:
\begin{align}
\underline{f}_\text{TI} &\propto [-\omega_n^2-(\hbar v_Fk)^2+m^2]\underline{1} + 2\i\omega_n\boldsymbolm\cdot\underline{\boldsymbolsigma} \notag\\
&+ 2\i\hbar v_F(\boldsymbolk_\perp \times \boldsymbolm)\cdot\underline{\boldsymbolsigma}.
\end{align}
The singlet amplitude is proportional to the unit matrix whereas the triplet amplitudes are proportional to $\hat{\boldsymbolsigma}$. As seen, the triplet component has both an \ow part $\propto$ $\omega_n$, appearing when $\boldsymbolm\neq 0$, and an \ew part.   Moving away from the Dirac point $\mu=0$, one finds an additional triplet component $\propto 2\mu\hbar v_F\boldsymbolk_\perp\cdot\underline{\boldsymbolsigma}$ that exists even in the absence of an exchange field. This observation is consistent with the SPOT constraint and {\em design rules} we discussed in the introduction. In this case the $ S = -1, P = +1, T = +1, O = +1$ pair is converted  into i)  $ S = +1, P = +1, T = -1, O = +1$ Berezinskii pair (term proportional to magnetization) and into a ii) $S = +1, P = -1, T = +1, O = +1$ triplet pairs.

In Ref. \cite{blackschaffer_prb_12}, the authors further developed the model of a superconductor-TI interface by taking into account the spatial dependence of the superconducting order parameter $\Delta$ near the interface region. In doing so, they identified an additional contribution to $\underline{f}_\text{TI}$ which existed without any magnetic field, namely an \ow triplet amplitude $\propto \partial_x\Delta\underline{\boldsymbolsigma}/\omega_n$. \Ow pairing will in fact be induced even without an interface so long as a gradient exists in the order parameter, \eg by applying a supercurrent. This result showed that the effective spin-orbit coupling $\boldsymbolk\cdot\underline{\boldsymbolsigma}$ on the TI surface induces \ow triplet pairing without requiring any magnetism. The $1/\omega_n$ dependence had also previously been reported theoretically for \ow pairing heavy fermion compounds \cite{coleman_prl_93}. Interestingly, this particular frequency dependence of the \ow superconducting correlations did not produce any low-energy states which, as discussed previously, usually have been considered one of the smoking gun signatures of \ow pairing. We return to this issue at the end of this subsection.\\

A full symmetry classification of the induced superconducting pairing amplitudes for a superconductor-TI bilayer were reported in \cite{blackschaffer_prb_13a}. This was accomplished using Bi$_2$Se$_3$ as a model TI, in which case the full Hamiltonian of the system takes the form:
\begin{align}
H = H_\text{SC} + H_\text{TI} + H_\text{t}
\end{align}
where $H_\text{SC}$ describes the superconducting part of the system
\begin{align}
H_\text{SC} = \sum_{\vk\sigma} \varepsilon_\vk c_{\vk\sigma}^\dag c_{\vk\sigma} + \frac{1}{2}\sum_{\vk\alpha\beta} [\Delta_{\vk,\alpha\beta} c_{\vk\alpha}^\dag c_{-\vk\beta}^\dag - \Delta_{-\vk,\alpha\beta}^* c_{\vk\sigma}c_{\vk\beta}].
\end{align}
The TI was modelled using its two Bi orbitals with a cubic lattice (lattice constant $a$):
\begin{align}
H_\text{TI} = \gamma_0 - 2\sum_{\vk j} \gamma_j \cos(k_ja) + \sum_{\vk\mu} d_\mu\Gamma_\mu,
\end{align}
where $d_0 = \epsilon-2\sum_jt_j\cos(k_ja)$, $d_j=-2\lambda_j\sin(k_ja)$, $\Gamma_0 = \tau_x \otimes \sigma_0$, $\Gamma_x = -\tau_z \otimes \sigma_y$, $\Gamma_y = \tau_z \otimes \sigma_x$, and $\Gamma_z = \tau_y \otimes \sigma_0$. The Pauli matrices in orbital and spin space are denoted $\tau_j$ and $\sigma_j$, respectively. The parameter values for $\gamma_j$ fitted to the Bi$_2$Se$_3$ dispersion are given in \cite{zhang_nphys_09, rosenberg_prb_12}.



Finally, the local tunneling Hamiltonian $H_t$ couples the superconductor with the TI through electron hopping:
\begin{align}
H_\text{t} = -\sum_{\vk\sigma} (t_1 c_{\vk\sigma}^\dag b_{1\vk\sigma} + t_2 c_{\vk\sigma}^\dag b_{2\vk\sigma} + \text{h.c.})
\end{align}
where $b_{a\vk\sigma}^\dag$ creates an electron in the orbital $a=1,2$ in the TI surface layer.

By performing an exact numerical diagonalization of the total Hamiltonian $H$, a comprehensive overview of different time-ordered pairing amplitudes $f^{ab}_{\alpha\beta}(\tau)$ arising in the TI surface layer were then obtained in Ref. \cite{blackschaffer_prb_13a} (see their Table I) and classified based on their symmetries in orbital and frequency space:
\begin{align}
f^{ab}_{\alpha\beta}(\tau) &= \frac{1}{2N_\vk} \sum_\vk S_{\vk\alpha\beta} \mathcal{T}_\tau \langle b_{a,-\vk,\beta}(\tau) b_{b\vk\alpha}(0) \notag\\
&\pm b_{b,-\vk,\beta}(\tau)b_{a\vk\alpha}(0)\rangle.
\end{align}
Above, $\pm$ refers to even/odd pairing in the orbital index, $N_\vk$ is the number of $\vk$ points in the Brillouin zone, and $\mathcal{T}$ is the time-ordering operator. This also included the case when the host superconductor was unconventional in itself, i.e. $p$- or $d$-wave. Moreover, we defined a symmetry factor $S_{\vk\alpha\beta} = \Delta_{\vk\alpha\beta}^*/\Delta_0$.

Later works studied further aspects of \ow pairing induced in TI structures via proximity to a host $s$-wave superconductor. Proximity-induced \ow pairing in the helical edge-states of a TI were studied in relation to crossed Andreev reflection in \cite{crepin_prb_15}, whereas the issue of \ow pairing in a quasiclassical framework using Eilenberger and Usadel equations was treated in \cite{hugdal_arxiv_16}. Finally, a microscopic calculation of the proximity effect between a superconductor and a TI was conducted in \cite{lababidi_prb_11}, but without considering the frequency-symmetry of the superconducting correlations.

\begin{figure}[t!]
\includegraphics[scale=0.7]{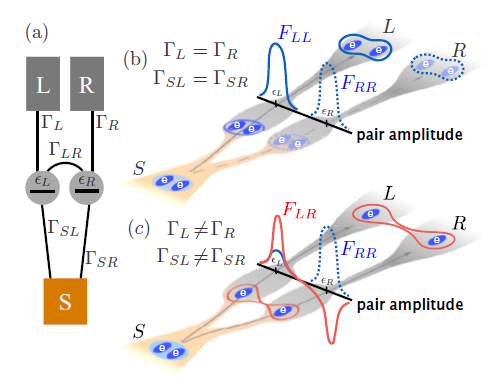}
\caption[fig]{\label{fig:burset}(Color online) Suggested expermiental setup for electrically controlled \ow pairing in a double-dot three-terminal device. (a) The quantum dots have level positions $\epsilon_{L,R}$ and are contacted by a superconducting lead S and two normal leads $L$ and $R$. (b) Illustration of a local Andreev reflection process where the Cooper pair electrons tunnel into a normal lead thorugh one dot. (c) Non-local Andreev reflection (AR) where the two electrons comprising the Cooper pair tunnel into different leads. The blue lines refer to the pair amplitudes $F_{LL}$ and $F_{RR}$ in the case of local AR whereas the red lines illustrate the non-local amplitude $F_{LR}$ which is \ow on the resonance point $\epsilon_L=\epsilon_R$. Figure adapted from \cite{burset_prb_16}.}
\end{figure}

When \ow superconductivity appears in quantum dots, it has the potential advantage that electric control of the \ow Cooper pairs is more feasible than in conventional metallic systems, such as those traditionally studied in superconductor-ferromagnet experiments. Sothmann \etal~\cite{sothmann_prb_14} proposed that \ow pairing triplet, as well as other types of unconventional superconductivity including higher order angular momentum pairing, would be controllable in a double quantum dot device hosting inhomogeneous magnetic fields. Burset \etal~\cite{burset_prb_16} realized that by utilizing a three-terminal device connected to a double-quantum dot, it was possible to control the \ow amplitude purely electrically without any need for magnetic fields. They showed that by tuning the quantum dot levels to resonance (see Fig. \ref{fig:burset}), Cooper pairs split into separate terminals via crossed Andreev reflections would be correlated exclusively with an \ow pairing symmetry.

We return now to the issue of the spectral signatures of \ow pairing mentioned above in relation to the $1/\omega_n$ dependence which did not produce any subgap states. This is in contrast to the numerous examples discussed so far in this review where \ow pairing seems to be generally accompanied by an enhancement of the electronic density of states at subgap energies. This is the case for \eg S/N structures \cite{rowell_prl_66, tanaka_prl_07a, eschrig_jltp_07}, S/F structures \cite{kontos_prl_01, yokoyama_prb_07, linder_prl_09, dahal_njp_09, dibernardo_natcom_15}, and vortex cores \cite{yokoyama_prb_08}. However, as noted in \cite{blackschaffer_prb_12}, \ow pairing does not necessarily enhance the low-energy density of states. At the same time, it was recently shown that there exists a connection between the local density of states and \ow triplet pairing in 2D topological insulators proximitized by a superconductor \cite{cayao_prb_17}.

One could then ask the question: is it possible to have a system with a fully gapped density of states that still has strong \ow superconducting correlations present? This issue was studied in \cite{linder_scirep_15} where an analytical criterion was derived for when \ow pairing can be present in a fully gapped system. This finding is of relevance for the experimental identification of \ow pairing, since STM-measurements of the density of states is a commonly used method for this purpose. For a single-band model, the proof of the criterion goes as follows \cite{linder_scirep_15}. Consider a system where both \ew and \ow correlations may exist. In the diffusive limit, it is convenient to use the quasiclassical Green function matrix $\hat{g}$ introduced in Sec. \ref{sec:FS}. It satisfies the normalization condition $\hat{g}^2=\hat{1}$ and may be written in the form:
\begin{align}
\hat{g} = \begin{pmatrix}
c_\uparrow & 0 & 0 & s_\uparrow \\
0 & c_\downarrow & s_\downarrow & 0 \\
0 & -s_\downarrow & -c_\uparrow & 0 \\
-s_\uparrow & 0 & 0 & -c_\uparrow \\
\end{pmatrix}
\end{align}
where $c_\sigma=\text{cosh}\theta_\sigma$ and $s_\sigma = \text{sinh}\theta_\sigma$ where $\theta_\sigma$ is a parameter which describes the spin-dependence of the superconducting correlations. In a BCS bulk superconductor, it is given by $\theta_\sigma = \text{atanh}[\sigma\Delta/(E+\i\eta)]$ where $\eta$ is the inelastic scattering rate. In that case, we see that $\theta_\uparrow = -\theta_\downarrow$, so that no \ow correlations $f_t = (s_\uparrow+s_\downarrow)/2=0$ exist. In the presence of \eg an exchange field $h$, $\theta_\uparrow\neq\theta_\downarrow$ so that $f_t\neq 0$. Using that the normalized density of states is $N(E)/N_F = \frac{1}{2} \sum_\sigma \text{Re}\{c_\sigma\}$ and that $\hat{g}^2=\hat{1}$, one finds
\begin{align}
\frac{N(E)}{N_F} = 2\text{Re}\Big\{\frac{f_tf_s}{c_\uparrow-c_\downarrow}\Big\}.
\end{align}
Assume now that the system is gapped so that $N(E)/N_0$ is zero for a range of energies $E$. This means that $c_\sigma$ must be a purely imaginary number. So long as $c_\uparrow \neq c_\downarrow$ (the system is not spin-degenerate), it follows that
\begin{align}\label{eq:enhancecond}
\frac{N(E)}{N_F} = 2\frac{\text{Im}\{f_tf_s\}}{\text{Im}\{c_\uparrow-c_\downarrow\}}=0.
\end{align}
The above equation expresses a crucial fact: when the \ew pair amplitude $f_s$ and the \ow pair amplitude $f_t$ are both real or both imaginary, hereafter referred to as in-phase, we see that that $N(E)=0$ regardless of the magnitude of $f_t$. In order for the presence of \ow pairing $f_t$ to produce an enhancement of the density of states, it thus needs to be out-of-phase with the singlet component $f_s$: otherwise, there are no subgap states available in spite of $f_t \neq 0$.

It should be noted that the above result does not mean that \ew singlet pairing $f_s$ must be present in general for \ow superconductivity $f_t$ to enhance the low-energy density of states. As discussed in Sec. \ref{sec:FS}, a system with pure \ow pairing \cite{linder_prl_09} can produce a strong zero-energy peak [in that system, $c_\uparrow=c_\downarrow$ in which case Eq. (\ref{eq:enhancecond}) cannot be used]. Nevertheless, the above derivation shows that the existence of \ow correlations is not equivalent to a non-gapped density of states: a large \ow amplitude $f_t$ can be present even if the system is fully gapped. A practical example of such a system where this occurs is a thin-film superconductor with an in-plane magnetic field \cite{linder_scirep_15}.

Closing this subsection, we note that \ow Berezinskii pairing has recently been discussed in the context of another class of insulating materials besides topological insulators, namely so-called Skyrme insulators \cite{erten_prl_17} existing on the brink of a superconducting phase, which could be an interesting topic to explore further.

\subsection{Andreev bound states and \ow pairing}\label{sec:ABS}

The equivalency between McMillan-Rowell resonances with energy $E<\Delta$ in ballistic NS junctions and the presence of odd-frequency
correlations was described in Sec. \ref{sec:NS} However, there is a fundamental equivalence not only between
\ow pairing and such spatially extended bound-states, existing throughout the N region, but also between \ow
pairing and so called zero-energy states bound to a superconducting interface. Such states play an important role
in the identification of unconventional types of superconductivity, where zero-energy states appear at certain
crystallographic orientations of a superconducting interfaces when the material has a non $s$-wave order parameter.
These zero-energy states (ZES) are also known as Andreev bound-states throughout the literature, even though Andreev bound states
need not in general reside at the Fermi level (zero energy).

\begin{figure}[h!]
\includegraphics[scale=0.9]{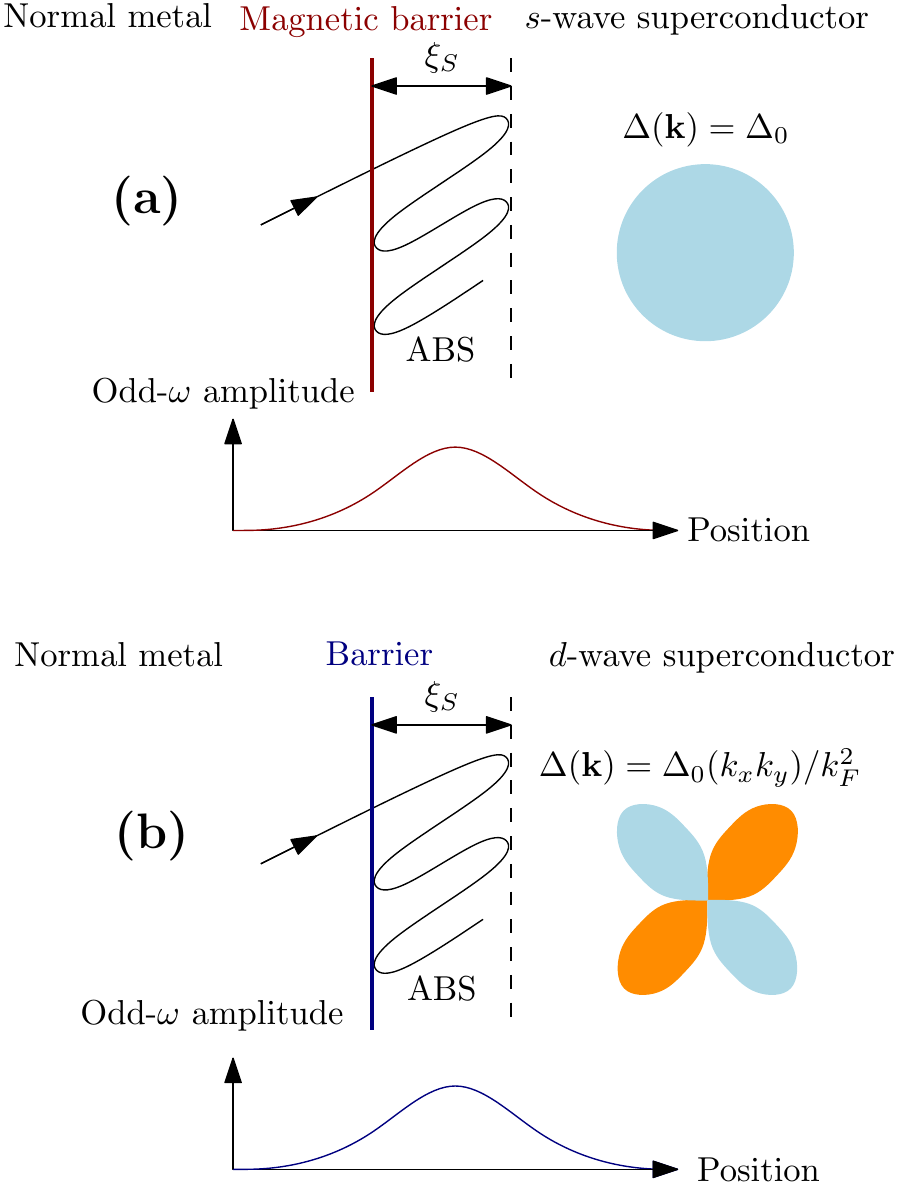}
\caption[fig]{\label{fig:abs_interface} (Color online) \textbf{(a)} Andreev-bound state (ABS) formed at the interface between a normal metal and $s$-wave superconductor separated by a magnetic barrier, \eg a ferromagnetic insulator. The spin-dependent phase-shifts arising due to the magnetic barrier give rise to an interface state which appears at zero-energy for strong enough phase-shifts. \textbf{(b)} ABS formed at the interface between a normal metal and a $d$-wave superconductor separated by a non-magnetic barrier, \eg an insulator. The electron- and hole-like excitations experience different signs of the pair potential $\Delta(\mathbf{k})$ upon scattering, leading to the formation of a zero-energy state. In both \textbf{(a)} and \textbf{(b)}, the bound state at the interface is accompanied by a strong increase in the magnitude of the \ow correlations, quantified via the anomalous Green function $f$. The dashed line indicates how quasiparticles are Andreev-reflected back toward the interface by the pair potential $\Delta(\mathbf{k})$ after penetrating a distance $\sim \xi_S$ into the superconductor. }
\end{figure}

An example of ZES appearing in unconventional superconducting systems \cite{tanaka_prl_95} is the high-$T_c$ cuprates which have a $d$-wave order parameter symmetry.
In the $ab$-plane of materials such as YBCO, experiments have shown that a $d$-wave superconducting order parameter emerges \cite{tsuei_rmp_00}.
Let a surface terminate the superconducting material so that $k_x$ is the component of the quasiparticle momentum perpendicularly to the surface whereas
$k_y$ is the component parallel to it. If the orientation of the surface is such that the order parameter satisfies the property
\begin{align}
\Delta(k_x,k_y) = -\Delta(-k_x,k_y),
\end{align}
a ZES appears at the surface for that particular value of $k_y$. In the $d_{xy}$-wave case $\Delta=\Delta_0(k_x k_y)/k_F^2$, this condition is met for all modes $k_y$,
leading to a large zero-bias conductance peak as observed in STM-measurements \cite{alff_prb_97, wei_prl_98}. Other types of unconventional pairing, such as chiral $p$-wave $\Delta=\Delta_0(k_x+\i k_y)/k_F$,
satisfies this condition only for specific values of $k_y$ ($k_y$=0 in the chiral $p$-wave case) which leads to an much less pronounced enhancement of the conductance at zero bias. The relation between zero-energy Andreev bound states and topology was examined in \cite{sato_prb_11}.

Coming back to the relation to \ow pairing, Tanaka \etal~\cite{tanaka_prl_07b} showed that when the criterion for
formation of ZES was satisfied, it was invariably accompanied by a strong enhancement of the \ow correlations at the interface, even exceeding the \ew correlations.
To see this analytically, one may derive an expression for the anomalous Green function induced at the interface separating a normal metal from an unconventional superconductor in the low-transparency limit. Neglecting the spatial
dependence of the pair potential near the interface, one obtains for a singlet $d_{xy}$ wave superconductor
\begin{align}
f = \frac{\i\Delta_0}{\omega_n}|\sin(2\theta)|\text{sgn}(\sin\theta)
\end{align}
whereas for a triplet $p_x$-wave superconductor the result is
\begin{align}
f = \frac{\i\Delta_0}{\omega_n}|\cos\theta|.
\end{align}
In both cases, the anomalous Green function is proportional to the inverse of $\omega_n$, reflecting precisely
the \ow symmetry. Importantly, there is a difference in parity with regard to the quasiparticle momentum direction $\theta$
in the two cases: the $p$-wave case results in an even-parity $f$ wheras the $d$-wave case results in an odd-parity $f$.
This causes the proximity effect to differ strongly between the two cases in the case where
the normal metal is diffusive, i.e. when impurity scattering is frequent, causing an isotropization of
quasiparticle trajectories equivalent to averaging $\int^{\pi/2}_{-\pi/2} d\theta \ldots$. The \ow Green function induced
from the $p$-wave superconductor survives due to its even parity, whereas it does not in the $d$-wave case. Hence, as noted in Ref. \cite{tanaka_prb_04, tanaka_prberratum_04},
the proximity effect and presence of ZES are antagonists in diffusive metals coupled to $d$-wave superconductors
whereas they can coexist in the $p$-wave case.

The presence of ZES, which we have argued above is accompanied by presence of strong \ow correlations and may thus be
interpreted as a manifestation of \ow superconductivity, does not necessarily require unconventional superconducting order such as
$p$- or $d$-wave. As discussed in Sec. \ref{sec:FS}, separating a conventional $s$-wave superconductor from a normal metal by a magnetic barrier (\eg a ferromagnetic insulator such as GdN or EuO), ZES would arise at the interface and manifest as a zero-energy peak both in the superconducting and normal metal region \cite{linder_prl_09, linder_prb_10}. Just as in the
case described above with unconventional superconductors, the ZES was again accompanied by \ow pairing and even completely suppressed \ew correlations at zero energy.

\begin{figure}[h!]
\includegraphics[scale=0.66]{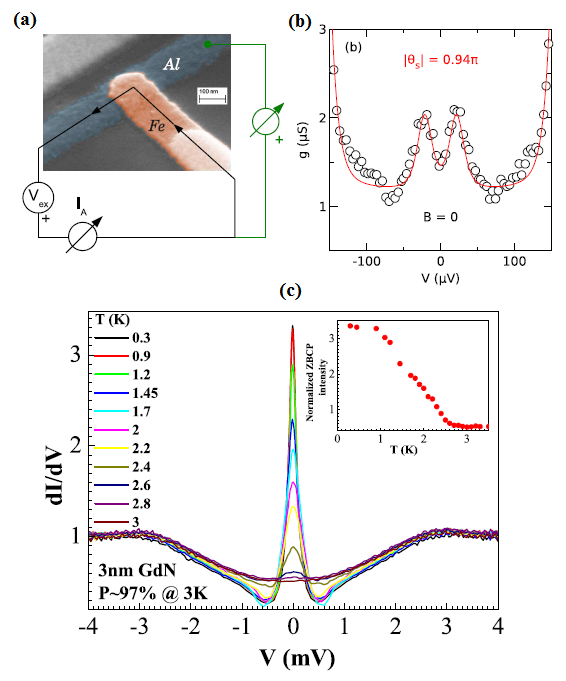}
\caption[fig]{\label{fig:hubler} (Color online) (a) Scanning electron microscopy image of the Al/AlO$_x$/Fe sample used in \cite{hubler_prl_12} along with the measurement scheme. (b) Differential conductance spectrum for the structure at zero magnetic field ($B=0$), together with a theoretical fit (red line). (c) Differential conductance (dI/dV) measurements \cite{pal_scirep_17} normalized to the normal-state value of a 100 nm NbN/3 nm GdN/30 nm TiN tunnel junction demonstrating the evolution of a zero-energy peak with decreasing temperature. Figures (a) and (b) adapted from \cite{hubler_prl_12} and Figure (c) adapted from \cite{pal_scirep_17}. }
\end{figure}

The first clear experimental observation of Andreev bound states close to zero energy due to a spin-active interface was reported by H{\"u}bler \etal~\cite{hubler_prl_12}. The authors reported on high-resolution differential conductance measurements on a nanoscale superconductor/ferromagnet tunnel junction with an oxide tunnel barrier, and saw evidence of a subgap surface state stemming from the spin-active interface [see Fig. \ref{fig:hubler}(a) and (b))]. A much stronger signature of an Andreev bound state at the Fermi level, manifested by a zero-energy peak several times larger than the normal-state value of the density of states, was
recently experimentally observed in an S/FI/N system comprised of NbN/GdN/TiN \cite{pal_scirep_17} [see Fig. \ref{fig:hubler}(c)].

The physical mechanism which allows for the appearance of ZES and \ow pairing
via a magnetic interface is spin-dependent scattering phase-shifts $\theta_\sigma$, $\sigma=\uparrow,\downarrow$ defined from the reflection coefficients $r_\sigma = |r_\sigma|\e{\i\theta_\sigma}$. When electrons scatter on a magnetic interface, transmitting or reflecting, both the magnitude of the scattering
coefficients and their phase depends on the electron spin. The difference between the spin-up and spin-down phases is thus in general finite, but it is particularly instructive to
consider the case where it is equal to $\pi$. The reason for this is that in this case, one can establish a perfect analogy to the ZES appearing due to higher angular momentum pairing
such as $p$-wave or $d$-wave. The phase-shifts then give rise to a sign change for each Andreev reflection process in the same way as the pairing potential itself provides this sign change
in the $p$- or $d$-wave case, as illustrated in Fig. \ref{fig:abs_interface}. As a result, a bound-state at zero energy arises even for a conventional $s$-wave superconductor in contact with a FI. For
an arbitrary value of the phase-shifts $\Delta\theta\equiv\theta_\uparrow-\theta_\downarrow$, the bound state energy in a ballistic S/FI/N junction occurs at
\begin{align}
E = \Delta_0\cos(\Delta\theta/2).
\end{align}
The theoretical fit to the experimental results thus suggested $\Delta\theta = 0.94\pi$ in \cite{hubler_prl_12} whereas $\Delta\theta = 0.98\pi$ in \cite{pal_scirep_17}.

It is worth to emphasize that there are other physical mechanisms that can provide zero-bias conductance peaks in fully conventional N/S junctions
without any occurrence of \ow pairing. One example of this is reflectionless tunneling \cite{volkov_physicac_93} which occurs for low-transparency junctions with a small Thouless energy $E_\text{Th}=\frac{D}{L^2}\ll\Delta$, which in essence consists of repeated attempts of electron transmission
through the barrier in the form of Andreev reflection due to backscattering from impurities. This phenomenon takes place in diffusive junctions even for $s$-wave superconductors
and thus leads to a zero-energy enhancement of the conductance without any presence of \ow correlations. This shows that it is important to distinguish between the conductance of a junction
and the local density of states: the two do not necessarily coincide. An enhancement of the local density of states in superconducting hybrid structures, in the form of \eg ZES at an interface,
will be accompanied by \ow correlations, whereas a zero-bias conductance enhancement in a voltage-biased N/S junction can occur due to Andreev reflection without any accompanying \ow
Cooper pairs.

\section{Berezinskii pairing for Majorana fermions and in non-superconducting systems}

\subsection{General definition of the pairing states and relation to \ow pairing}\label{sec:genpairing}

It is useful to place \ow states in a broader context of the pairing states beyond superconductivity. To be general we define a "pairing state" as a state where the thermodynamic ground state is represented by a behavior of the matter field operator ${\hat O}$ such that the expectation value $\langle{\hat{O}}\rangle = 0$, yet the pair field operator $\langle\hat{O}(1)\hat{O}(2)\rangle$ has a long range order where $1,2$ label states (be it space, time, spin, orbital and other indices). Inclusion of  time seem to be a natural extension needed to consider dynamic orders.  This is a natural generalization of the definition given in \cite{yang_rmp_62} for off-diagonal long-range order. We will call this state a pairing state in the sense that pairing correlations develop. In principle any field, be it bosonic or fermionic, can develop pairing correlations. Specific examples of a pairing state include the following cases.
\begin{itemize}
\item Fermions, where ${\hat O} = \psi$ and $\psi$ is a fermion operator. In this case, the pairing state can be (but does not have to be) a superconducting state. Certainly, any superconducting order is an example of the pairing state. In the case of fermions we get $\langle\psi\rangle = 0$, yet $\langle\psi(1)\psi(2)\rangle$ has a long range order in the superconducting state.
\item Bosons, with ${\hat{O}} = b$ and $b$ is a boson operator. In this case one can envision the states like Bose-nematic \cite{balatsky_arxiv_14} or spin nematic \cite{balatsky_prl_95}. For spin bosons ${\hat O} = {\bf S}$, we obtain a paramagnetic state with no single spin expectation value yet with finite nematic order \cite{andreev_jetp_84,balatsky_prl_95}.
\item Majorana fermions, ${\hat O} = \gamma$. One can easily extend the Berezinski symmetry classification to Majorana states and one arrives at
\begin{align}
M_{ab}(1,2) &= - M_{ba}(2,1), \notag\\
 M_{ab}(1,2) &= -\langle\mathcal{T}\gamma_a(1)\gamma_b(2)\rangle
 \label{eq:Majoranaclass}
\end{align}
The proof goes essentially along same lines as in case of Dirac fermions and is discussed below and  in detail in Ref. \cite{huang_prb_15}.
\end{itemize}

One has to distinguish a pairing state from a true superconducting state, for they are in general different. In the superconducting case one has a whole set of attributes such as the Meissner effect, phase stiffness, superflow, flux quantization, and so forth. A pairing state, while looking similar to the superconducting state at first glance, \textit{does not not have to} possess any of these features. In this sense, a pairing state is a simpler phenomenon than superconductivity. The second reason is that a pairing state can occur for states that are localized, like the case of localized Majorana modes or for neutral bosons like the case of spin nematic. Neither of these states will be able to carry any charge current.

This general introduction to pairing states now prepares us to later in this section go beyond traditional superconducting states and discuss pairing states in novel settings.

\subsection{Majorana fermions as a platform for  \ow pairing}\label{sec:majorana}

The possibility to create and manipulate Majorana fermions in condensed matter systems is currently subject to intense research \cite{alicea_rpp_12}. Noted to exist at the edge of spinless $p$-wave superconductors \cite{kitaev_pu_01}, the interest in solid-state Majorana excitations took off on a spectacular level in 2008 when it was predicted that they would appear in heterostructures comprised of topological insulators and superconductors \cite{fu_prl_08}. Soon after, it was also predicted that Majorana fermions (more accurately referred to as a Majorana bound-state as it is typically bound to interfaces or vortex cores) should exist in heterostructures comprised of semiconducting and superconducting structures \cite{lutchyn_prl_10, oreg_prl_10} as well as in superfluids with Rashba spin-orbit coupling and a Zeeman-field \cite{sato_prl_09}. Recent experiments \cite{mourik_science_12, nadjperge_science_14, albrecht_nature_16} have reported measurements which are largely consistent with the theoretical predictions.

We start with the question about the pairing state of Majorana fermions when they are considered to be free particles. In other words, first we assume that there are Majorana fermionic excitations and that they can exist as independent particles. The question we are asking is: what are the symmetries of possible pairing states that emerge? We point out that Majorana fermions basically realize the \ow pairing from the outset.

A Majorana fermion is its own antiparticle, a property expressed through $c=c^\dag$ in a second quantized language. The general symmetry of any pairing state of Majorana fermions is given by Eq. (\ref{eq:Majoranaclass}). We will see from this classification  there is an important  relationship between Majorana fermions and \ow pairing. Majorana fermion operators are real and they represent particle creation and annihilation operator at the same time. Hence, any particle-hole propagator $G = -\langle \mathcal{T}_{\tau}\gamma^\dag(\tau)\gamma(0)\rangle$ is at the same time a particle-particle propagator $F(\tau) = -\langle \mathcal{T}_{\tau} \gamma(\tau) \gamma(0) \rangle$. For the single zero energy mode we thus obtain:
\begin{eqnarray}\label{eq:typo}
G(\omega_n) =  F(\omega_n)  = \frac{1}{i\omega_n}
\end{eqnarray}
This observation is at the core of the growing list of examples of the \ow state in Majorana fermions \cite{asano_prb_13,huang_prb_15}. It is appropriate to mention here the early works by Coleman, Miranda, and Tsvelik \cite{coleman_prl_93, coleman_prb_94, coleman_prl_95} who discussed \ow Berezinskii pairing in a model with Majorana fermions.
Since the Majorana phase is topological, the structure of the propagators may change but the basic property that pairing correlator $F$ is an odd function of frequency/time will remain. To illustrate the utility of Majorana states as a platform for \ow pairing state we consider the case of i) free Majorana fermions and ii) the case of zero energy Majorana modes at the ends of a wire, in effect bound states.

\begin{figure}[t!]
	\includegraphics[scale=0.75]{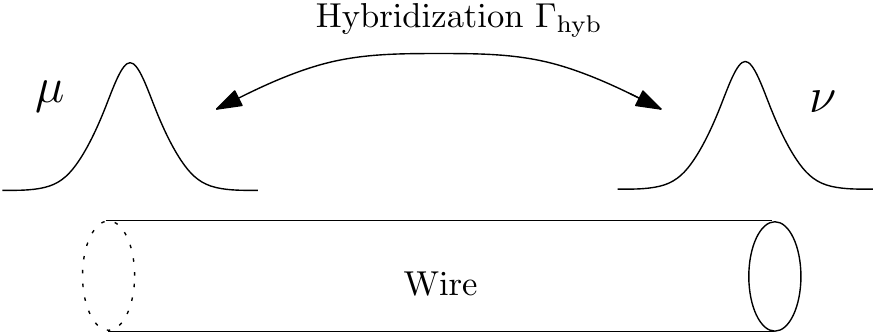}
	\caption[fig]{\label{fig:majoranawire} (Color online) Two Majorana modes localized at the end of a wire with a finite hybridization $\Gamma_\text{hyb}$, as considered in Ref. \cite{huang_prb_15}.}
\end{figure}

\textit{Case i):} The free Majorana theory has a Lagrangian:

\begin{equation}
L = \sum_{\bk} (i \gamma_{\bk}^{\dag} \partial_{\tau} \gamma_{\bk} - E_{\bk}\gamma_{\bk}^{\dag} \gamma_{\bk})
\label{eq:Majorana1}
\end{equation}

with the condition that Majorana fermions obey the reality conditions for the fermion operator: $\gamma_\bk = \gamma{\dag}_{-\bk}$ and $\gamma^{\dagger}({\bf r}) = \gamma({\bf r})$. Here, $E(\bk)$ is the dispersion of the Majorana mode whose detailed shape is not important for this discussion. The Green function (particle-hole Majorana fermion propagator) $G({\bf r}, \tau) =  -\langle \mathcal{T}_{\tau} \gamma^{\dagger}({\bf r}, \tau) \gamma(0,0)\rangle$ is then identical to anomalous Greens function (particle-particle) $F({\bf r}, \tau) =  -\langle\mathcal{T}_{\tau} \gamma({\bf r}, \tau) \gamma(0,0)\rangle$. Thus, the free Majorana fermion propagator has the form:
\begin{equation}
G(\bk,i\omega_n) = F(\bk,i\omega_n) =  1/(i \omega_n - E_\bk).
\end{equation}
Interestingly, Majorana fermions realize a mixed pairing state. From the above equations we deduce that $F$ describes a pairing state that has both even-frequency and \ow components:
\begin{eqnarray}
F_{even} \sim \frac{E_{\bk}}{(i \omega_n)^2 + E^2_\bk}, \  F_{odd} \sim \frac{i\omega_n}{(i \omega_n)^2 + E^2_\bk}
\label{eq:Foe}
\end{eqnarray}

This conclusion could have been drawn in 1937 when Majorana fermions were proposed for the first time \cite{majorana_nc_37, wilczek_natphys_09}.   Unfortunately, this connection to pairing was not possible at the time as the notion of the anomalous propagators (Gor'kov $F$ function)  as a key element for microscopics of superconductivity was not invented yet. With all its simplicity this relation between $F$ and $G$ in case of Majorana fermions projects a very important general message: Majorana fermions as a many body system is conducive to form \ow pairing states. This conclusion is universal. We give few specific examples below.

\textit{Case ii):} We next proceed with the case of two Majorana zero energy modes. The scheme to realize zero energy modes located at the ends of a superconducting wire is shown in Fig. \ref{fig:majoranawire}. For the case of two modes at the ends of the wire ($\mu, \nu$) with no hybridization between them the two energy modes correspond to $E_\bk = 0$ in Eq. (\ref{eq:Foe})  and one has two \ow pairing correlations for $\mu$, $\nu$ fermions. Upon turning on the hybridization $\Gamma_\text{hyb}$ between ends of the wire,
the Lagrangian of the system becomes:

\begin{eqnarray}
L = i \mu \partial_{\tau}\mu + i \nu \partial_{\tau} \nu - i\Gamma_\text{hyb} \mu \nu
\label{eq:Lmunu}
\end{eqnarray}
In matrix form, one has for a Majorana spinor $\Psi = (\mu, \nu)^\text{T}$  that $L = \Psi i\partial_{\tau} \Psi - i \Gamma_\text{hyb}\Psi \sigma_y \Psi$ which leads to
\begin{eqnarray}
\hat{G}(i\omega_n) = \frac{i\omega_n + \Gamma_\text{hyb} \sigma_y}{(i \omega_n)^2 +\Gamma_\text{hyb}^2}
\label{eq:GFmajorana1}
\end{eqnarray}
Again, we see that hybridized Majorana wire contains both \ew and \ow components:
\begin{eqnarray}
G^\text{odd}_{\mu \mu} =\frac{i\omega_n \delta_{\mu \nu} }{(i \omega_n)^2 +\Gamma_\text{hyb}^2}  \\
G^\text{even}_{\mu \nu} =\frac{\Gamma_\text{hyb} \sigma_{y, \mu \nu} }{(i \omega_n)^2 +\Gamma_\text{hyb}^2}
\label{eq:GFmajorana2}
\end{eqnarray}

The  Berezinskii component will be odd under $\mu-\nu$ permutations and the \ow component is explicitly even under orbital index permutation, consistent with the general SPOT constraint,  required for the pairing matrix $M$ in Eq. (\ref{eq:Majoranaclass}) \cite{huang_prb_15}.

Both the examples illustrate unique utility of Majorana states as a platform to realize \ow pairing. The field of pairing states of Majorana fermions is in its infancy  and it is poised to generate new results, hopefully with surprises along the way. So far, we have been addressing the issue of the pairing states of Majorana fermions that hold regardless of their precise origin.

With the experimental realization of the Majorana fermions in the wires, we now can address the {\em pairing states} of Majorana fermions in the case where we have a large number of them. Going beyond single Majorana fermions, Huang \etal~\cite{huang_prb_15} studied the interaction between different Majorana fermions located at the opposite ends of \eg a topological wire. From such interactions, pairing between Majorana fermions can be envisioned to occur which prompts the question: what type of instability occurs when pairs of Majorana fermions condense? When studying pairing instabilities within an effective Hamiltonian framework, one usually considers a time independent scenario whereby the instabilities are implicitly assumed to be dominated by their equal time behavior. This must necessarily be described by the \ew component of the pairing amplitude. However, since Majorana fermions are their own antiparticles, one should be careful with regard to any time (or frequency) dependence. This can be seen by considering a pairing correlator of the type $f_\tau = \mathcal{T}_\tau\langle \gamma(\tau)\gamma(0)\rangle$ where $\gamma$ is a Majorana operator $\tau$ is the (Matsubara) time. Such a correlator must be odd in $\tau$ by the mathematical properties of $\gamma$, thus forcing $f_\tau$ to vanish at equal times.

\begin{figure}[t!]
\includegraphics[scale=0.55]{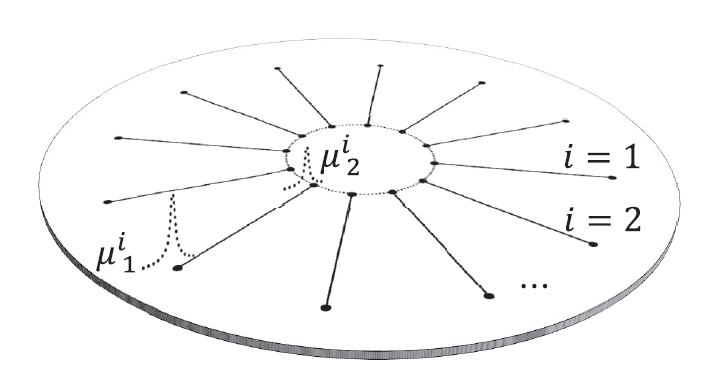}
\caption[fig]{\label{fig:majoranawires} (Color online) Schematic setup of multiple superconducting wires hosting Majorana fermions on their edges. The fermions are represented by operators $\mu^i_a$ (denoted $\gamma^i_a$ in the main text) where $i$ is the wire index and $a=1,2$ label the two edges of each wire. The dashed line in the figure illustrates possible couplings between Majorana fermions on neighboring wires. Figure adapted from \cite{huang_prb_15}. }
\end{figure}

Following \cite{huang_prb_15}, consider the model in Fig. \ref{fig:majoranawires} where different Majorana modes can pair up due to an interaction induced via coupling to an external boson. Unlike same-mode pairing, which must be \ow due to fermionic statistics, there is no such requirement on the frequency depednence for cross-mode Majorana pairing. At the same time, the \ow solution has a lower free energy than \ew solution for such pairing and indicates that the former is the most stable. When considering a pairing amplitude of the type $f_\tau$ described above, one usually associates it with some form of long-range order such as superconductivity or superfluidity. However, it should be remarked that the existence of $f_\tau\neq0$ does not automatically guarantee for instance U(1) gauge symmetry breaking related to phase coherence. It is still of interest to discuss such a pairing correlator such as $f_\tau$ as they may be important indicators of the existence of \eg superstates.

Let $a=1,2$ denote the two edges for each wire in Fig. \ref{fig:majoranawires} and let $i$ denote the wire index, so that $\gamma^i_a = (\gamma^i_a)^\dag$ represents a Majorana operator at edge $a$ of wire $i$. The pair amplitude satisfies:
\begin{align}\label{eq:fmajorana}
f_{ab}^{ij}(\tau) = \mathcal{T}\langle \gamma^i_a(\tau)\gamma^i_b(0)\rangle = -f_{ba}^{ji}(-\tau)
\end{align}
which follows simply from the definition of the time-ordering operator as long as there is only a dependence on the relative time-coordinate $\tau$ (as assumed here). This is the same type of antisymmetry under an exchange of particle indices as encountered in the standard Pauli principle for Dirac, rather than Majorana, fermions. Majorana fermion pairing can in fact be viewed as an analogue to equal-spin Dirac fermion pairing.

If one initially considers same-wire pairing ($i=j$) in the absence of any interactions, it follows that
\begin{align}
f_{ab}^{ii}(\tau) = f_{ab}^{ii}(\tau)\delta_{ab}
\end{align}
where the $\delta_{ab}$ dependence arises due to the absence of any interactions between the edges. To satisfy Eq. (\ref{eq:fmajorana}), it is clear that $f_{ab}^{ii}(\tau) = -f_{ab}^{ii}(-\tau)$ which means that the only pairing channel available for a single Majorana fermion is the \ow one. In fact, this analysis shows that a Majorana state at zero energy is simply a realization of an \ow pairing state. This makes sense physically, since the Majorana fermion is both a particle and a hole, so that its single-particle propagators is simultaneously a pair propagator. The case of interacting Majorana fermions on a \textit{single} wire is more complicated and has been covered in detail in \cite{huang_prb_15}. The key result in this case is that an Berezinskii state is stabilized when the coupling strength $g$ between the Majorana modes exceeds a critical value, as shown in Fig. \ref{fig:majoranaoddstate}.

\begin{figure}[h!]
\includegraphics[scale=0.5]{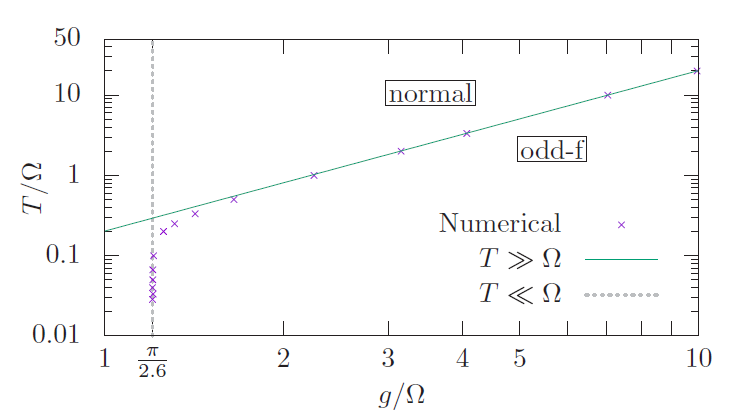}
\caption[fig]{\label{fig:majoranaoddstate} (Color online) Phase diagram of the normal and pairing state of collection of Majorana states is shown. For the large coupling between the ends of the wire, the system will have a strong pairing fluctuations in the \ow channel. The phase diagram is drawn for the mean field solution of the pairing state. As explained, Majorana states have a strong propensity to form the \ow state. The green solid line shows the critical temperature $T_c \propto g^2$. From Ref. \cite{huang_prb_15}. }
\label{fig:majoranaoddstate}
\end{figure}

\begin{figure}[t!]
\includegraphics[scale=0.6]{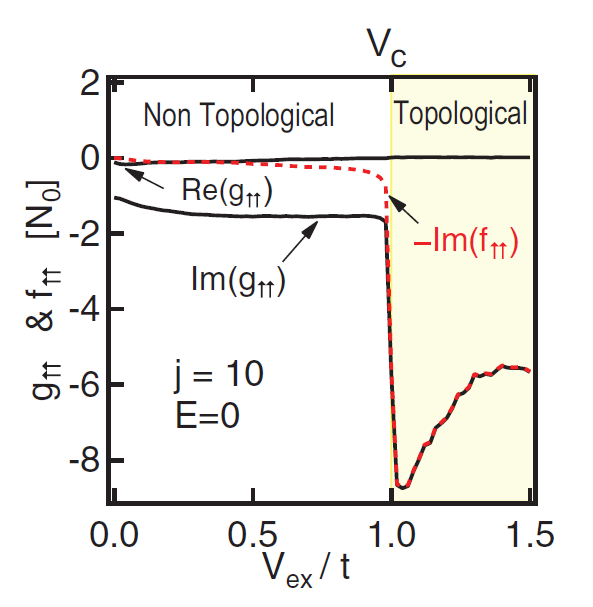}
\caption[fig]{\label{fig:topotrans} (Color online) The normal ($g_{\uparrow\uparrow}$) and anomalous $(f_{\uparrow\uparrow})$ Green functions plotted at the superconducting interface (lattice position $j=10$ in this model) for $E=0$ as a function of the exchange field strength $V_\text{ex}$ normalized to the hopping amplitude $t$. At the topological phase transition $V_\text{ex}=V_c$ where the Majorana fermion emerges, the \ow amplitude has a sharp increase. Figure adapted from \cite{asano_prb_13}. }
\end{figure}

A complementary approach taken in the literature is to connect the properties of the original fermion superconducting states with zero energy states and with a Majorana fermion description.  Asano and Tanaka \cite{asano_prb_13} investigated this issue by considering topologically non-trivial NS and SNS junctions. Here, N is a nanowire with strong spin-orbit coupling subject to either an external magnetic field or with an proximity-induced exchange field from \eg a magnetic insulator. Their main finding was that the \ow correlation function amplitude abruptly increased upon transitioning from the topologically trivial and non-trivial states, and that \ow superconductivity arose at the precise locations of the Majorana fermions. For a quantitative analysis, it is useful to note that the physics in the topologically non-trivial state of the nanowire is essentially the same as that of a spinless 1D $p_x$-wave superconductor \cite{kitaev_pu_01}. It is in this framework that the relationship between Majorana fermions and \ow Cooper pairs can be brought out most clearly. Following \cite{asano_prb_13}, consider a semi-infinite $p_x$ superconducting wire occupying the region $x>0$ which is known to host a Majorana fermion at its edge. The Majorana fermion resides at the Fermi level $E=0$. By solving the Bogolioubov-de Gennes equation, the wavefunction $\phi_0(x)$ for the Majorana surface states is obtained as
\begin{align}
\phi_0(x) = C(x) \begin{pmatrix}
\chi\\
\chi^*\\
\end{pmatrix}
\end{align}
where we defined the quantities:
\begin{align}
C(x) = \sqrt{2/\xi}\e{-x/2\xi_0}\sin(kx),\; \chi=\e{\i\pi/4}\e{\i\phi/2}.
\end{align}
and $\xi$ is the coherence length. Introduce also the retarded Green functions in the standard way:
\begin{align}\label{eq:defretarded}
G(x,t,x',t') &= -\i\Theta(t-t')\langle \{\psi(x,t),\psi^\dag(x',t')\}\rangle,\notag\\
F(x,t,x',t') &= -\i\Theta(t-t')\langle \{\psi(x,t),\psi(x',t')\}\rangle,
\end{align}
where $\psi(x)$ is the annihilation operator of a spinless electron. In the low-energy regime $|E|\ll\Delta$, the electron operator representing the surface state reads
\begin{align}\label{eq:fieldmajorana}
\psi(x) = \psi_0(x) = C(x)\chi(\gamma_0 + \gamma_0^\dag)
\end{align}
where $\gamma_0$ is a fermion annihilation operator. Inserting Eq. (\ref{eq:fieldmajorana}) into Eq. (\ref{eq:defretarded}) after converting the Green functions to a spectral representation, one obtains for $|E|\ll\Delta$:
\begin{align}\label{eq:GandF}
G(x,x',E) &= \frac{2C(x)C(x')}{E+\i\delta},\notag\\
F(x,x',E) &= \frac{2C(x)C(x')}{E+\i\delta}\i\e{\i\varphi},
\end{align}
so that the relation $G(x,x',E) = -\i\e{-\i\varphi}F(x,x',E)$ is satisfied. To extract the $s$-wave pairing amplitude described by this anomalous Green function $F$, we set $x=x'$ to consider local pairing. Doing so, it follows from Eq. (\ref{eq:GandF}) that the real part of $-\i\e{-\i\varphi}F$ is an odd function of energy $E$ whereas the imaginary part is an even function of $E$. As shown in \cite{asano_prb_13}, this is the defining mathematical property of \ow superconductivity. This work established that $p$-wave superconducting pairing in the case of topologically nontrivial case produces Majorana fermions and relates them to the appearance of the  \ow pairing of the original pairing states ($\Psi$). Going back to the original nanowire-superconductor heterostructure, a numerical computation of the Green functions using a tight-binding Hamiltonian confirmed the sharp increase in the \ow amplitude at the topological transition point, as shown in Fig. \ref{fig:topotrans}).

Another intimate link between \ow superconductivity and Majorana fermions has recently been further explored \cite{lee_arxiv_16, kashuba_arxiv_16}. Lee \etal~showed that by coupling $s$-wave superconductors to spin-orbit coupled semiconducting wires, \ow superconductivity was induced in the wires and provided a paramagnetic Meissner effect \cite{lee_arxiv_16}, similarly to the system considered in \cite{espedal_prl_16}. Kashuba \etal~proposed to use an STM-tip with a Majorana bound-state at the tip as a probe for \ow superconductivity in materials. The reasoning behind this idea is that, as noted in \cite{huang_prb_15}, the Majorana bound-state is the smallest unit that by itself shows \ow pairing due its particle-antiparticle equality. Therefore, a supercurrent can only flow between the Majorana STM-tip and the material being probed if \ow superconductivity is present in the material itself. The authors applied this idea to the tunneling problem between a Majorana STM and a quantum dot coupled to a conventional superconductor, as shown in Fig. \ref{fig:majoranastm}. By applying an external field, the effective superconducting pairing in the quantum dot can be tuned between \ew and \ow with a resulting clear signature provided in the STM-tunneling spectra.

\begin{figure}[h!]
\includegraphics[scale=0.55]{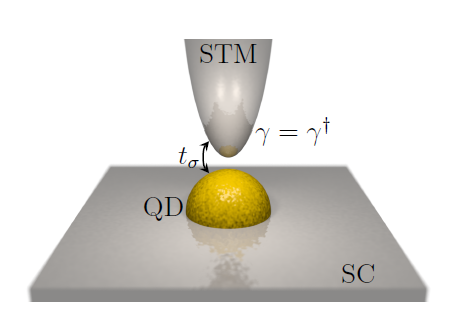}
\caption[fig]{\label{fig:majoranastm} (Color online) Schematic usage of the Majorana STM. The tip contains a Majorana bound state $\gamma$ which probes \ow superconductivity in the quantum dot (QD) via a tunnel coupling. Superconductivity exists in the QD via proximity to a host $s$-wave superconducting material, and the pairing can be tuned between \ew and \ow via an external magnetic field. Figure adapted from \cite{kashuba_arxiv_16}. }
\label{fig:majoranastm}
\end{figure}


\subsection{Berezinskii pairing in non-superconducting systems}\label{sec:nonsc}

There is a priori no reason to expect that the Berezinskii states are confined to only superconducting states. Hence the exploration of other \ow Berezinskii state is only natural.  In this chapter we review the work that takes a broader view on \ow state and goes beyond superconductivity.
There is a good motivation to work on non-superconducting \ow states. This effort, while small in scale, has a potential to open connections to {\em hidden} orders. Namely, the orders where conventional equal time correlations vanish and one has to expand the search to allow for composite or strongly time dependent correlations.

\subsubsection{Ultracold Fermi gases}\label{sec:fermigas}

Whereas electrons comprise the Cooper pairs in superconductors, the superfluid state in fermionic cold atom systems exhibits conceptually the same type of pairing between atoms. This means that all previously discussed symmetry classifications of the pairing correlation functions in this review carry over to the cold atom case. A particularly interesting scenario occurs if not only fermions are present, but if instead a binary mixture of bosonic and fermionic cold atoms coexist. In such a case, one might expect the standard fermion pairing mediated by the phonon field of the boson gas to take place and send the system into a superfluid phase. However, it turns out that Berezinskii  pairing shows up in this context as well, underscoring the ubiquity of this type of order in a wide variety of systems.

The possibility of realizing \ow superfluidity in a boson-fermion mixture of cold atoms, experimentally possible to achieve in atomic traps, was discussed in \cite{kalas_prb_08}. Due to interactinos with the phonon excitations in the bosonic subsystem, the fermionic atoms were shown to exhibit \ow pairing at low temperatures if the coupling $\gamma$ between the fermions and phonons exceeded a threshold value $\gamma_c$. Starting out with a Hamiltonian density describing the fermion-boson mixture:
\begin{align}
H = H_B^0 + H_F^0 + \frac{\lambda_\text{BB}}{2}|\psi_B^\dag\psi_B|^2 + \lambda_\text{BF}\psi^\dag_B\psi_B\psi_F^\dag\psi_F
\end{align}
where $H^0_{B,F}$ are the Hamiltonians for non-interacting bosons and fermions whereas $\lambda_{BB}$ and $\lambda_{BF}$ are the boson-boson and boson-fermion coupling constants. Direct fermion coupling was neglected in \cite{kalas_prb_08} by assuming a magnetic trap with fully spin-polarized fermions.

\begin{figure}[h!]
\includegraphics[scale=0.66]{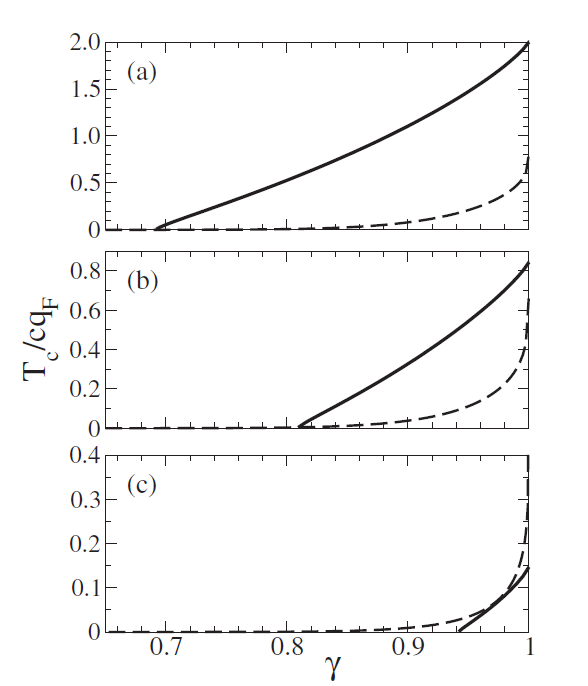}
\caption[fig]{\label{fig:ultracold} (Color online) Critical temperature $T_c$ vs. the scaled fermion-phonon coupling $\gamma$. Solid line: critical temperature for $s$-wave \ow pairing. Dashed line: critical temperature for $p$-wave pairing. We have defined $c_S$ as the phonon speed of sound and $\xi=\sqrt{\xi_0^2 + \gamma/12q_F^2}$ where $\xi_0$ is the boson coherence length. Figure adapted from \cite{kalas_prb_08}. }
\end{figure}

As usual, the onset of a pairing instability is accompanied by a non-zero anomalous correlation function $f(\omega_n,\boldsymbolq)$ which is related to the normal Green function $g(\omega_n,\boldsymbolq)$ via a linearized selfconsistency equation derived within the Eliashberg formalism:
\begin{align}
&g^{-1}(\omega_n,\boldsymbolq)g^{-1}(-\omega_n,\boldsymbolq) f(\omega_n,\boldsymbolq) = T_\text{temp}\sum_{\omega_{n'},\boldsymbolq'} f(\omega_{n'},\boldsymbolq')\notag\\
&\times\frac{\lambda_{BF}^2}{2}[D(\omega_n-\omega_{n'},\boldsymbolq-\boldsymbolq') - D(\omega_n+\omega_{n'},\boldsymbolq+\boldsymbolq')],
\end{align}
where $\omega_n=\pi T_\text{temp}(2n+1)$ is the Matsubara frequency and $D$ is the renormalized phonon propagator. A key observation is that the above equation does not permit standard $s$-wave equal-time pairing, due to the effective spinless nature of the fermions in the system under consideration. The renormalized propagators $g$ and $D$ can be obtained via the Dyson equation and the resulting critical temperatures $T_c$ for the $s$-wave \ow  and $p$-wave superfluid states, respectively, as a function of the scaled fermion-phonon coupling parameter $\gamma \equiv \lambda_{BF}^2 q_F^2/(2\pi^2\lambda_{BB} v_F)$ is shown in Fig. \ref{fig:ultracold}, where $q_F$ and $v_F$ is the Fermi momentum and Fermi velocity. As seen, the \ow superfluid transition is possible above a critical strength $\gamma_c$ for the scaled fermion-phonon coupling, which turns out to be close to the coupling strength at which the mixture phase-separates \cite{kalas_prb_08}.

\subsubsection{Bose-Einstein condensates}\label{sec:bec}

Up to now, we have treated \ow pairing between fermions in the context of superconductors and superfluids. Such states are characterized by a finite expectation for two-fermion correlation functions of the type $\langle c c\rangle \neq 0$ where $c$ describes the annihilation of a fermion. In contrast, the superfluid ground-state in Bose-Einstein condensates is characterized by a finite expectation value for a single particle boson field $b$, so that $\langle b \rangle \neq 0$. In the steady-state, no time dependence needs to be invoked.

However, there are other scenarios where the single particle expectation value is zero at the same time as there exists a non-trivial ordering in the system. Spin-nematics, to be treated in more detail in the next section, is an example of this where
\begin{align}
\langle \boldsymbolS (\boldsymbolr) \rangle = 0
\end{align}
while the two-spin correlator:
\begin{align}\label{eq:spinnematic}
\langle S^i(\boldsymbolr)S^j(\boldsymbolr')\rangle = Q^{ij}(\boldsymbolr,\boldsymbolr') = Q(n^in^j - \delta_{ij}/3)
\end{align}
is finite and describes a non-trivial spin texture via the nematic vector $\boldsymboln$. If we now generalize Eq. (\ref{eq:spinnematic}) to include the time-coordinate as well, it is possible to obtain even- and odd-time magnetic correlations in the spin system, analogously to \ow pairing.

Based on this reasoning, it should in principle be possible to introduce an \ow two-particle Bose-Einstein condensate as proposed in \cite{balatsky_arxiv_14}. Consider the correlation function
\begin{align}
D_{ab}(\boldsymbolr-\boldsymbolr',\tau-\tau') = \mathcal{T}_\tau \langle b_a(\boldsymbolr,\tau) b_b(\boldsymbolr',\tau')\rangle
\end{align}
as relevant for a translationally invariant, equilibrium state with no center-of-mass space or time dependence. We have attached an index $a$ to the boson-operators to characterize their quantum state, encompassing \eg spin, orbital index, or band. If the system is such that
\begin{align}
\langle b_a(\boldsymbolr,\tau=0) = 0
\end{align}
while at the same time
\begin{align}
D_{ab}(\boldsymbolr-\boldsymbolr',\tau-\tau') \neq 0
\end{align}
we have established a situation where there is no single-particle condensate, whereas there still exists a non-trivial boson condensate ($D_{ab} \neq 0$). This condensate consists of pairs of bosons which have an \ow symmetry if $D_{ab}$ is an odd function of time, meaning $D_{ab}(\tau-\tau') = -D_{ab}(\tau'-\tau)$.

A symmetry classification for the possible two-boson condensates described by the correlator $D_{ab}$ differs from the fermionic case treated earlier in this review, since Bose statistics dictates that
\begin{align}
D_{ab}(\boldsymbolr,\tau) = D_{ba}(-\boldsymbolr,-\tau).
\end{align}
It is useful to draw upon the operators $P,T,O$ introduced previously in this review. Let $a$ denote orbital index for concreteness and define the orbital permutation as $OD_{ab} = D_{ba}$. The difference between bosons and fermions is reflected in a new $SPOT = +1$ rule {\em for bosons}.  The complete list of possible non-trivial condensates with $D_{ab} \neq 0$ is summarized in Tab. \ref{tab:bosesym}.

\begin{center}
\begin{table}[]

\caption{Symmetry properties of the two-boson correlator $D_{ab}$ under the operators $PTO$. The \ow states are those where $TD_{ab} = - D_{ab}$. Adapted from \cite{balatsky_arxiv_14}. \\}

\label{tab:bosesym}

\begin{tabular}{cccc}

\hline

$P$ & $T$ & $O$ & Total \\

\hline

+1 & +1 & +1 & +1\\

+1 & -1 & -1 & +1\\

-1 & +1 & -1 & +1\\

-1 & -1 & +1 & +1\\

\end{tabular}

\end{table}
\end{center}

If the condensate has an \ow symmetry, it follows that the equal-time correlator must vanish so that $D_{ab}(\boldsymbolr,0)=0$. In that case, there is no finite expectation value for either single- or two-particle correlation functions. In order to define an order parameter for the \ow two-particle Bose-Einstein condensate which exists at equal-times, one possibility is to use the time derivative of $D_{ab}$. For small enough $\tau$, we can write $D_{ab}(\boldsymbolr,\tau) = d_{ab}(\boldsymbolr)\tau$ so that
\begin{align}
d_{ab}(\boldsymbolr) = \partial_\tau D_{ab}(\boldsymbolr,\tau)|_{\tau=0}
\end{align}
serves as a \textit{bona fide} order parameter for the condensate.

The question is nevertheless if it is possible to realize experimentally such an \ow two-boson Bose-Einstein condensate. A main challenge is the composite nature of such a condensed state and the fact that a single-particle condensate should not simultaneously exist. One possibility could nevertheless be to use a Bose-Einstein condensate proximity effect, where the presence of a medium with additional low energy excitations could dress the bosons in the conventional Bose-Einstein condensate via tunneling and possibly develop an \ow component. A fully microscopic model supporting a two-boson Bose-Einstein condensate as its ground-state remains an open problem.

\subsubsection{Chiral spin-nematic}\label{sec:chiral}

It is also possible to introduce a magnetic analogue of \ow superconducting order \cite{balatsky_prl_95}. The generalization of \ow ordering to a spin system requires consideration of the symmetry equation describing the dynamic correlation function for the spin density $S_i(\boldsymbolr,t)$ ($i=1,2,3$). The spin-spin correlation function may be written as
\begin{align}
\Lambda_{ij}(\boldsymbolr,\boldsymbolr',t) = \mathcal{T}_t \langle S_i(\boldsymbolr,t)S_j(\boldsymbolr',0)\rangle
\end{align}
where $t$ as before is the relative time coordinate between the spin operators. Due to the properties of the time-ordering operator $\mathcal{T}_t$ alone, it follows that
\begin{align}
\Lambda_{ij}(\boldsymbolr,\boldsymbolr',t) = \Lambda_{ji}(\boldsymbolr',\boldsymbolr,-t)
\end{align}
which is valid for any rank spin $S$. At a mathematical level, this establishes the possibility to have \ow magnetic states characterized by a spin-spin correlation function that is an odd function of the relative time $t$. Interestingly, not only is the chiral spin liquid state recovered as one classifies magnetic states that have \ow magnetic correlations, but a new state is predicted as well which is the odd-in-time analogue of a spin nematic state. Similarly to the spin nematic state, first considered in \cite{andreev_jetp_84}, the new state has nematic ordering in spin space but additionally breaks time inversion and parity symmetry. This state was dubbed a chiral spin nematic in \cite{balatsky_prl_95}.

A spin nematic state displays spontaneous breaking of the O(3) spin rotation group without any average microscopic expectation value of a single-spin operator, i.e. $\langle S_i(\boldsymbolr,t)\rangle=0$. As in the Bose-Einstein case considered in Sec. \ref{sec:bec} and also in the case of \ow charge- and spin-density waves discussed in \cite{pivovarov_prb_01}, a possible choice for \ow order parameter is the time-derivative evaluated at the relative time $t=0$:
\begin{align}\label{eq:chiral1}
\partial_t \Lambda_{ij}(\boldsymbolr,\boldsymbolr',t)|_{t=0} = \mathcal{T}_t \langle \partial_t S_i(\boldsymbolr,t) S_j(\boldsymbolr',0)\rangle_{t=0}.
\end{align}
The equation of motion for the spin operator $S_i$ takes the form:
\begin{align}\label{eq:chiral2}
\partial_t S_i(\boldsymbolr,t) = \i[H,S_i(\boldsymbolr,t)] = \epsilon_{ijk} S_j(\boldsymbolr,t)M_k(\boldsymbolr,t).
\end{align}
where the quantity $M_k(\boldsymbolr,t)$ can be thought of as the molecular field for the Hamiltonian $H$ of the system. In the event that $H$ is bilinear in the spin operators, the general form of $M_k$ is:
\begin{align}\label{eq:chiral3}
M_k(\boldsymbolr) = \int d\boldsymbolr' K_{kn}(\boldsymbolr,\boldsymbolr')S_n(\boldsymbolr'),
\end{align}
where the kernel $K_{kn}$ explicitly depends on the two coordinates $\boldsymbolr$ and $\boldsymbolr'$. In particular, assuming that the Hamiltonian can be generally written as:
\begin{align}
H = -\sum_{mn} \int d\boldsymbolr d\boldsymbolr' S_m(\boldsymbolr) L_{mn}(\boldsymbolr,\boldsymbolr') S_n(\boldsymbolr'),
\end{align}
the kernel takes the form $K_{kn}(\boldsymbolr,\boldsymbolr') = 2L_{kn}(\boldsymbolr,\boldsymbolr')$. A key observation at this stage is that a contribution from the kernel to the time derivative of the \ow correlator [via Eqs. (\ref{eq:chiral1})-(\ref{eq:chiral3})] only occurs if $K(\boldsymbolr,\boldsymbolr')$ contains a spatially odd component, i.e. antisymmetric under exchange of $\boldsymbolr$ and $\boldsymbolr'$. This places severe constraints on the type of possible spin exchange models that can support an \ow spin-nematic state. An example of such a $H$ is nevertheless
\begin{align}
H = -\frac{\alpha}{2} \sum_{\langle i,j\rangle} [\boldsymbolS_1\times \boldsymbolS_2\cdot\boldsymbolS_3]_{P_i}[\boldsymbolS_4 \times \boldsymbolS_5\cdot \boldsymbolS_6]_{P_j}
\end{align}
where $\alpha>0$. The sum is taken over nearest neighbor plaquettes $P_i$ (containing spins 1,2,3) and $P_j$ (containing spins 4,5,6) on a triangular lattice. This particular Hamiltonian has a chiral spin-liquid ground state.

The new chiral spin-nematic state that arises when accounting for an \ow spin-spin correlation function $\Lambda_{ij}$ that is also odd under a parity transformation $\boldsymbolr\leftrightarrow\boldsymbolr'$. One possible way in which to generate this state could be to consider the quadrupolar interaction in the chiral spin-liquid state \cite{balatsky_prl_95}. The relation between a spin-nematic order parameter and an \ow spin-density wave state was discussed in \cite{pivovarov_prb_01}.






\section{Conclusions and outlook}\label{sec:conclusions}

We close this review by offering a perspective on directions that in our opinion will be important for further progress in the field of \ow Berezinskii superconductivity. As we discussed, \ow states can be spontaneously  generated in the bulk or can be induced  in the heterostructures as a result of scattering of conventional  Cooper pairs. The guiding principle here is the SPOT constraint that, together with the Table I and 2, predicts the possible pathways to induce \ow state. Most of the literature on  how to generate \ow states falls into these two broad categories.   We expect interesting future developments in the field of \ow states both in the case of bulk states and in heterostructures.

On the fundamental physics side, perhaps the most interesting  question is if a bulk \ow superconducting state can be realized experimentally and, if so, what the underlying microscopic mechanism for such a state is. The debate regarding the thermodynamical stability of a bulk \ow superconducting state has, as has been disseminated in this review, been intense. At present, there is no consensus on the Meissner stiffness of \ow Berezinskii superconductors. On the one hand, early works \cite{coleman_prl_93, abrahams_prb_95} concluded that stability requires a
staggered composite order while later works \cite{belitz_prb_99, solenov_prb_09, kusunose_jpsj_11b} concluded that a thermodynamically stable \ow superconducting bulk state featuring a diamagnetic Meissner effect is in principle possible even without a staggered order parameter. On the other hand, Fominov \etal \cite{fominov_prb_15} claimed that a realization of a diamagnetic \ow Berezinskii state implies the absence of a mean-field Hamiltonian description of such a system. These two viewpoints have yet to be reconciled.

Although it is too early to claim that a general consensus has been reached, particularly in view of \cite{fominov_prb_15}, several works on the topic do conclude that a thermodynamically stable \ow superconducting bulk state featuring a diamagnetic Meissner effect is possible. However, it is unclear what microscopic Hamiltonian would support this state.

We also reviewed a rapidly growing list of the \ow Berezinskii components induced in a bulk superconductor either due to multiband effects \cite{blackschaffer_prb_13b}, \eg in Sr$_2$RuO$_4$ \cite{komendova_prl_17} or MgB$_2$ \cite{aperis_prb_15},  due to interfacial coupling with the topological states, e.g. \cite{blackschaffer_prb_12} and due to conventional dc Josephson effect  between two conventional superconductors. The work on the induction of \ow components in the {\em bulk} of superconductors only recently started and this direction of research is likely to continue to grow.

A qualitatively new approach to generate Berezinskii states dynamically has emerged recently. The inherent dynamic nature of the \ow Berezinskii state, where the internal time dependence of the pair correlation should be kept explicitly, in hindsight, was always pointing to its origin as a dynamic order \cite{triola_prb_16, triola_prb_17}. The view that the Berezinskii state is a dynamic order offers a possible connection to the ongoing discussion on time crystals \cite{wilczek_prl_12, zhang_nature_17, choi_nature_17}. We hope that this intriguing connection will be further explored. In that regard, the dynamic Rabi-like oscillations revealed in the \ow channel in conventional Josephson junction are suggestive, as discussed in Sec. \ref{sec:josephson}.

It is clear that the concept of \ow pairing has implications that reach well beyond superconductivity. As we have discussed, \ow pairing may well lie at the root of different types of order which do require considering non-local correlations in time, whether these are correlations in the spin, charge, or another type of channel. One example is the extension of the Berezinskii pairing to the case of  Majorana fermions \cite{huang_prb_15, gnezdilov_prb_19}. We  discussed the early stages of an understanding of how an \ow state in a Majorana system is realized in a collection of Majorana fermions. In principle, the question about the proof of principle that an \ow state can be realized in the bulk is thus answered. The setup required to produce this state in collection of Majorana fermions is a complicated one, but once we attain the many-body Majorana state we can see that \ow correlations are expected in the ground state.

We believe the heterostructures and applications of \ow states to spintronics will remain an active area.  Existence of \ow pairing is by now well established both theoretically and experimentally in hybrid structures. Therefore, it is possible to turn the gaze toward possible applicational aspects of this type of superconductivity. In other words, can \ow superconductivity offer a new type of functionality which conventional BCS superconductivity cannot, for instance in superconducting electronics? In this regard, the prospect of utilizing \ow spin-polarized Cooper pairs in diffusive heterostructures has garnered the most attention so far \cite{linder_nphys_15}. In fact, such Cooper pairs demonstrate a resilience both toward the Pauli limiting field and impurity scattering simultaneously, in contrast to conventional Cooper pairs which only are robust toward impurity scattering according to Anderson's theorem. The fact that \ow triplet superconductivity is so robust makes it an attractive candidate for possible applications involving the merging of magnetic and superconducting elements. Therefore, this direction will continue to stimulate further experiments  towards  practical utilization of
spin-polarized \ow Berezinskii Cooper pairs in spintronics devices.

 An unique feature of \ow pairing  aside from being a  novel pairing state is in creating previously unattainable synergy between magnetic and superconducting materials which pertains specifically to the frequency-symmetry and not the spin-polarization of the Cooper pairs. This is the paramagnetic Meissner response that \ow Cooper pairs can feature. The recent experimental demonstration \cite{dibernardo_prx_15} of an inverted electromagnetic response in a Au/Ho/Nb trilayer open interesting perspectives for new paths in the utilization
of hybrid systems comprising magnets and superconductors. These devices defy the conventional paradigm where a magnetic field is viewed as exclusively harmful for superconducting order.


If the study of \ow superconductivity over the last decades has demonstrated anything, it is that it occurs ubiquitously. At the same time, an intrinsic odd-frequency superconducting condensate has yet to be realized and its discovery remains to this date as one of the main aspirations in this field. More often than not, any system with a superconducting component will feature some form of \ow pairing. This fact points toward the importance of considering other symmetry-allowed temporal correlations, albeit unconventional, guided by the $SPOT$ constraint, in different settings beyond superconductivity. Allowing for non-trivial dynamic correlations will lead to  outcomes that  could be surprising. We have strived to give a sense of some future directions of development in the field that we foresee. At the same time we hope there are new and unexpected ideas and experiments  that will propel the field  of \ow states further. We believe that the outlook for research on \ow pairing, in superconducting systems and otherwise, is brimming with exciting possibilities and new physics to be discovered.


\section{Acknowledgments}

We are indebted to several colleagues and friends for discussions, and would like to thank E. Abrahams, M. Amundsen, N. Banerjee, S. Bergeret, A. Black-Schaffer, M. Blamire,  J. Bonca, P. Coleman, H. Dahal, A. Di Bernardo, K. Efetov,  V. Emery, M. Eschrig, Y. Fominov, M. Geilhufe, K. Halterman, S. Jacobsen, Y. Kedem, S. Kivelson, A. Krikun, E. Langman,  O. Millo, V. Mineev, D. Mozyrski, C. Nayak, J. A. Ouassou, S. Pershoguba, V. Risingg{\aa}rd, J. Robinson, J.R. Schrieffer, D. Scalapino, A. Sudb{\o}, Y. Tanaka, C. Triola, A. Tsvelik, G. Volovik, and T. Yokoyama. A.V.B. is particularly grateful to E. Abrahams who introduced the concept of Berezinskii pairing and to M. Geilhufe for help with writing the section on symmetry classification of the \ow states. We are grateful to  A. Black-Schaffer for carefully reading through the manuscript and a helpful feedback.

J.L. acknowledges funding via the Outstanding Academic
Fellows program at NTNU, the NV-Faculty, and the Research
Council of Norway Grant number 240806. This work was partly supported by the Research Council of Norway through its Centres of Excellence funding scheme, project number 262633, “QuSpin”. A.V.B. acknowledges financial support from US DOE BES E3B7, ERC Synergy HERO (810451),  KAW 2013-0096, KAW 2018-0104 and  VILLUM FONDEN via the Centre of Excellence for Dirac Materials (Grant No. 11744).

\appendix

\section{List of symbols}

\begin{tabular}{cp{0.6\textwidth}}
  $S$ & Spin permutation operator\\
	$P$ & Spatial    parity operator\\
$P^{*}$ & Spatial  permutation   operator\\
	$O$ & Orbital index permutation operator\\
  $T$ & Time-reversal operator \\
  $T^{*}$ & Time- permutation operator\\
  $\omega, \omega_n$ & Fermionic Matsubara frequency \\
	$\Omega, \Omega_n$ & Bosonic Matsubara frequency \\
  $\Gamma_\text{hyb}. \Gamma_\vk$ & Hybridization parameter \\
  $T_\text{temp}$ & Temperature \\
	$T_c$ & Critical temperature \\
	$T_\text{int}$ & Tunneling interface transparency \\
	$T_\text{tun}$ & Tunneling matrix element \\
	$a,b,\ldots \text{(subscript)}$ & Orbital and band indices \\
	$\alpha,\beta,\ldots \text{(subscript)} $ & Spin indices \\
	$\psi, c$ & Fermion operators \\
	$\vk,\vp,\vq$ & Momenta \\
	$S(\vecr)$ & Spin operators \\
	$G$ & Normal Green function (propagator) \\
	$f,F$ & Anomalous Green function (propagator) \\
	$\hat{g}, \underline{g}, \underline{f}$ & Quasiclassical Green functions\\
	$T_\text{cm}$ & Center of mass time \\
	$\vecR$ & Center of mass coordinate \\
  $t$ & Time coordinate \\
	$\vecr$ & Spatial coordinate \\
	$L$ & Angular momentum \\
	$\mathcal{T}$ & Time-ordering operator \\
	$E$ & Quasiparticle energy \\
	$\beta$ & Inverse temperature \\
	$\Delta$ & Superconducting order parameter \\
	$\boldsymbol{\sigma}$ & Vector of Pauli-matrices \\
	$\hat{\sigma}^j,\hat{\sigma}_j$ & Pauli-matrix $j$ \\
	$\boldsymbol{d}(\vk)$ & Triplet $d$-vector \\
	$g$ & Coupling constant \\
	$N_F$ & Fermi level density of states \\
	$\chi$ & Susceptibility \\
	$S_\text{spin}$ & Spin quantum number \\
	$P_\text{parity}$ & Parity eigenvalue \\
	$\Gamma$ & Interband scattering \\
	$\vecj$ & Electric current \\
	$\vecA$ & Magnetic vector potential \\
	$\varphi$ & Superconducting phase \\
	$\epsilon_\vk, \epsilon(\vk), \xi_\vk$ & Normal-state electron dispersion \\
	$V_{\vk,\vk'}, V(\vk,\vk')$ & Pairing interaction\\
	$E_F$ & Fermi energy \\
	$\Sigma$ & Self-energy \\
	\end{tabular}
	\begin{tabular}{cp{0.6\textwidth}}
	$Z$ & Dimensionless barrier strength \\
	$\xi,\xi_S$ & Superconducting coherence length \\
	$D$ & Diffusion coefficient \\
	$E_\text{th}$ & Thouless energy \\
	$h$ & Exchange energy (magnetic)\\
	$\vecm,\vecM$ & Magnetization vector \\
	$l_\text{mfp}$ & Mean free path\\
	$\mu$ & Chemical potential \\
	$\tau$ & Matsubara-time \\
	$\Theta(t)$ & Heaviside step-function \\

\end{tabular}

\bibliography{oddrmp_revised}

\end{document}